# Electron transmission through molecules and molecular interfaces

Abraham Nitzan

School of Chemistry, the Sackler Faculty of Science, Tel Aviv University,
Tel Aviv, 69978, Israel

## Table of Contents

# Electron transmission through molecules and molecular interfaces

Abraham Nitzan
School of Chemistry, the Sackler Faculty of Science, Tel Aviv University,
Tel Aviv, 69978, Israel

## Abstract

Electron transmission through molecules and molecular interfaces has been a subject of intensive research due to recent interest in electron transfer phenomena underlying the operation of the scanning tunneling microscope (STM) on one hand, and in the transmission properties of molecular bridges between conducting leads on the other. In these processes the traditional molecular view of electron transfer between donor and acceptor species give rise to a novel view of the molecule as a current carrying conductor, and observables such as electron transfer rates and yields are replaced by the conductivities, or more generally by current-voltage relationships, in molecular junctions. Such investigations of electrical junctions, in which single molecules or small molecular assemblies operate as conductors constitutes a major part of what has become the active field of molecular electronics.

In this paper I review the current knowledge and understanding of this field, with particular emphasis on theoretical issues. Different approaches to computing the conduction properties of molecules and molecular assemblies are reviewed, and the relationships between them are discussed. Following a detailed discussion of static junctions models, a review of our current understanding of the role played by inelastic processes, dephasing and thermal relaxation effects, is provided. The most important molecular environment for electron transfer and transmission is water, and our current theoretical understanding of electron transmission through water layers is reviewed. Finally, a brief discussion of overbarrier transmission, exemplified by photoemission through adsorbed molecular layers or low energy electron transmission through such layers is provided. Similarities and differences between the different systems studied are discussed.

# I. Introduction

Electron transfer, a fundamental chemical process underlying all redox reactions, has been under experimental and theoretical study for many years.[1-6] Theoretical studies of such processes seek to understand the ways in which their rate depends on donor and acceptor properties, on the solvent and on the electronic coupling between the states involved. The different roles played by these factors and the way they affect qualitative and quantitative aspects of the electron transfer process have been thoroughly discussed in the past half-century. This kind of processes, which dominates electron transitions in molecular systems, are to be contrasted with electron transport in the solid state, i.e. in metals and semiconductors. Electrochemical reactions, which involve both molecular and solid state donor/acceptor systems, bridge the gap between these phenomena.[6] Here electron transfer takes place between quasi-free electronic states on one side and bound molecular electronic states on the other.

The focus of the present discussion is another class of electron transfer phenomena: electron transmission between two regions of free or quasi-free electrons through molecules and molecular layers. Examples for such processes are photoemission (PE) through molecular overlayers, the inverse process of low energy electron transmission (LEET) into metals through adsorbed molecular layers and electron transfer between metal and/or semiconductor contacts through molecular spacers. Figure 1 depicts a schematic view of such systems. the 'standard' electron transfer model in Fig. 1a shows donor and acceptor sites connected by a molecular bridge. In Fig. 1b the donor and the acceptor are replaced by continua of electronic states representing free space or metal electrodes. (This replacement can occur on one side only, representing electron transfer between a molecular site and an electrode). In Fig. 1c the molecular bridge is replaced by a molecular layer. A schematic view of the electronic states involved is shown in Fig. 2. The middle box represents the bridging molecule or molecular layer and a set of levels $\{n\}$ represents the relevant molecular orbitals. In a 'standard' electron transfer system (Fig. 2a) this bridge connects the donor and acceptor species, now represented by potential surfaces associated with the vibronic structure of the corresponding inramolecular and solvent nuclear motions. When the bridge connects two metal electrodes (or separates a metal substrate from vacuum) these nuclear baths are replaced by manifolds of electronic states $\{\ell\}$ and $\{r\}$ that represent continua of free or quasi-free electron states in the substrates (or, depending on the process, in vacuum). In addition, coupling to the thermal environment (represented by the box $\Theta$) may affect transmission through the bridge. The double arrows in the Figure represent the couplings between these different subsystems.

The first two of the examples given above, PE and LEET, involve electrons of positive energy (relative to zero kinetic energy in vacuum), and as such are related to normal scattering processes. The third example, transmission between two conductors through a molecular layer, involves negative energy electrons and as such is closely related to regular electron transfer phenomena. The latter type of processes has drawn particular attention in recent years due to the growing interest in conduction properties of individual molecules and of molecular assemblies. Such processes have become subjects of intensive research due to recent interest in electron transfer phenomena underlying the operation of the scanning tunneling microscope (STM) on one hand, and in the transmission properties of molecular bridges between conducting leads on the other. In the latter case the traditional molecular view of electron transfer between donor and acceptor species give rise to a novel view of the molecule as a current carrying conductor, and observables such as electron transfer rates and yields are replaced by the conductivities, or more generally by current-voltage relationships, in molecular junctions. Of primary importance is the need to understand the interrelationship between the molecular structure of such junctions and their function, i.e. their transmission and conduction properties. Such investigations of electrical junctions, in which single molecules or small molecular assemblies operate as conductors connecting 'traditional' electrical components such as metal or semiconductor contacts, constitute a major part of what has become the active field of molecular electronics.[7-17] Their diversity, versatility and amenability to control and manipulation make molecules and molecular assemblies potentially important components in nano-electronic devices. Indeed basic properties pertaining to single electron transistor behavior and to current rectification have already been demonstrated. At the same time major difficulties lie on the way to real technological applications.[18] These difficulties stem from problems associated with the need to construct, characterize, control and manipulate small molecular structures at confined interfaces with a high degree of reliability and reproducibility, and from issues of stability of such small junctions.

It should be obvious that while the different processes outlined above correspond to different experimental setups, fundamentally they are controlled by similar physical factors. Broadly speaking we may distinguish between processes for which *lifetimes* or *rates* (more generally the time evolution) are the main observables and those that monitor *fluxes* or *currents*. In this review we focus on the second class, which may be further divided into processes that measure current-voltage relationships, mostly near equilibrium, and those that monitor the non-equilibrium electron flux, e.g. in photoemission experiments.

*Notations.* A problem characteristic to an interdisciplinary field such as the one we are covering is that notations that became standard in particular disciplines overlap similarly standard notations of other disciplines. The T operator of scattering theory and the temperature constitute one example; the β parameter of of bridge mediated electron transfer theory and the inverse (temperature×Boltzmann constant) is another. I have therefore used non-standard notations for some variables in order to avoid confusion. Following is a list of the main notations used in this article.

| Notation | Variable |
|---|---|
| $T$ | Scattering operator |
| $\mathcal{T}$ | Transmission coefficient |
| $\Theta$ | Temperature |
| $\beta$ | $(k_B \Theta)^{-1}$ |
| $\beta'$ | Range parameter in electron transfer rate theory |
| $g$ | Conduction |
| $\sigma$ | Used in different contexts for conductivity and for the reduced system's density operator. |
| $I$ | Current |
| $J$ | Flux |
| $\rho$ | Used in different contexts for charge density and for the density operator of the total system |
| $E_F$ | Fermi energy ($E_{FL}$ and $E_{FR}$ sometimes used for 'left' and 'right' electrodes) |
| $\mu$ | Electron electrochemical potential ($\mu_L$ and $\mu_R$ sometimes used for 'left' and 'right' electrodes) |
| $\mathcal{F}$ | the thermally averaged and Franck Condon (FC) weighted density of nuclear states |
| $F$ | System-thermal bath interaction. In specific cases we also use $H_{el-ph}$ |
| $V$ | Electronic coupling between zero order molecular states |
| $H$ | System's Hamiltonian |
| $H_B$ | Bridge Hamiltonian |
| $H_\Theta$ | Hamiltonian of the thermal bath (in some specific cases we also use $H_{ph}$ |
| $Z$ | Overlap Matrix: $Z_{i,j}=\langle i|j\rangle$ |
| $\mathbb{H}$ | $EZ - H$ |
| $\mathcal{H}$ | Combined system+thermal bath Hamiltonian |
| $S$ | S matrix |
| $\upsilon$ | Speed |
| $\Phi$ | Potential or potential difference |
| $\Sigma$ | Self energy |
| $\Gamma$ | Width (decay rate) |
| **Acronyms** | |
| MMM | Metal-Molecule-Metal (junction) |
| MIM | Metal-Insulator-Metal (junction) |
| EH | Extended Huckel |
| HF | Hartree Fock |
| FC | Franck-Condon |
| STM | Scanning tunneling microscope |
| LEET | Low energy electron transmission |
| PE | Photo-emission |

## 2. Theoretical approaches to molecular conduction

The focus of this section is electron transfer between two conducting electrodes through a molecular medium. Such processes bear strong similarity to the more conventional systems that involve at least one molecular species in the donor/acceptor pair. Still, important conceptual issues arise from the fact that such systems can be studied as part of complete electrical circuits, providing current-voltage characteristics that can be analyzed in terms of molecular resistance, conductance and capacitance.

2.1. *Standard electron transfer theory*

To set the stage for our later discussion we first briefly review the rate expressions for 'standard' electron transfer processes (Figs. 1a, 2a, 3a). We focus on the particular limit of non-adiabatic electron transfer, where the electron transfer rate is given (under the Condon approximation) by the golden rule based expression

$$k_{et} = \frac{2\pi}{\hbar} |V_{DA}|^2 \mathcal{F} \tag{1}$$

where $V_{DA}$ is the coupling between the donor (D) and acceptor (A) electronic states and where

$$\mathcal{F} = \mathcal{F}(E_{AD}) = \sum_{\nu_D}\sum_{\nu_A} P_{th}\left(\varepsilon_D(\nu_D)\right)|\langle \nu_D | \nu_A \rangle|^2 \delta\left(\varepsilon_A(\nu_A) - \varepsilon_D(\nu_D) + E_{AD}\right) \tag{2}$$

is the thermally averaged and Franck Condon (FC) weighted density of nuclear states. In Eq. (2) $\nu_D$ and $\nu_A$ denote donor and acceptor nuclear states, $P_{th}$ is the Boltzamnn distribution over donor states, $\varepsilon_D(\nu_D)$ and $\varepsilon_A(\nu_A)$ are nuclear energies above the corresponding electronic origin and $E_{AD} = E_A - E_D$ is the electronic energy gap between the donor and acceptor states. In the classical limit $\mathcal{F}$ is given by

$$\mathcal{F}(E_{AD}) = \frac{e^{-(\lambda+E_{AD})^2/4\lambda k_B \Theta}}{\sqrt{4\pi\lambda k_B \Theta}} \tag{3}$$

where $k_B$ is the Boltzmann constant and $\Theta$ is the temperature, and where $\lambda$ is the reorganization energy, a measure of the electronic energy that would be dissipated after a sudden jump from the electronic state describing an electron on the donor to that associated with electron on the acceptor. If the donor (say) is replaced by an electrode,[6, 19, 20] we have to sum over all occupied electrode states

$$|V_{DA}|^2 \mathcal{F} \Rightarrow \sum_k f(\varepsilon_k)\mathcal{F}(\varepsilon_k - e\Phi)|V_{kA}|^2 = \int d\varepsilon f(\varepsilon)\mathcal{F}(\varepsilon - e\Phi)\sum_k \delta(\varepsilon - \varepsilon_k)|V_{kA}|^2 \tag{4}$$

where

$$f(\varepsilon) = \frac{1}{1+e^{\varepsilon/k_B\Theta}} \tag{5}$$

is the Fermi-Dirac distribution function with $\varepsilon$ measured relative to the electron chemical potential $\mu$ in the electrode, and $\Phi$, which determines the position of the acceptor level relative to $\mu$ is the *overpotential.* Defining

$$\sum_k \delta(\varepsilon - \varepsilon_k)|V_{kA}|^2 \equiv |V(\varepsilon)|^2 \tag{6}$$

the eletcron transfer rate takes the form

$$k_{et} = \frac{2\pi}{\hbar}\int d\varepsilon \frac{e^{-(\lambda-e\Phi+\varepsilon)^2/4\lambda k_B\Theta}}{\sqrt{4\pi\lambda k_B\Theta}}|V(\varepsilon)|^2 f(\varepsilon) \tag{7}$$

Note that the reorganization energy that appears in Eq. (7) is associated with the change in the redox state of the molecular species only. The nominal change in the 'oxidation state' of the macroscopic electrode does not affect the polarization state of the surrounding solvent because the transferred electron or hole do not stay localized.

Much of the early work on electron transfer have used expressions like (3) and (7) with the electronic coupling term $V_{DA}$ used as a fitting parameter. More recent work has focused on ways to characterize the dependence of this term on the electronic structure of the donor/acceptor pair and on the environment. In particular, studies of bridge mediated electron transfer, where the donor and acceptor species are rigidly separated by molecular bridges of well defined structure and geometry have been very valuable for characterizing the interrelationship between structure and functionality of the separating environment in electron transfer processes. As expected for a tunneling process, the rate is found to decrease exponentially with the donor-acceptor distance

$$k_{et} = k_0 e^{-\beta' R_{DA}} \qquad (8)$$

where $\beta'$ is the range parameter that characterizes the distance dependence of the electron transfer rate. The smallest values for $\beta'$ are found in highly conjugated organic bridges for which $\beta'$ is in the range 0.2–0.6Å$^{-1}$[21-32]. In contrast, for free space, taking a characteristic ionization barrier $U_B = 5eV$ we find $\beta' = \sqrt{8mU_B/\hbar^2} \approx 2.4 \text{Å}^{-1}$ ($m$ is the electron mass) Lying between these two regimes are many motifs, both synthetic and natural, including cytochromes and docked proteins,[33-41] DNA,[42-50] and saturated organic molecules.[51-57] Each displays its own characteristic range of $\beta'$ values, and hence its own timescales and distance dependencies of electron transfer. A direct measurement of $\beta'$ along a single molecular chain was recently demonstrated.[58]

In addition to bridge assisted transfer between donor and acceptor species, electron transfer has been studied in system where the spacer is a well characterized Langmuir-Blodgett film.[59-61] The scanning tunneling microscope provides a natural apparatus for such studies.[58, 62-76] Other approaches include break junctions[77-79] and mercury drop contacts.[80-84]

Simple theoretical modeling of $V_{DA}$ usually relies on a single electron (or hole) picture in which the donor-bridge-acceptor (DBA) system is represented by a set of levels: $|D>, |A>, \{|1>,...|N>\}$ as depicted in Fig. 3. In the absence of coupling of these bridge states to the thermal environment, and when the energies $E_n$ ($n=1,...,N$) are high relative to the energy of the transmitted electron (the donor/acceptor orbital energies in Figs. 1a, 2a and 3a or the incident electron energy in Figs. 1b-c, 2b and 3b), this is the super-exchange model for electron transfer. [85] Of particular interest are situations where {n} are localized in space, so that the state index n corresponds to position in space between the donor and acceptor sites (Fig. 3a) or between the two electron reservoirs (Fig. 3b). These figures depict generic tight binding models of this type, where the states n=1,...,N are the bridge states, here taken degenerate in zero order. Their localized nature makes it possible to assume only nearest neighbor coupling between them, i.e., $V_{n,n'} = V_{n,n\pm 1}\delta_{n',n\pm 1}$. We recall that the appearance of $V_{DA}$ in Eq. (1) is a low-order perturbation theory result. A more general expression is obtained by replacing $V_{DA}$ by $T_{DA}$ where the T operator is defined by $T(E) = V + VG(E)V$, with $G(E) = (E - H + (1/2)i\Gamma)^{-1}$ and where $\Gamma$ stands for the inverse lifetime matrix of bridge levels. Assuming for simplicity that the donor level |D> is coupled only to bridge state |1> and that the acceptor level |A> is coupled only to bridge level N, the effective coupling between donor and acceptor is given by

$$T_{DA}(E) = V_{DA} + V_{D1}G_{1N}(E)V_{NA} \tag{9}$$

This naturally represents the transition amplitude as a sum of a direct contribution, $V_{DA}$, which is usually disregarded for long bridges, and a bridge mediated contribution. In using $T_{DA}$ instead of $V_{DA}$ in Eq. (1) the energy parameter E in (9) should be taken equal to $E_D=E_A$ at the point where the corresponding potential surfaces cross (or go through an avoided crossing). For the level structure of Fig. 3a that corresponds to the DBA system in Fig. 1a, making the tight binding approximation and in the weak coupling limit, $\max|V| \ll \min(E_B - E)$,[a] the Green's function element in (9) is given by

$$G_{1N}(E) = \frac{1}{E - E_N}\prod_{n=1}^{N-1}\frac{V_{n,n+1}}{E - E_n} \tag{10}$$

For a model with identical bridge segments $E_n$ and $V_{n,n+1}$ are independent of *n* and will be denoted $E_n=E_B$ and $V_{n,n+1}=V_B$. Using this in Eq. (1) leads to

$$k_{et} = \frac{2\pi}{\hbar}\left|\frac{V_{1D}V_{NA}}{V_B}\right|^2 \left(\frac{V_B}{\Delta E_B}\right)^{2N} \mathcal{F} \tag{11}$$

where $\Delta E_B = E_B - E$. Similarly, for a bridge assisted transfer between a molecule and an electrode, Eq. (7) applies with $|V(\varepsilon)|^2$ given by

$$|V(\varepsilon)|^2 = \left(\frac{V_B}{\Delta E_B}\right)^{2N}\sum_k \delta(\varepsilon-\varepsilon_k)\left|\frac{V_{1k}V_{NA}}{V_B}\right|^2 \tag{12}$$

These results imply a simple form for the distance parameter $\beta$ of Eq. (8)

---

[a] For a generalization of Eq. (10) that does not assume weak coupling see Ref. [86] and [87]. See also [88].

$$\beta' = \frac{2}{a} \ln\left(\frac{\Delta E_B}{V_B}\right) \qquad (13)$$

where $a$ measures the segment size, so that the bridge length is $Na$. The exponential dependence on the bridge length is a manifestation of the tunneling character of this process. For typical values, e.g. $\Delta E_B / V_B = 10$ and $a$=5Å, Eq. (13) gives $\beta'$=0.92Å$^{-1}$. More rigorous estimates of the electronic coupling term in electron transfer processes involve electronic structure calculation for the full DBA system. Such calculations, in the context of molecular conduction, will be discussed later.

2.2. *Transmission between conducting leads*

Eqs. (1), (7) and (11) are expressions for the *rate* of electron transfer between donor and acceptor molecules or between a molecule and a metal electrode. As already mentioned, for electron transfer in metal-molecule-metal (MMM) junctions, the primary observable is the *current-voltage characteristics* of the system. Putting another way, while the primary observable in 'standard' charge transfer processes involving molecular donors and/or acceptors is a *transient* quantity,[b] in MMM junctions we focus on the *steady state* current through the junction for a given voltage difference between the two metal ends.

Consider first a simple model for a metal-insulator-metal (MIM) system, where the insulator is represented by a continuum characterized by a dielectric constant $\varepsilon$.[89] For specificity assume that the electrode surfaces are infinite parallel planes perpendicular to the x direction. In this case the transmission problem is essentially 1-dimensional and depends only on the incident particle velocity in the x direction, $v_x = \sqrt{2E_x/m}$. In the WKB approximation the transmission probability is given by

$$\mathcal{T}(E_x) = \exp\left[-\frac{4\pi}{\hbar} \int_{s_1}^{s_2} \left[2m(U_B(x) - E_x)\right]^{1/2} dx\right] \qquad (14)$$

where $U_B(x)$ is the barrier potential that determine the turning points $s_1$ and $s_2$ and m is the mass of the tunneling particle. The tunneling flux is given by $\mathcal{T}(E_x) n(E_x) \sqrt{2E_x/m}$, where $n(E_x)$ is the density per unit volume of electrons of energy $E_x$ in the x direction. $n(E_x)$ is obtained by integrating the Fermi-Dirac function with respect to $E_y$ and $E_z$. When a potential $\Phi$ is applied so that the right electrode (say) is positively biased, the net current density is obtained in the form[89]

$$J = \int_0^\infty dE_x \mathcal{T}(E_x)\xi(E_x) \qquad (15)$$

where

$$\xi(E_x) = \frac{2m^2 e}{(2\pi\hbar)^3} \int_{-\infty}^\infty dv_y \int_{-\infty}^\infty dv_z \left[f(E) - f(E+e\Phi)\right] = \frac{4\pi me}{(2\pi\hbar)^3} \int_0^\infty dE_r \left[f(E) - f(E+e\Phi)\right] \qquad (16)$$

and where $E_r = E - E_x = (1/2)m(v_y^2 + v_z^2)$ is the energy in the direction perpendicular to $x$. In obtaining this result it is assumed that the electrodes are chemically identical. At zero temperature and when $\Phi \to 0$, $f(E) - f(E+e\Phi) = e\Phi \delta(E - E_F)$. Eqs. (15) and (16) then lead to an expression for the conduction per unit area, i.e. the conductivity per unit length

$$\sigma_x = \frac{4\pi me^2}{(2\pi\hbar)^3} \int_0^{E_F} dE_x \mathcal{T}(E_x) \qquad (17)$$

---

[b] In addition to *rates*, other observables are the *yields* of different products of the electron transfer reaction. Furthermore, for light induced electron transfer processes, the steady state current under a constant illumination can be monitored.

For finite $\Phi$ these expressions provide a framework for predicting the current-voltage characteristics of the junction; explicit approximate expressions were given by Simmons.[89] Here we only emphasize,[89] that the dependence on $\Phi$ arises partly from the structure of Eqs. (15) and (16), for example, at zero temperature

$$J = \frac{4\pi m^2 e^2}{(2\pi\hbar^3)}\left[ e\Phi \int_0^{E_F - e\Phi} dE_x T(E_x) + \int_{E_F - e\Phi}^{E_F} dE_x (E_F - E_x)\mathcal{T}(E_x) \right], \qquad (18)$$

but mainly from the voltage dependence of $\mathcal{T}$. The simplest model for a metal-vacuum-metal barrier between identical electrodes without an external field is a rectangular barrier of height above the Fermi energy given by the metal workfunction. When a uniform electric field is imposed between the two metals a linear potential drop from $E_F$ on one electrode to $E_F$-$e\Phi$ on the other is often assumed (see fig. 4). In addition, the image potential experienced by the electron between the two metals will considerably modify the potential barrier. For a point charge $e$, located at position $x$ between two conducting parallel plates that are a distance $d$ apart, the image potential is

$$V_I = \left(-\frac{e^2}{4\pi\varepsilon}\right)\left[\frac{1}{2x} + \sum_{n=1}^\infty \left\{\frac{nd}{\left[(nd)^2 - x^2\right]} - \frac{1}{nd}\right\}\right] \qquad (19)$$

where ε is the dielectric constant of the spacer. For *x=d/2* this becomes

$$V_I = -\frac{e^2 \ln 2}{2\pi\varepsilon d} \qquad (20)$$

This negative contribution to the electron's energy reduces the potential barrier (Fig. 4), and has been invoked to explain the lower than expected barrier observed in STM experiments.[90, 91] Some points should however be kept in mind. First, the classical result (19) fails close to the metal surface where quantum mechanical and atomic size effects change both the position of the reference image plane and the functional form of the image potential.[92-96] Second, consideration of the dynamic nature of the image response should be part of a complete theory.[97-99] [100(a)] The timescale of electronic response of metals can be roughly estimated from the plasma frequency to be $\sim 10^{-16}$s. This should be compared to the time during which a tunneling particle can respond to interactions localized in the barrier. For transmitted particles this is the traversal time for tunneling[101, 102] (see Section 3.1) that, for an electron traversing a 10Å wide 1eV barrier is of the order of ~1fs. This comparison would justify the use of the static image approximation in this context, but this approximation becomes questionable for deeper tunneling or narrower barriers.

The planar geometry implied by the assumption that transmission depends only on the energy of the motion parallel to the tunneling direction, as well as the explicit form of Eq. (14) are not valid for a typical STM configuration that involves a tip on one side and a structured surface on the other. To account for these structures Tersoff and Hamman[103] have applied the Bardeen's formalism[104] which is a perturbative approach to tunneling in arbitrary geometries. The Bardeen's formula for the tunneling current is[c]

$$I = \frac{4\pi e}{\hbar} \sum_{l,r} \left[ f(E_l)(1 - f(E_r + e\Phi)) - (1 - f(E_l))f(E_r + e\Phi) \right] |M_{lr}|^2 \delta(E_l - E_r) =$$
$$\frac{2\pi e}{\hbar} \sum_{l,r} \left[ f(E_l) - f(E_r + e\Phi) \right] |M_{lr}|^2 \delta(E_l - E_r) \qquad (21)$$

where

$$M_{lr} = \frac{\hbar^2}{2m} \int d\vec{S} \cdot \left( \psi_l^* \nabla \psi_r - \psi_l \nabla \psi_r^* \right) \qquad (22)$$

---

[c] This is just the Golden rule rate expression (multiplied by the electron charge e), with M playing the role of coupling. In Ref.[103] only the first term in the square brackets of the first line appears. This gives the partial current from the negative to the positive electrode. The net current is obtained by subtracting the reverse current as shown in Eq. (21). Also, compared with Ref[103], Eq. (21) contains an additional factor of 2 that accounts for the spin multiplicity of the electronic states.

is the transition matrix element for the tunneling process. In these equations $\psi_\ell$ and $\psi_r$ are electronic eigenstates of the negatively biased (left) and positively biased (right) electrodes, respectively, $\Phi$ is the bias potential and the integral is over any surface separating the two electrodes and lying entirely in the barrier region. The wavefunctions appearing in Eq. (22) are eigenfunctions of Hamiltonians that describe each electrode in the absence of the other, i.e interfaced with an infinite spacer medium. These functions therefore decay exponentially in the space between the two electrodes in a way that reflects the geometry and chemical nature of the electrodes and the spacer. For $\Phi \to 0$ Eq. (21) yields the conduction

$$g \equiv \frac{I}{\Phi} = \frac{4\pi e^2}{\hbar} \sum_{l,r} |M_{lr}|^2 \delta(E_l - E_F)\delta(E_r - E_F) \qquad (23)$$

Tersoff and Hamman[103] have used substrate wavefunctions that correspond to a corrugated surface of a generic metal while the tip is represented by a spherical s orbital centered about the center $\mathbf{r}_0$ of the tip curvature. In this case they find

$$I \propto \sum_{\nu} |\psi_\nu(\mathbf{r}_0)|^2 \delta(E_\nu - E_F) \qquad (24)$$

The r.h.s. of (24) is the local density of states of the metal. While this result is useful for analysis of spatial variation of the tunneling current on a given metal surface, the contributions from the coupling matrix elements in (23) can not be disregarded when comparing different metals and or different adsorbates.[20]

2.3. *The Landauer Formula*

The results (14)-(17) and (21)-(23) are special cases of a more systematic representation of the conduction and the current-voltage characteristic of a given junction due to Landauer.[105,106] Landauer's original result was obtained for a system of two 1-dimensional leads connecting two macroscopic electrodes ('electron reservoirs') via a scattering object or a barrier characterized by a transmission function $\mathcal{T}(E)$. The zero temperature conductance, measured as the limit $\Phi \to 0$ of the ratio $I/\Phi$ between the current and the voltage drop between the reservoirs, was found to be[d]

---

[d] The corresponding resitance, $g^{-1}$, can be represented as a sum of the intrinsic resistance of the scatterer itself, $\left[(e^2/\pi\hbar)(\mathcal{T}/(1-\mathcal{T}))\right]^{-1}$ and a contribution $\left(e^2/\pi\hbar\right)^{-1}$ from two contact resistances between the leads and the reservoirs. See Chapter 5 of [107] for a discussion of this point.

$$g = \frac{e^2}{\pi\hbar}\mathcal{T}(E_F) \qquad (25)$$

This result is obtained by computing the total unidirectional current carried in an ideal lead by electrons in the energy range $(0,E)=(0,\hbar^2 k_E^2/(2m))$. In a 1-dimensional system of length $L$ the density of electrons, including spin, with wavevectors in the range between $k$ and $k+dk$ is $n(k)dk = 2(1/L)(L/2\pi)f(E_k)dk = f(E_k)dk/\pi$. The corresponding velocity is $v = \hbar k/m$. Thus

$$I(E) = e\int_0^{k_E} dk\,v(k)n(k) = e\int_0^{k_E} dk\,(\hbar k/m)f(E_k)/\pi = \frac{e}{\pi\hbar}\int_0^E dE'\,f(E') \qquad (26)$$

At zero temperature, the net current carried under bias $\Phi$ is

$$I = \frac{e}{\pi\hbar}\int_0^\infty dE\,(f(E)-f(E+e\Phi)) \xrightarrow{\Theta\to 0} \frac{e^2}{\pi\hbar}\Phi \qquad (27)$$

Thus the conductance of an ideal 1-dimensional lead is $I/\Phi = e^2/\pi\hbar = (12.9 K\Omega)^{-1}$. In the presence of the scatterer this is replaced by

$$I = \frac{e}{\pi\hbar}\int_0^\infty dE\,\mathcal{T}(E)(f(E)-f(E+e\Phi)) \xrightarrow{\Theta\to 0,\Phi\to 0} \frac{e^2}{\pi\hbar}\mathcal{T}(E_F)\Phi \qquad (28)$$

which leads to (25). This result is vallid for 1-dimensional leads. When the leads have finite size in the direction normal to the propagation so that they support traversal modes, a generalization of (25) to this case yields[108e]

$$g = \frac{e^2}{\pi\hbar}\sum_{i,j}\mathcal{T}_{ij}(E_F) = \frac{e^2}{\pi\hbar}\mathbf{Tr}(SS^\dagger)_{E_F} \qquad (29)$$

where $\mathcal{T}_{ij}=|S_{ij}|^2$ is the probability that a carrier coming from the left (say) of the scatterer in transversal mode $i$ will be transmitted to the right into transversal mode $j$ ($S_{ij}$, an element of the S matrix, is the corresponding amplitude). The sum in (29) is over all traversal modes whose energy is smaller than $E_F$. More generally, the current for a voltage difference $\Phi$ between the electrodes is given by

$$I = \int_0^\infty dE\,[f(E)-f(E+e\Phi)]\frac{g(E)}{e} \qquad (30)$$

---

[e] The analog of Eq. (29) for the microcanonical chemical reaction rate was was first written by Miller[109]. Similarly, Eq. (34) was first written in a similar context in Ref. [110].

$$g(E) = \frac{e^2}{\pi \hbar} \sum_{i,j} \mathcal{T}_{ij}(E) \tag{31}$$

As an example consider the case of a simple planar tunnel junction (see Eqs. (14)-(17)), where the scattering process does not couple different transversal modes. In this case the transmission function depends only on the energy in the tunneling direction

$$\sum_{i,j} \mathcal{T}_{ij}(E) = \sum_i \mathcal{T}_{ii}(E) = \frac{L_y L_z}{(2\pi)^2} \int dk_y \int dk_z \mathcal{T}\left[E - \left(\hbar^2/2m\right)\left(k_y^2 + k_z^2\right)\right] =$$
$$= \frac{L_y L_z}{(2\pi)^2} \frac{2\pi m}{\hbar^2} \int_0^E dE_r \mathcal{T}(E - E_r) \tag{32}$$

$E_r$ is defined below Eq. (16). Using this in Eq. (29) yields the conductivity per unit length

$$\frac{g}{L_y L_z} \equiv \sigma = \frac{4\pi m e^2}{h^3} \int_0^{E_F} dE_x \mathcal{T}(E_x) \tag{33}$$

in agreement with Eq. (17).

Similarly, Eqs. (21) and (23) are easily seen to be equivalent to (25) or (31) if we identify $M_{lr}$ with $T_{lr}$ in Eq. (37) below. An important difference between the results (29) and (31) and results based on the Bardeen's formalizm, Eqs. (21)-(23), is that the former are valid for any set of transmission probabilities, even close to 1, while the latter yields a weak coupling result. Another important conceptual difference is the fact that the sums over $\ell$ and $r$ in Eqs.(21)-(23) is over zero order states defined in the initial and final subspaces, while the sums in Eqs. (29)-(31) is over scattering states, i.e. eigenstates of the exact system's Hamiltonian. It is the essesnce of Bardeen's contribution[104] that in the weak coupling limit (i.e. high/wide barrier) it is possible to write the transmission coefficient $\mathcal{T}_{ij}$ in terms of a golden rule expression for the transition probability between the zero order standing wave states $|l>$ and $|r>$ localized on the left and right electrodes, thus establishing the link between the two representations. (For an alternative formulation of this link see Ref. [111])

To explore this connection on a more formal basis, we can replace the expression based on transmission coefficients $\mathcal{T}$ by an equivalent expression based on scattering amplitudes, or T matrix elements, between zero order states localized on the electrodes. This can be derived directly from Eqs. (29) or (31) by using the identity

$$\sum_{i,j} \mathcal{T}_{ij}(E) = 4\pi^2 \sum_{l,r} |T_{lr}|^2 \delta(E - E_l)\delta(E - E_r) \tag{34}$$

On the left side of (34) a pair of indices (*i,j*) denote an exact scattering state of energy *E*, characterized by an incoming state *i* on the left (say) electrode and an outgoing state *j* on the right electrode. On the right, *l* and *r* denote zero order states confined to the left and right electrodes, respectively. *T* is the corresponding transition operator whose particular form depends on the details of this confinement. Alternatively we can start from the golden-rule-like expression

$$I = e\frac{4\pi}{\hbar}\sum_{l,r}\left[f(E_l)(1-f(E_r+e\Phi))-f(E_r+e\Phi)(1-f(E_l))\right]|T_{lr}|^2\,\delta(E_l-E_r) =$$
$$= \frac{4\pi e}{\hbar}\sum_{l,r}\left[f(E_l)-f(E_r+e\Phi)\right]|T_{lr}|^2\,\delta(E_l-E_r) \quad (35)$$

(An additional factor of 2 on the r.h.s. accounts for the spin degeneracy). It is convenient to recast this result in the form

$$I = \frac{4\pi e}{\hbar}\int_0^\infty dE\left[f(E)-f(E+e\Phi)\right]\sum_{l,r}|T_{lr}|^2\,\delta(E-E_l)\delta(E-E_r) =$$
$$= \int_0^\infty dE\left[f(E)-f(E+e\Phi)\right]\frac{g(E)}{e} \quad (36)$$

where

$$g(E) \equiv \frac{4\pi e^2}{\hbar}\sum_{l,r}|T_{lr}|^2\,\delta(E-E_l)\delta(E-E_r) \quad (37)$$

Note that Eqs. (34) and (37) imply again Eq. (31). For $\Phi\to 0$ Eqs. (36) and (37) lead to $I = g\Phi$ with

$$g = g(E_F) \quad (38)$$

The analogy of this derivation to the result (23) is evident.

2.4. *Molecular conduction*

Eqs. (36)-(38) provide a convenient starting point for most treatments of currents through molecular junctions where the coupling between the two metal electrodes is weak. In this case it is convenient to write the system's Hamiltonian as the sum, $H = H_0 + V$, of a part $H_0$ that represents the uncoupled electrodes and spacer and the coupling *V* between them. In the weak coupling limit the T operator

$$T(E) = V + VG(E)V \;;\quad G(E) = (E-H+i\varepsilon)^{-1} \quad (39)$$

is usually replaced by its second term only. The first 'direct' term *V* can be disregarded if we assume that *V* couples the states $\ell$ and *r* only via states of the molecular spacer. Consider now a simple model where this spacer is an N-site bridge connecting the two electrodes so that site 1 of the bridge is attached to the left electrodes and site N - to the right electrode, a variant of Fig. 3b. In this case we have $T_{lr} = V_{l1}G_{1N}V_{Nr}$, so that at zero temperature[112, 113]

$$\sum_{i,j} \mathcal{T}_{ij}(E) = |G_{1N}(E_F)|^2 \, \Gamma_1^{(L)}(E_F)\Gamma_N^{(R)}(E_F) \tag{40}$$

and (using Eqs. (36) and (37))

$$I(\Phi) = \frac{e}{\pi\hbar} \int_{E_F - e\Phi}^{E_F} dE \, |G_{1N}(E,\Phi)|^2 \, \Gamma_1^{(L)}(E)\Gamma_N^{(R)}(E + e\Phi) \tag{41}$$

with

$$\Gamma_1^{(L)}(E) = 2\pi \sum_l |V_{l1}|^2 \, \delta(E_l - E) \quad ; \quad \Gamma_N^{(R)}(E) = 2\pi \sum_r |V_{Nr}|^2 \, \delta(E_N - E) \tag{42}$$

The Green's function in Eq. (40) is itself reduced to the bridge's subspace by projecting out the metals' degrees of freedom. This results in a renormalization of the bridge Hamiltonian: in the bridge subspace

$$(E - H + i\eta)^{-1} \to (E - H_B - \Sigma_B(E))^{-1} \tag{43}$$

where $H_B = H_B^0 + V_B$ is the Hamiltonian of the isolated bridge entity with

$$H_B^0 = \sum_{n=1}^{N} E_n \, |n\rangle\langle n| \quad ; \quad V_B = \sum_{n=1}^{N}\sum_{n'=1}^{N} V_{n,n'} \, |n\rangle\langle n'| \tag{44}$$

and where in the basis of eigenstates of $H_B^0$

$$\Sigma_{nn'}(E) = \delta_{n,n'} \left(\delta_{n,1} + \delta_{n,N}\right) \left[\Lambda_n(E) - (1/2)i\Gamma_n(E)\right] \tag{45}$$

$$\Gamma_n(E) = 2\pi \sum_j |V_{nj}|^2 \, \delta(E - E_j) \tag{46}$$

$$\Lambda_n(E) = \frac{PP}{2\pi} \int_{-\infty}^{\infty} dE' \frac{\Gamma_n(E')}{(E - E')} \tag{47}$$

In Eq. (46) the sum is over both the right and the left manifolds, i.e., $j$ goes over all states $\{l\}$ and $\{r\}$ in these manifolds) so that $\Gamma_n = \Gamma_n^{(L)} + \Gamma_n^{(R)}$; $n = 1, N$. The transmission problem is thus reduced to evaluating a Green's function matrix element and two width parameters. The first calculation is a simple inversion of a finite (order N) matrix. The width $\Gamma$ and and the associated shift $\Lambda$, represent the finite lifetime of an electron on a molecule adsorbed on the metal surface, and can be estimated, for example,[113] using the Newns-Anderson model of chemisorption.[114] In the simple tight binding model of the bridge and in the weak coupling limit, $G_{1N}$ is given by Eq. (10) modified by the inclusions of the self energy terms

$$G_{1N}(E) = \frac{V_{1,2}}{(E - E_1 - \Sigma_1(E))(E - E_N - \Sigma_N(E))} \prod_{j=2}^{N-1} \frac{V_{j,j+1}}{E - E_j} \qquad (48)$$

Eqs. (40)-(48) thus provide a complete simple model for molecular conduction, equivalent to similar approximations used in theories of molecular electron transfer.(e.g. [115, 116] and references therein) For applications of variants of this formalism to electron transport in specific systems see Refs. [86, 87, 117, 118]. Below we discuss more general forms of this formulation.

2.5. *Relation to electron transfer rates*

It is interesting to examine the relationship between the conduction of a molecular species and the electron transfer properties of the same species.[f] We should keep in mind that because of tunneling there is always an Ohmic regime near zero bias, with conduction given by the Landauer formula. Obviously this conduction may be extremely low, indicating in practice an insulating behavior. Of particular interest is to estimate the electron transfer rate in a given donor-bridge-acceptor (DBA) system that will translate into a measurable conduction of the same system when used as a molecular conductor between two metal leads. To this end consider a DBA system, with a bridge that consists of N identical segments (denoted 1,2,...N) with nearest neighbor coupling $V_B$. The electron transfer rate is given by Eq. (11) that we rewrite in the form

$$k_{D \to A} = \frac{2\pi}{\hbar} |V_{D1} V_{NA}|^2 |G_{1N}(E_D)|^2 \mathcal{F} \qquad (49)$$

where, in the weak coupling limit, $|V_B| \ll |E_B - E|$ (cf. Eq. (10))

$$G_{1N}(E) = \frac{|V_B|^{N-1}}{(E_B - E)^N} \qquad (50)$$

and where $\mathcal{F}$ is the Franck-Condon-weighted density of nuclear states, given in the classical limit by Eq. (3). The appearance of $\mathcal{F}$ in Eq. (49) indicates that the process is dominated by the change in the nuclear configuration between the two localization states of the electron. Suppose now that the same DBA complex is used to connect between two metal contacts such that the donor and acceptor species are chemisorbed on the two metals (denoted 'left' and 'right' respectively). We wish to calculate the conduction of this junction and its relation to $k_{D \to A}$. First note that the conduction process does not involve localized states of the electron on the donor or the acceptor, so the factor $\mathcal{F}$ will be absent. (We will disregard for the moment energy loss arising from transient distortions of the nuclear configuration associated with transient populations of electronic states of the DBA complex). Assuming as before that states of the molecular complex are coupled to the metal only via the D and A orbitals, and that the latter are coupled only to their adjacent metal contacts, the conduction is given by an equation similar to (40), except that the bridge (1,...,N) is replaced by the complex DBA=(D,1,...,N,A)

$$g(E) = \frac{e^2}{\pi \hbar} |G_{DA}(E)|^2 \, \Gamma_D^{(L)}(E) \Gamma_A^{(R)}(E) \tag{51}$$

where, in analogy to Eq. (48)

---

[f] Nitzan A. To be published. Such an estimate was given before in Ref.[119], but the procedure given there is limited to a 1-dimensional model, and has disregarded the Franck-Condon factor in the electron transfer rate. The procedure outlined here is more general.

$$G_{DA}(E) = \frac{V_{D1} V_{NA}}{(E - E_D - \Sigma_D(E))(E - E_A - \Sigma_A(E))} G_{1N}(E) \tag{52}$$

Since the donor and acceptor species are chemisorbed on their corresponding metal contacts, their energies shift closer to the Fermi energies. We assume that this shift occurs uniformly in the DBA complex, without distorting its internal electronic structure (strictly speaking this can happen only in the symmetric case of identical donor and acceptors and identical metal electrodes, but the result of Eq. (53) below is probably a good approximation for more general cases because $G_{1N}(E)$ is often not strongly dependent on E). Assuming therefore that the denominator in (52) is dominated by the imaginary parts of the self energies $\Sigma$, we get

$$g = g(E_F) = \frac{16 e^2}{\pi \hbar} \frac{|V_{D1} V_{NA}|^2}{\Gamma_D^{(L)}(E_F) \Gamma_A^{(R)}(E_F)} |G_{1N}(E_F)|^2 \; ; \; E_F = E_D = E_A \tag{53}$$

Comparing to Eq. (49) we get

$$g = \frac{e^2}{\pi \hbar} \frac{k_{D \to A}}{\mathcal{F}} \frac{8 \hbar}{\pi \Gamma_D^{(L)} \Gamma_A^{(R)}} \tag{54}$$

It has been argued that, provided the energy spacing $E_B - E_F$ between the bridge levels and the Fermi energy is large relative to $k_B T$, Eq. (54) holds also when the electron transfer process involves thermal activation into the bridge states (and not only for the bridge assisted tunneling implied by Eq. (53)).[120] Using the classical expression for $\mathcal{F}$, Eq. (3), we have in the present case $\mathcal{F} = \left(\sqrt{4\pi \lambda k_B T}\right)^{-1} \exp(-\lambda / 4 k_B T)$. For a typical value of the reorganization energy ~0.5eV, and at room temperature this is ~ $0.02(\text{eV})^{-1}$. Taking also $\Gamma_D^{(L)} = \Gamma_A^{(R)}$ ~0.5eV leads to $g \sim \left(e^2/\pi\hbar\right)\left(10^{-13}k_{D\to A}(s^{-1})\right) \cong \left[10^{-17}k_{D\to A}(s^{-1})\right]\Omega^{-1}$. This sets a criterion for observing Ohmic behavior for small voltage bias in molecular junctions: With a current detector sensitive to pico-amperes, $k_{D\to A}$ has to exceed $\sim 10^6 s^{-1}$ (for the estimates of $\mathcal{F}$ and $\Gamma$ given above) before measurable current can be observed at 0.1V voltage across such a junction.

## 2.6. *Quantum chemical calculations*

The simple models discussed above are useful for qualitative understanding of molecular conductivity, however the Landauer formula or equivalent formulations can be used as a basis for more rigorous molecular calculations using extended Huckel calculations[70-72, 79, 121-135] or Hartree Fock[136-139]. These approaches follow similar semiempirical and ab-initio calculations of electron transfer rates in molecular systems,[140] however instead of focusing on the computation of the electronic coupling $V_{DA}$ needed in Eqs. (1), the sum in Eq. (34) is calculated directly. Structural stability considerations suggest that useful metal-molecule-metal bridges should involve strong chemisorption bonding between the molecule and the metal substrate, implying large electronic coupling between them.[141] It is therefore preferable to use a 'supermolecule' approach, in which the quantum chemical calculations are carried for a species that comprises the molecule and two clusters of metal atoms, so that the reduction that introduces the self energy $\Sigma$ (Eq. (43)) is done at some deeper metal-metal contact. Such atomic level calculations usually start from a (non-orthogonal) basis set of atomic orbitals, so the formalism described above has to be generalized for this situation.[g] We also relax the assumption that the molecule-metal contact is represented by coupling to a single molecular orbital. Defining the operator

$$\mathbb{H}(E) = EZ - H \qquad \text{with } Z_{ij} = <i|j> \tag{55}$$

the Green's function is $G(E)=\mathbb{H}(E)^{-1}$. In Eq. (55), $i$ and $j$ denote atomic orbitals that may be assigned to the supermolecule (M), the left metal (L) and the right metal (R) subspaces. Denoting formally the coupling between the subspace $M$ and the subspaces $K=L,R$ by the corresponding submatrices $\mathbb{H}_{MK}$, the Green's function for the supermolecule subspace is

$$G^{(M)}(E) = \left(\mathbb{H} - \Sigma^{(L)} - \Sigma^{(R)}\right)^{-1} \tag{56}$$

---

[g] Alternatively, it has been shown by Emberly and Kirczenow[142,143] that one can map the problem into a new Hilbert space in which the basis states are orthogonal.

with[h]

$$\Sigma^{(K)} = \mathbb{H}_{MK}(\mathbb{H}^{-1})_{KK}\mathbb{H}_{KM} \qquad (57)$$

Using also

$$T_{lr} = \sum_{n,n'} \mathbb{H}_{ln} G_{nn'} \mathbb{H}_{n'r} \qquad (58)$$

(*l* and *r* in the metal L and R subspaces, respectively; *n,n'* in the supermolecule subspace) in Eq. (37) leads to

$$g(E) = \frac{e^2}{\pi\hbar} Tr\left[ G^{(M)}(E)\Gamma^{(R)}(E)G^{(M)}(E)\Gamma^{(L)} \right] \qquad (59)$$

where, e.g. for the 'left' metal

$$\Gamma_{n,n'}^{(L)} = 2\pi \sum_l \mathbb{H}_{nl}\mathbb{H}_{ln'} \delta(E-E_l) \quad (n \text{ and } n' \text{ in the molecular subspace}) \qquad (60)$$

In practice, $\Sigma$ and $\Gamma = -2\text{Im}(\Sigma)$ can be computed by using closure relations based on the symmetry of the metal lattice.[132] The trace in Eq. (59) is over all basis states in the (super)molecular subspace. The evaluation of the Green's function matrix elements and of this trace is straightforward in semi-empirical single electron representations such as the extended Huckel (EH) approximation, and can be similarly done at the Hartree-Fock (HF) level using, after convergence, the Fock rather then the Hamiltonian matrix in Expressions (55)-(60).[i]

An important attribute of the approach described above is that, within the approximation used, it provides the total current carried by the system, both through the unoccupied molecular levels (electron conduction) and the occupied ones (hole conduction). This results from the fact that the trace in (59) is over all the atomic orbitals that comprise the (super)molecular basis set, that upon diagonalization in the (super)molecular Hamiltonian will yield both occupied and unoccupied molecular orbitals. In a 1-electron theory such as the extended Huckel approximation both types of orbitals contribute in the same way. For example, the terms in Eq. (59) that describe an electron moving from the highest occupied molecular orbital (HOMO) into empty states of the anode, followed by an electron moving from the cathode into the HOMO ("hole transport"), and an electron moving from the cathode to the lowest unoccupied molecular orbital (LUMO) then moving on into the cathode ("electron transport") are similar (their values depend on the energies of the molecular orbitals involved with respects to the Fermi energies), irrespective of whether the corresponding orbitals are occupied or not. The same is true in the HF calculation if the Koopmans' theorem,[144] stating that the HF orbital energies represent the actual energies involved in removing an electron from an occupied orbital or putting an electron into an unoccupied one, holds. The Koopman's theorem is accurate only for large systems, and the approximation involved in applying it to small systems is one reason why HF is not necessarily superior to EH for calculating the conduction properties of small molecular junctions.[j]

---

[h] $\Sigma^{(K)}$ is a matrix in the molecular subspace and Eq. (57) is a compact notation for $\left(\Sigma^{(K)}\right)_{n,n'} = \sum_{k,k'} \mathbb{H}_{nk}(\mathbb{H}^{-1})_{kk'}\mathbb{H}_{k'n'}$ where *k* and *k'* are states in the metal K subspace.

[i] Note that the Fock operator depends on the ground state electronic configuration. The latter is taken in Refs. [136-139] to be that of the isolated supermolecule, assuming that the contact with the bulk electrodes does not affect it appreciably. In particular, the supermolecule is usually assumed neutral in these calculations.

In spite of these limitations, EH and HF based calculations have provided important insight into the conduction properties of molecular junctions. Fig. 5 shows a remarkable example. The (EH) calculation is done for a single $\alpha,\alpha'$-xylyl dithiol molecule adsorbed between two gold contacts. The experiment monitors the current between an STM tip (obtained by cutting a Pt/Ir wire) and a monolayer of such molecules deposited on gold, and it is assumed that lateral interaction between the molecules is unimportant. Two unknown parameters are used for fitting. The first is the position of the metals Fermi energy in the unbiased junction relative to the molecular energy levels expressed by $E_{FH} \equiv E_F - E_{HOMO}$. The second describes the electrostatic potential profile along the junction, represented by a parameter $\eta$ that expresses the distribution of the voltage drop between the two metal leads (see Fig. 6 and Eq. (71) below). As seen in Fig. 5, good agreement between theory and experiment is obtained for $E_{FH}$=0.9eV and $\eta$=0.5.

---

[j] This is true particularly for LUMO dominated conduction, because the HF is notoriously inadequate for electron affinities.[145-148]. See [139] for further discussion of this point. Another potential (but in principle avoidable) problem in these calculations is associated with the finite, relatively small basis of atomic orbitals used. Close to resonance, when the electrode electrochemical potential $\mu = E_F + e\Phi$ approaches the HOMO or LUMO energies, the corresponding HOMO or LUMO orbitals dominate the electron transfer and a small basis that describe correctly these orbitals is sufficient. When $E_F$ is a distance $\Delta E$ away from $E_{HOMO}$ or $E_{LUMO}$, all molecular orbitals in the range $\Delta E$ below $E_{HOMO}$ and in a similar range above $E_{LUMO}$, can contribute to the transmission probability and cannot be ignored, implying the need for a larger molecular basis.[111,129]. We note in passing that the recently discussed transmission antiresonances[142,150] associated with the non-orthogonality of the atomic orbital basis sets, have been shown[111] to be sometimes artifacts of a small basis calculation.

In view of the other unknowns, associated both with the uncertainty about the junction structure and with the simplified computation, the main value of these results should be placed not in the absolute numbers obtained but rather in highlighting the importance of these parameters in determining the junction conduction behavior. We return to the issue of the junction potential profile below. Other qualitative issues that were investigated with these types of calculations include the effect of the nature (length and conjugation) of the molecular bridge,[128, 129] the effect of the molecule-electrode binding and of the molecular binding site,[132] the relation of conductance spectra to molecular electronic structure[79] and the effect of bonding molecular wires in parallel.[130] (See also Ref. [151])

2.7. *Spatial-grid based pseudopotential approaches*

Another way to evaluate the expressions appearing in Eqs. (34) and (37) as well as related partial sums is closely related to the discrete variable representation of reaction probabilities as formulated by Seideman and Miller.[152-154] We have already seen that the sum

$$s(E) \equiv \sum_{l,r} |T_{lr}|^2 \delta(E - E_l) \delta(E - E_r) \tag{61}$$

which is related to the conduction by $g(E) = (4\pi e^2 / \hbar) s(E)$ (c.f Eq. (37)) can be represented by (c.f. Eq. (59))

$$s(E) = \frac{1}{4\pi^2} Tr\left[ G^{(M)}(E) \Gamma^{(R)}(E) G^{(M)*}(E) \Gamma^{(L)} \right] \tag{62}$$

If instead of considering transitions from 'left' to 'right' electrode we think of Eq. (61) as expressing a sum over transition probabilities from all initial (*i*) states of energy $E$ in the reactant space to all final (*f*) states of the same energy in the product space, $s(E)$ is also associated with the so called *cumulative reaction probability*,[152-154] which in terms of the reaction S matrix is defined by $N(E) = \sum_{i,f} |S_{if}(E)|^2 = 4\pi^2 s(E)$, i.e. $N(E) = \sum_{i,f} \mathcal{T}_{if}(E)$. Eq. (62) now expresses the important observation that the cumulative reaction probability for a reactive scattering process can be expressed as a trace over states, *defined in a finite subspace that contains the interaction region*, of an expression that depends on the reduced Green's function and the associated self energy defined in that subspace. Following Seideman and Miller we can use a spatial grid representation for the states in this subspace, so that the trace in (62) becomes a sum over grid points. Also, in this representation the overlap matrix **Z** is zero. In general, any subspace of position space that separate reactants from products (i.e. that encompasses the entire interaction region; the molecular bridge in our application) can be used in (62), provided that the consequences of truncating the "rest of the universe", expressed by the corresponding $\Sigma$ and $\Gamma$ can be computed. The absorbing boundary condition Green's function (ABCGF) method of Seideman and Miller is based on the recognition that if this subspace is taken large enough so that its boundaries are far from the interaction region, the detailed forms of $\Sigma$ and $\Gamma$ are not important; the only requirement is that scattered waves that approach these boundaries will be absorbed and not reflected back into the interaction zone. In the ABCGF method this is accomplished by taking $\Sigma = -(1/2)i\Gamma = -i\varepsilon(\mathbf{r})$, a local function in position space, taken to be zero in the interaction region and gradually increasing from zero when approaching the subspace boundaries. Its particular form is chosen to affect complete absorption of waves approaching the boundary to a good numerical accuracy. Eq. (62) then becomes

$$s(E) = 4Tr\left[G^{ABC}(E)\varepsilon_R G^{ABC*}(E)\varepsilon_L\right] \tag{63}$$

where $G^{ABC}(E) = (E - H + i\varepsilon)^{-1}$; $\varepsilon = \varepsilon_R + \varepsilon_L$ and where $\varepsilon_R$ and $\varepsilon_L$ are different from zero only on grid points near the the right side (more generally the product side) and the left (reactant) side of the inner subspace, respectively.

A similar development can be done for the partial sum

$$s_l(E) \equiv \sum_r |T_{lr}|^2 \delta(E - E_r) \tag{64}$$

which, provided that $l$ is taken as an eigenstate of the Hamiltonian describing the left electrode (or the reactant sunspace), is related the 'one to all' rate, $k_l(E)$ to go from an initial state of energy $E$ on the left electrode (or in reactant space) to all possible states on the right one (product space) according to $k_l = (2\pi/\hbar)s_l$.[k] We use the same definition of the coupling $V$ between our subspace (bridge) and the reactant and product (electrode) states. Putting $T=VGV$ in (64) we get

$$s_l(E) = \frac{1}{2\pi} <l|VG^{(M)}\Gamma^{(R)}G^{(M)*}V|l> \tag{65}$$

---

[k] The "microcanonical rate" is defined by $k(E) = \rho_L^{-1}(E)\sum_l k_l \delta(E - E_l) = \left(2\pi\hbar\rho_L(E)\right)^{-1} 4\pi^2 s(E) = \left(2\pi\hbar\rho_L(E)\right)^{-1} N(E).$

Using again a position grid representation of the intermediate states used to evaluate this expression, and applying the same methodology as above, Eq. (65) can be recast in the form[1]

$$s_l(E) = \frac{1}{\pi} <l|V G^{ABC}(E) \varepsilon_R G^{ABC*}(E) V|l>$$
$$= \frac{1}{\pi} <l|\varepsilon_L G^{ABC}(E) \varepsilon_R G^{ABC*}(E) \varepsilon_L|l> \quad (66)$$

The results (63) and (66) are very useful for computations of transmission probabilities in models where the interaction between the transmitted particle and the molecular spacer is given as a position dependent pseudo-potential. Applications to electron transmission through water and other molecular layers are discussed in Section 4.

## 2.8. *Density functional calculations*

Density functional methods provide a convenient framework for treating metalic interfaces.[100(b)] Applications of this methodology to the problem of electron transport through atomic and molecular bridges have been advanced by several workers. In particular, Lang's approach[90, 155-161] is based on the density functional formalism[162, 163] in which the single electron wavefunctions $\psi_0(r)$ and the electron density $n_0(r)$ for two bare metal (jellium) electrodes is computed, then used in the Lippman-Schwinger equation

$$\psi(\mathbf{r}) = \psi_0(\mathbf{r}) + \int d\mathbf{r}' d\mathbf{r}'' G^0(\mathbf{r},\mathbf{r}') \delta V(\mathbf{r}',\mathbf{r}'') \psi(\mathbf{r}'') \quad (67)$$

to get the full single electron scattering wavefunctions $\psi(r)$ in the presence of the additional bridge. (Lang's earlier calculations[90, 164] use a related density functional approach to calculate the tunneling current between an atomic tip and a jellium electrode without an atomic or molecular bridge). In Eq. (67) $G^0$ is the Green's function of the bare electrode system and $\delta V$ is the difference between the potential of the full system containing an atomic or a molecular spacer and that of the bare electrodes. In atomic units ($|e|,\hbar,m=1$) it is

$$\delta V(\mathbf{r},\mathbf{r}') = V_{ps}(\mathbf{r},\mathbf{r}') + \left[V_{\mathbf{xc}}(n(\mathbf{r})) - V_{\mathbf{xc}}(n_0(\mathbf{r})) + \int d\mathbf{r}'' \frac{\delta n(\mathbf{r}'')}{|\mathbf{r}-\mathbf{r}''|}\right] \quad (68)$$

---

[1] The second part of Eq. (66) is obtained by using the identity $\varepsilon_R|l>=0$ to write $\varepsilon_R G^*V|l> = \varepsilon_R(1+G^*V)|l> = \varepsilon_R G^*(G^{*-1}+V)|l>$, which, together with $G^{*-1} = E-H_0-V+i\varepsilon$, $(E-H_0)|l>=0$ and $\varepsilon|l>=\varepsilon_l|l>$, yields the desired result.

where $V_{ps}$ is the sum of non-local pseudopotentials representing the cores of the spacer atoms and $V_{xc}$ is the LDA of the exchange correlation potential. $n$ is the electron number density for the full system (electrodes and atoms) and $\delta n = n - n_0$. Eq. (67) yields scattering states that can be labeled by their energy $E$, momentum $\mathbf{k}_\parallel$ in the direction (yz) parallel to the electrodes and spin. In addition, Lang distinguishes between wavefunctions that in the electrode regions carry positive (+) or negative (-) momentum in the tunneling direction. Denoting by $\mu_L$ and $\mu_R$ the electron electrochemical potential in the left and right electrode, respectively, the zero temperature electrical current density from left to right (for $\mu_L > \mu_R$) is then

$$J(\mathbf{r}) = -2 \int_{\mu_L}^{\mu_R} dE \int d^2 K_\parallel \, \text{Im}\{\psi_+^* \nabla \psi_+\} \tag{69}$$

The factor 2 accounts for the double occupancy of each orbital. This approach was used Recently[165] to calculate current through a molecular species, Benzene 1,4-dithiolate molecule (as used in the experiment of Ref.[67]), between two jellium surfaces. The result demonstrates the large sensitivity of the computed current to the microscopic structure of the molecule-metal contacts.

A similar density functional approach, using an atomic-level description of the electrodes, was described by Di ventra and Pantelides.[166] These authors use density-functional based ground state molecular dynamics[167] in order to get the relaxed structure of the metal-atomic system-metal junction, then evaluated the current through the relaxed structure.

The density-functional based calculations described above where done for small applied potential bias between the electrodes. In contrast, the density functional approach of Hirose and Tsukada[168] calculates the electronic structure of a metal-insulator-metal system under strong applied bias. The main difference from the density functional approaches described above comes in the way the effective 1-electron potential is calculated. The potential used in this work contains the usual contributions from the Coulomb and the exchange-correlation interactions as well as from the ionic cores. However the Coulomb (Hartree) contribution is obtained from the solution of a Poisson equation

$$\nabla^2 V_H(\mathbf{r}) = -4\pi [\rho(\mathbf{r}) - \rho_+(\mathbf{r})] \tag{70}$$

in the presence of the applied potential boundary conditions. $\rho_+(\mathbf{r})$ is the fixed positive charge density, and the electron density $\rho(\mathbf{r})$ is constructed by summing the squares of the wavefunctions over the occupied states. At the same time the exchange-correlation potential is calculated in the standard local density approximation, neglecting the effect of the finite current that exists in the steady state system. The resulting formalism thus accounts approximately for non-equilibrium effects within the density functional calculation.[m]

To end this brief overview of density functional based computations of molecular conduction we should note that this approach suffers in principle from problems similar to those encountered in using the Hartree Fock approximation, namely the inherent inaccuracy of the computed LUMO energy and wavefunctions. The errors are different, for example HF overestimates the HOMO-LUMO gap (since the HF LUMO energy is too high[145-149, 170, 171]) while DFT underestimates it.[163, 172] Common to both approaches is the observation that processes dominated by the HOMO level will be described considerably better by these approaches than processes controlled by coupling to the LUMO.[139, 173]

## 2.9. *Potential profiles*

The theoretical and computational approaches described above are used to compute both the Ohm-law conduction, $g(E_F)$ of a molecular bridge connecting two metals, Eq. (37) or (59), and the current-voltage characteristics of the junction, also beyond the Ohmic regime, Eq. (36). We should keep in mind that these calculations usually disregard a potentially important factor – the possible effect of the imposed electrostatic field on the nuclear structure as well as on the electronic structure of the bridge. A change in nuclear configuration under the imposed electrostatic field is in fact not very likely for stable chemisorbed molecular bridges. On the other hand, the electronic wavefunctions can be distorted by the imposed field, and this in turn may affect the electrostatic potential distribution along the bridge,[n] the electronic coupling between bridge segments and the position of the molecular energy levels vis-a-vis the metals Fermi energies. These effects were in fact taken approximately into account by Hirose and Tsukada [168] and by Mujica, Roitberg and Ratner[169] by solving simultaneously the coupled Schrödinger and Poisson equations. The latter yields the electrostatic potential for the given electron density and under the imposed potential boundary conditions.

The importance of the electrostatic potential profile on the molecular bridge in determining the conduction properties of a metal-molecule-metal junction was recently discussed by Tian et al[72] in conjunction with the current-voltage characteristics of a junction comprised of an STM tip, a gold substrate and a molecule with two bonding sites (e.g. α,α'-xylyl dithiol) connecting the two. Fig. 6 displays several possible potential profiles between electrode 1 (substrate) and electrode 2 (tip) when the potential bias is $\mu_1-\mu_2=\Phi$. The linear ramp, $A_1A_2$ represents a commonly made assumption for metal-molecule-metal junctions with a strong chemical bonding of the molecule to both metals. The assumption that the electrostatic potential on the molecule is pinned to that of the substrate so that all the potential drop occures between the molecule and the tip (profile $A_1CA_2$) is often made in analyzing STM experiments where the molecule interacts strongly with the substrate and weakly with the tip. Because the molecule is a polarizable object we expect that the linear ramp potential should be replaced by the dashed line in the figure, that is actually better approximated by the profile $A_1B_1B_2A_2$.[72] Indeed, the recent model calculation by Mujica et al[169] suggests that this is indeed a good approximation. In this model, the conduction properties of the junction are determined by the position of the molecular bridge states relative to the metals equilibrium Fermi energy, and by the *voltage division factor* $\eta$ that determines the voltage drops at the molecule-substrate and the molecule-tip contacts by

$$\frac{\overline{A_1B_1}}{\overline{A_2B_2}} = \frac{\eta}{1-\eta} \qquad (71)$$

---

[m] A simplified version of the same methodology has recently been presented by Mujica et al. [169].

[n] In a single electron description this local electrostatic potential will be an input, associated with the underlying many electron response of the molecular bridge, to the position dependent energies of the bridge electronic states in the site representation.

If $\eta=0$, all the potential drop occurs at the molecule–tip interface. In this case changing the voltage across the junction amounts to changing the energy difference between the molecular levels and the tip electrochemical potential. Enhanced conduction is expected when the latter matches either the HOMO (when the tip is positively biased) or the LUMO (when the tip is negatively biased) energies. However, because the HOMO and the LUMO states are usually coupled differently to the metals (for example, in the aromatic thiols the HOMO is a sulfur-based orbital that couples strongly to the metal while the HOMO is a ring-based orbital that couples weakly to it), this implies strong asymmetry, about zero voltage, in the current-voltage dependence, i.e. rectification. In contrast, the observed dependence is essentially symmetric about $\Phi=0$, a behavior obtained from Eq. (36) for a symmetric voltage division factor $\eta=0.5$.[72]

Lamoen et al[173] have carried out a density-functional based calculation of the equilibrium structure and the electrostatic properties of Pd-doped porphyrin and perylene molecules adsorbed on gold slabs under an external electric field imposed in the direction of the molecule-metal axis. The observed behavior is qualitatively similar to a rounded, one-sided version, of the dashed line of Fig. 6. To what extent the electrostatic potential calculated in this work and in Ref.[169] are relevant for single electron models of molecular junctions still remains to be clarified. In particular, in other treatments of excess electrons at the insulator side of a metal-insulator interface the image potential attracting the electron to the interface plays an important role if the insulator dielectric constant is not too large[174-178] and experimental implications of this potential are well known.[179-184] The observation[72] that details of the electrostatic potential distribution across a metal-molecule-metal junction can significantly affect qualitative aspects of the junction electrical properties, makes further theoretical work in this direction highly desirable.

2.10. *Rectification*

The possibility to construct molecular junctions with rectifying behavior has been under discussion ever since Aviram and ratner[185] suggested that an asymmetric donor-bridge-acceptor system connecting two metal leads can rectify current. The proposed mechanism of operation of such a device is shown in Fig. 7. When the left electrode is negatively biased, i.e., the corresponding electrochemical potentials satisfy $\mu_L > \mu_R$ as shown, electrons can move from this electrode to the LUMO of molecular segment A as well as from the HOMO of molecular segment D to the right electrode. Completion of the transfer by moving an electron from A to D is assisted by the intermediate bridge segment B. When the polarity of the bias is reversed the same channel is blocked. This simple analysis is valid only if the molecular energy levels do not move together with the metal electrochemical potentials, and if the coupling through the intermediate bridge is weak enough so that the orbitals on the D and A species maintain their local nature. Other models for rectification in molecular junctions have been proposed.[186] As discussed above, the expected rectifying behavior can be very sensitive to the actual potential profile in the ABD complex, which in turn depend on the molecular response to the applied bias.[72, 187] This explains why rectification is often not observed even in asymmetric molecular junctions.[187] Still, rectification has been observed in a number of metal-molecule-metal junctions as well as in several STM experiments involving adsorbed molecules,[61, 65, 188-192].

2.11. *Carrier-carrier interactions*

The models and calculations discussed so far focus on processes for which the probability that a charge carrier populates the bridge is low so that carrier-carrier interactions can be disregarded. Electron-electron interactions were taken into account only in so far that they affected the single electron states, either in constructing the molecular spectrum (in the ab-initio HF or DFT calculations) or in affecting the junction electrostatic potential through the electronic polarization response of the molecule or the metal contacts. When the density of carriers in the space between the metal contacts becomes large, Coulomb interactions between them have to be taken into account explicitly. Here we briefly discuss the effect of such interactions.

In classical (hopping) transport of carriers through insulating films separating two metals, inter-carrier interactions appear as suppression of current due to film charging.[193, 194] In nano-junctions involving double barrier structures, increased electron population in the intermediate well under resonance transmission should affect the transport process for similar reasons. For example, consider a small metal sphere of radius R in the space between two metal electrodes (Fig. 8), and assume that both sphere and electrodes are made of the same metal of workfunction W. Neglecting possible proximity effect of these electrodes, the classical energy for removing an electron from the sphere to infinity is $W+e^2/2R$ and the classical energy for the opposite process is $W-e^2/2R$.[o] Here the sphere plays the role of a molecular bridge in assisting electron tunneling between the two electrodes, and these energies now play the same role as the corresponding HOMO and LUMO energies of the bridge. This implies that a finite voltage difference is needed before current can flow in this sphere assisted mode between the two metals, a phenomenon known as Coulomb blockade. For a larger potential bias, other conduction channels, corresponding to more highly charged states of the sphere give rise to the phenomenon of Coulomb steps.[196] For experimental manifestations of such and related phenomena see, e.g. Refs[197-203]. The possibility to observe such phenomena in electrochemical systems was discussed by Kuznetsov and Ulstrup[204] and possibly demonstrated by Fan and Bard.[205]

When the junction consists of a molecule or a few molecules connecting two metal leads, such Coulomb blockade phenomena are not expected to appear so clearly. The first Coulomb threshold is replaced, as just described, by the gap associated with the position of the metals Fermi energies relative to the molecular HOMO and LUMO levels (modified by appropriate electron correlations). However, the discreteness (in the sense that $\Delta E \gg k_B T$) of the molecular spectrum implies that for any given charging state of the molecule, e.g., a molecule with one excess electron or one excess hole, there will be several distinct conduction channels that will appear as steps in the current vs. voltage plot. It will be hard to distinguish between this structure and between 'genuine' Coulomb blockade structure. It should be emphasized that for potential applications, e.g. using the molecular junction in single electron transistor devices, the distinction between the origins of these conduction structure is in principle not important.

---

[o] From experimental and theoretical work on ionization potentials of small metal clusters[195] we know that the actual energies are approximately $W+0.4e^2/R$ and $W-0.6e^2/2R$, respectively; with the differences arising from quantum size effects).

Still, understanding the role played by electron-electron interactions (in particular correlation effects beyond the HF approximation) remains an important challenge in the study of molecular nano-junctions. Several recent theoretical works have addressed this problem within the Hubbard model with[206-208] or without[209] the mean field approximation. In particular, Malysheva and Onipko have derived a tight binding analog of the model for negative differential resistance originally proposed by Davydov and Ermakov[210] (see also [211] and [212, 213]). Numerical simulations[214] can assist in gauging the performance of the mean field approximations used in these calculations. Such models may be relevant to the understanding of recent experimental observation of negative differential resistance in a metal-self assembled monolayer-metal junction with the SAM containing a nitroamine redox center.[215]

We conclude this discussion by emphasizing again that understanding correlated carrier transport in molecular junctions continues to be an important experimental and theoretical challenge. Recent work by Gurvitz et al[216-218], using exactly solvable models of electron transport in two and three barrier structures, have indicated that new phenomenology may arise from the interplay of inelastic transitions and inter-carrier interactions in the barrier. In fact, dephasing transitions in the barrier may prove instrumental in explaining the charge quantization that give rise to the single electron transport behavior of such junctions ([219], Section 6.3).

2.12. *Some open issues*

This section discusses some subtle difficulties that are glossed over in most of the treatments of electron transmission using the formalisms described above. These should be regarded as open theoretical issues that should be addressed in future developments. The source of these problems is our simplified treatment of what is actually a complex many body open system. In particular, common ways of incorporating many body effect using single body effective potentials become questionable in particular limits of timescales and interactions strengths.

One such issue, already mentioned, is the use of static image to account for the effect of metal polarizability (namely the response of the metal electrons) on charge transfer processes at metal surfaces. The timescales estimated in Section 3.1 below are of the same order as metal plasma frequencies that measure the electronic reponse time of metals. Still static image theory has been used in the analysis of Section 2.2 and in other treatments of electron injection from metals into insulating phases.[220] To what extent dynamic image effects are important is not known, though theories that incorporates such effects have been developed. [97-99] [100(a)]

Assuming that image interactions at metal surfaces should be accounted for in the static limit, namely that the metal responds instantaneously to the tunneling charge, opens other questions. Many calculations of electronic processes near metal surfaces (e.g. [89] (See Sect. 2.2 above) *assume* that the metal electrons respond instantaneously to the *position* of the tunneling electron. Other calculations use *atomic or molecular orbitals*,[p] or more general electronic charge distributions, and computing these under the given potential boundary conditions (see, e.g. Ref [169]) implies that the corresponding orbitals or charge distributions are well defined on timescales shorter than the metal response times.[q] Examination of the energies and timescales involved suggests that assuming instantaneous metal response to the electron position is more suitable in most situations than taking instantaneous response to the charge distribution defined by a molecular orbital, but the corresponding timescales are not different enough to make this a definite statement.

A similar issue appears in attempts to account for the electronic polarizability of a solvent in treating fast electronic processes involving solute molecules or excess electrons in this solvent. For example, in treating electron transmission in MIM junctions, the potential barrier that enters into expressions like (14) depends on the electronic structure of the insulating spacer. For vacuum tunneling a rectangular barrier, whose height above the metal Fermi energy is the metal workfunction, modified by image interactions as discussed above and in Section 2.2, seems appropriate. For a dielectric spacer the barrier should be further modified by the fast (electronic) dielectric response of this spacer in the same way that it is modified by the electronic response of the metal, raising issues similar to those discussed above. We return to this point in Section 4.

---

[p] Computing molecular orbitals self-consistently with image interactions is the common practice in quantum chemistry calculations for solvated molecules using reaction field (cavity) models. Again we have a choice: either imposing the reaction field on the electronic Hamiltonian in the position representation, thus modifying all Coulomb interaction terms, then calculate the electronic wavefunctions under the new potential, or compute the electronic wavefunctions with the original Hamiltonian under the imposed dielectric boundary conditions. The fact that the two representations are not equivalent is associated with the approximate nature of the approach which replaces a detailed treatment of the electronic structure of the solvent by its electronic dielectric response. (See also footnote q).

[q] These two approaches are not equivalent, because the Schrödinger equations derived from them are non-linear in the electronic wavefunctions.

Finally, an interesting point of concern is related to the way the Fermi distribution functions enter into the current equations. For example, the Bardeen's transmission formula (21) is based on weak coupling between states localized on the two electrodes, the partial or unidirectional currents contain a product, $f(1-f)$, i.e. the probability that the initial state is occupied multiplied by the probability that the final state is not. In this viewpoint the transitions occur between two weakly coupled systems, each of them in internal thermal equilibrium, which are out of equilibrium with each other because of the potential bias.

Alternatively, we could work in the basis of exact eigenstates of the whole system comprising the two electrodes and the spacer between them. This system is in an internal non-equilibrium state in which transmission can be described as a scattering problem. The relevant eigenstates correspond to incident (incoming) waves in one electrode and transmitted waves in the other. The flux associated with those scattering states arising from an incident state in the negatively biased electrode is proportional to $f(E)$, while that associated with incoming waves in the positively biased electrode is proportional to $f(E+e\Phi)$. The net flux is therefore found again to proportional to the difference $f(E)-f(E+e\Phi)$. This argument cannot be made unless the process can be described in terms of coherent scattering states defined over the whole systems. When inelastic scattering and dephasing processes take place the description in terms of exact scattering states of the whole system becomes complicated,[219, 221] although kinetic equations for electron transport can be derived for relatively simple situations.[216-218] On the other hand, it appears that for weakly coupled contacts the perturbative approach that leads to Eq. (21) is valid. This approach describes the transmission in terms of electron states localized on the two electrodes where unidirectional rates appear with *f*(1-*f*) factors, and can in principle be carried over to the inelastic regime. (See also Sect 3.4). The exact correspondence between these different representation needs further study.

## 3. Dephasing and relaxation effects

The theoretical treatments of electron transmission and conduction through insulating barriers reviewed in the last section have assumed that the barrier nuclear configuration is static. The conduction of such junctions was thus assumed to be determined by the electronic structure of static interfacial configurations. Nuclear reorganization does play a dominant role in the analogous theory of electron transfer in molecular systems, however here again the electronic coupling itself is computed for static structures, while coupling to nuclear motion is assumed to be associated with the initial and final *localized* states of the transferred electron. As discussed in Sect. 2.5, the corresponding nuclear reorganization energies are unimportant in an MMM junction, because the transferred electron does not stay localized on the molelular species. Disregarding thermal interactions also during the transmission process therefore leads to a rigid junction model. While we cannot rule out the possible validity of such a model, it is important to consider possible scenarios where thermal relaxation on the bridge is important for two reasons. First, dephasing processes associated with electron-phonon coupling are the primary source for converting the transmission process from coherent transfer to incoherent hopping. Therefore ignoring nuclear dynamics disregards a potentially important transfer mechanism. Second, as discussed in the introduction, an important factor in desiging molecular conductors is their structural stability, therefore understanding heat generation and dissipation in molecular conductors is an important issue.[222, 223] This naturally motivates a study of inelastic effect and thermal relaxation during electron transmission. Indeed, the effect of dephasing and relaxation on carrier transport through molecular junctions (as well as other microscopic charge transport devices), on its temperature and system-size dependence and on possible interference effects has recently attracted much attention.

3.1. *Tunneling traversal times*

The underlying assumption in the treatments of electron transfer and transmission described in Section 2 is that the junction nuclear structure is rigid. The validity of this assumption should be scrutinized. Obviously, whether the barrier appears rigid to the tunneling electron, and to what extent inelastic transitions can occur and affect transmission and conductance depend on the relative scales of barrier motions and the transmission traversal time, properly defined.

A framework for discussing these issues is the theory of tunneling traversal times. 'Straightforward' timescales for tunneling, such as the rate for probability buildup on one side of a barrier following a collision of an incoming particle on the other side, or the time associated with the tunneling splitting in a symmetric double well potential, are important measures of the *tunneling rate*. Following the work of Landauer and Buttiker[101, 102, 224-227] and others,[228-230] it has been recognized that other timescales may be relevant for other observables associated with the tunneling process. The question 'how long does the tunneling particle actually spends in the classically forbidden region of the potential' is of particular interest. This *traversal time for tunneling* is useful in estimates of the relative importance of processes that may potentially occur while the particle is in the tunneling region. Energy exchange with other degrees of freedom in the barrier and interaction with external fields focused in the barrier region (e.g.

deflection of a tunneling electron by an electrostatic field induced by a heavy ion) are important examples.

The Büttiker-Landauer approach to tunneling timescales is based on imposing an internal clock on the tunneling system, for example a sinusoidal modulation of the barrier height.[101] At modulation frequencies much smaller than the inverse tunneling time the tunneling particle sees a static barrier that is lower or higher than the unperturbed barrier, depending on the phase of the modulation. At frequencies much higher than the inverse tunneling time the system sees an average perturbation and so no effective change in the barrier height, but inelastic tunneling can occur by absorption or emission of modulation quanta. The inverse of the crossover frequency separating these regimes is the estimated traversal time for tunneling. For tunneling through the 1-dimensional rectangular barrier

$$V(x) = \begin{cases} U_B & ; \quad x_1 \leq x \leq x_2 \\ 0 & otherwise \end{cases} \qquad (72)$$

and provided that $d=x_2-x_1$ is not too small and that the tunneling energy $E$ is sufficiently below $U_B$, this analysis gives

$$\tau = \frac{d}{v_I} = \sqrt{\frac{m}{2(U_B - E_0)}}\, d \qquad (73)$$

for a particle of mass $m$ and energy $E_0 < U_B$. $v_I$, defined by (73), is the imaginary velocity for the under-barrier motion. A similar result is obtained by using a clock based on population transfer between two internal states of the tunneling particle induced by a small barrier localized coupling between them.[102] Using the same clock for electron transfer via the super-exchange mechanism in the model of Fig. 3 (equal donor and acceptor energy levels, $E_A=E_D$, coupled to opposite ends of a molecular bridge described by an N-state tight binding model with nearest-neighbor coupling $V_B$, with an energy gap $\Delta E_B = E_B - E_D \gg V_B$), yields[231]

$$\tau = \frac{\hbar N}{\Delta E_B} \qquad (74)$$

Nitzan et al have shown[231] that both results (73) and (74) are limiting cases (wide and narrow band limits) of a more general expression:

$$\tau = \frac{\hbar N}{2V_B \sqrt{\frac{\Delta U_B}{V_B} + \left(\frac{\Delta U_B}{2V_B}\right)^2}} \qquad (75)$$

where $\Delta U_B \equiv E_B - 2V_B - E_D$ is difference between the initial energy $E_D$ and the bottom of the conduction band, $E_B$-$2V$, see Fig. 9. When $V_B \rightarrow 0$, $\Delta U_B \rightarrow \Delta E_B$ and the r.h.s of Eq. (75) becomes that of Eq. (74). In the opposite limit, $V_B \rightarrow \infty$ with $\Delta U_B$ kept constant, Eq. (75) becomes

$$\tau = \frac{\hbar N}{2\sqrt{V_B \Delta U_B}} \tag{76}$$

if we express $V_B$ in terms of the effective mass for the band motion, $m = \hbar^2/2V_B a^2$ with $a = d/N$, Eq. (76) yields the Büttiker Landauer result, Eq. (73).

The interpretation of $\tau$ defined above as a characteristic time for the tunneling process should be used with caution. An important observation made by Buttiker,[102] is that the tunneling time is not unique, but depends on the observable used as a clock. Still, as shown in Ref.[101], for a proper choice of clock the traversal time provides a useful measure for the adiabaticity or non-adiabaticity of the interaction of the tunneling particle with barrier degrees of freedom. The calculation that leads to Eqs. (74)-(76) uses a clock based on two internal states, |1> and |2>, of the tunneling particle with a small barrier-localized coupling, $\lambda(|1><2|+|2><1|)$, between them. The incident particle is in state |1>. The population of state |2> in the transmitted wavefunction can be related to the duration of the interstate coupling, i.e. to the traversal time. Writing the transmitted state in the form $c_1|1>+c_2|2>$ this procedure yields

$$\tau = \lim_{\lambda \rightarrow 0}\left(\frac{\hbar}{|\lambda|}\left|\frac{c_2}{c_1}\right|\right) \tag{77}$$

For the 1-dimensional rectangular barrier model, Eq. (72), and in the limit $\kappa d >> 1$, this leads again to Eq. (73). Galperin et al[232] have applied the same approach to compute traversal times through water layers (see Sect. 4).

For tunneling through a molecular spacer modeled as a barrier of width ~10Å (N=2-3) and height $U_B$-$E \cong \Delta E$ ~ 1eV, Eqs. (73) and (74) yield $\tau \cong 0.2$fs and $\tau \cong 2$fs, respectively, both considerably shorter than the vibrational period of molecular vibrations. When the barrier is lower or when tunneling is affected or dominated by barrier resonances, the traversal time becomes longer, and competing relaxation and dephasing processes in the barrier may become effective. This is expected to be the rule for resonance transmission through molecular bridges, because the bandwidth associated with the bridge states (i.e. the electronic coupling between them; see Fig. 9) is considerably smaller than in metals. As a consequence thermal relaxation and dephasing are expected to dominate electron transport at and near resonance. This issue is discussed next.

3.2. *Nuclear relaxation during electron transmission*

It has long been recognized that tunneling electrons interact, and may exchange energy, with nuclear degrees of freedom in the tunneling medium. One realization of such processes is inelastic electron tunneling spectroscopy,[220, 233] where the opening of inelastic channels upon increasing the electrostatic potential difference between the source and sink metals is manifested as a peak in the second derivative of the tunneling current with respect to this potential drop. Recent applications of this phenomenon within scanning tunneling spectroscopy hold great promise for making the STM a molecular analytical tool.[234] Inelastic electron tunneling may also cause chemical bond breaking and chemical rearrangement in the tunneling medium, either by electron induced consecutive excitation or via transient formation of a negative ion.[r] [235-238]

As discussed by Gadzuk,[239], the phenomenology of inelastic electron transmission is also closely related to other electronic processes in which transient occupation of an intermediate state drives a phonon field. Intramolecular vibrational excitation in resonant electron scattering,[240] phonon excitation in resonant electron tunneling in quantum-well heterostructures[241] and electron induced desorption[242, 243] can all be described using similar models. A prototype Hamiltonian describing these models is (see Fig. 3b)

$$H = H_{el} + H_{ph} + H_{el-ph} \tag{78}$$

where $H_{el}$ is the electronic Hamiltonian

$$H_{el} = \sum_n E_n c_n c_n + \sum_{n,n'(n \neq n')} V_{n,n'} c_n c_{n'} + \sum_k E_k c_k c_k + \sum_k \sum_n \left( V_{k,n} c_k c_n + V_{n,k} c_n c_k \right) \tag{79}$$

$H_{ph}$ is the Hamiltonian of the phonon bath

$$H_{ph} = \sum_\nu \hbar \omega_\nu b_\nu b_\nu \tag{80}$$

and $H_{el-ph}$ is the electron-phonon interaction, usually written in the form

$$H_{el-ph} = \sum_n \sum_\nu \lambda_{n\nu} c_n c_n \left( b_\nu + b_\nu \right) \tag{81}$$

---

[r] While our language refer to electron transport and electron tunneling, hole transport and nuclear excitation via transient positive ion formation are equally possible.

Here $c_j$ and $c_j$ ($j=n,n',k$) create and annihilate an electron in electronic state $j$, while $b_\nu$ and $b_\nu$ similarly create and annihilate a phonon of mode $\nu$, of frequency $\omega_\nu$. In Eq. (79) the states ($k$) are taken to be different manifolds of continuous scattering states, denoted by a continuous index $k$ (Fig. 3b shows two such manifolds, k={$\ell$}, {$r$}), while the set of states {$n$} are discrete electronic states of the observed molecular system. The electronic Hamiltonian (79) can describe a scattering process in which the electron starts in one continuous manifold and ends in another and the states {n} belongs to the target the causes the scattering process. These states may be the eigenstates of the target Hamiltonian, in which case $V_{n,m}$ in Eq. (79) vanishes, or some zero-order representation in which the basis states are mutually coupled by the exact target Hamiltonian. Eqs. (80) represents the thermal environment as a harmonic phonon bath. The coupling between the electronic system and this bath is assumed in Eq. (81) to originate from a target-state dependent shift in the equilibrium position of each phonon mode. An exact solution to this scattering problem can be obtained for the particular case where the target is represented by a single state $n=1$ and the phonon bath contains one oscillator of frequency $\omega$. In this case it is convenient to consider the oscillator as part of the target which is therefore represented by a set of states $|m\rangle$ with energies $E_1 + m\hbar\omega$ (the zero point energy can be set to 0). If the oscillator is initially in the ground state (m=0) the cross-section for electron tunneling (or scattering) from the left to the right side in Fig. 1 is given by[239, 240, 244]

$$\mathcal{T}(E_i, E_f) \sim \Gamma^{(L)}\Gamma^{(R)} \sum_{m'=0}^{\infty} \delta\left(E_i - E_f - m'\hbar\omega\right) \sum_{\tilde{m}=0}^{\infty} \frac{<m'|\tilde{m}><\tilde{m}|0>}{E_i - E_{\tilde{m}} - \Lambda_{\tilde{m}}(E_i) + (i/2)\Gamma_{\tilde{m}}(E_i)} \quad (82)$$

where $|\tilde{m}\rangle$ are states of the shifted harmonic oscillator that corresponds to the temporary negative ion (electron residing on the target) and $E_{\tilde{m}} = E_1 + m\hbar\omega - \lambda^2/\hbar\omega$. $\Lambda_{\tilde{m}}$ and $\Gamma_{\tilde{m}}$ are the shifts and widths of the dressed target states associated with their coupling to the continuous manifolds and

$$\Gamma^{(K)}(E) = 2\pi \sum_k |V_{k,1}|^2 \, \delta(E - E_k) \quad ; \quad K = L, k = l \ \ or \ \ K = R, k = r. \quad (83)$$

The exact solution (82) can be obtained because of the simplicity of the system, which was chracterized by a single intermediate electronic state and a single phonon mode. In more realistic situations characterized by many bridge electronic states and many phonon modes one need to resort to approximations or to numerical simulations. We discuss such systems next.

To get the proper perspective on the nature of this problem consider again the standard electron transfer process in a donor-bridge-acceptor (DBA) system without metal electrodes. As already emphasized (see Section 2.5), nuclear dynamics and conversion of electronic energy to nuclear motions, resulting from solvent reorganization about the donor and acceptor sites upon changing their charge state, are essential ingredients of this process. The reason for the prominent role of nuclear dynamics in this case is that the transferred charge is localized on the donor/acceptor orbitals, consequently affecting distortion of their nuclear environments (represented by the parabolas in Figs 2a and 3a). Standard electron transfer theory assumes that nuclear motion is coupled to the donor and acceptor electronic states only, and the electronic coupling itself is taken independent of the nuclear configuration (the Condon approximation). This assumption is sometimes questionable, in particular when intermediate electronic states are involved, as in Figures 1-3. The possible role of nuclear motion on such intermediate electronic potential surfaces has been discussed by Stuchebrukhov and coworkers.[245, 246] Focusing on bridge assisted electron transfer processes, these authors separate the nuclear degrees of freedom into two groups. The first include those nuclear modes that are strongly coupled to the donor-acceptor system (solvent polarization modes and vibrational modes of the donor and acceptor species). In the absence of the other modes this coupling leads to the standard electron transfer rate expression due to Marcus (c.f. Eqs. (1), (3) and (9))

$$k_{et} = \frac{2\pi}{\hbar} |T_{DA}|^2 \frac{e^{-(\lambda+E_{AD})^2/4\lambda k_B \Theta}}{\sqrt{4\pi\lambda k_B \Theta}} \qquad (84)$$

where $\lambda$ is the reorganization energy, $E_{AD}$ is (free) energy difference between the initial (electron on donor) and final (electron on acceptor) equilibrium configurations and $T_{DA}$ is the non-adiabatic electronic coupling matrix element that incorporates the effect of the bridge via, e.g. Eqs. (9) and (10). The other group of degrees of freedom, 'bridge modes', are coupled relatively weakly to the electron transfer process, and it is assumed that their effect can be incorporated using low-order perturbation theory. This is accomplished by considering the modulation of the electronic coupling $T_{DA}$ by these motions, $T_{DA} = T_{DA}(\{x_v\})$, where $\{x_v\}$ is the set of the corresponding nuclear coordinates. It is important to note that the separation of nuclear modes into those coupled to the donor and acceptor states (schematically represented by the Marcus parabolas in Fig. 2a and 3a) and those associated with electronic coupling between them is done for convenience only, and is certainly not a rigorous procedure. Within this picture the electron transfer rate is obtained[245] as a convolution

$$k = \int d\varepsilon \rho_B(\varepsilon) k_0 (E_{AD} + \varepsilon) \qquad (85)$$

where

$$\rho_B(\varepsilon) = \int dt\, e^{i\varepsilon t} \frac{<T_{DA}(t)T_{DA}(0)>}{<T_{DA}^2>} \tag{86}$$

and

$$T_{DA}(t) = e^{iH_B t/\hbar} T_{DA} e^{-iH_B t/\hbar} \tag{87}$$

where $H_B$ is the bridge Hamiltonian including the thermal environment (Θ of Fig. 2). Calculations based on this formalism indicate[245] that inelastic contributions to the total electron transfer flux are substantial for long (>10 segments) bridges.

It should be emphasized that dynamical fluctuations in the bridge can considerably affect also the elastic transmission probability. For example, a substantial effect of the bridge nuclear motion on the electron transfer rate has been observed in simulations of electron transfer in azurin[246, 247] in agreement with earlier theoretical predictions.[248, 249] There are some experimental indications that electron transfer rate in proteins is indeed substantially affected by the protein nuclear motion.[250]

The Medvedev-Stuchebrukhov theory[245] corresponds to the lowest order correction, associated with intermediate state nuclear relaxation, for bridge mediated electron transfer rate. On the other extreme side we find sequential processes that are best described by two or more consecutive electronic transitions. For this to happen two conditions have to be satisfied. First, the intermediate state(s) energy should be close to that of the donor/acceptor system, so these states are physically populated either directly or by thermal activation. Second, nuclear relaxation and dephasing should be fast enough so that the bridging states can be treated as well-defined thermally averaged electronic configurations. Obviously, intermediate situations can exist. Bridge mediated electron transfer can be dominated by two (donor-acceptor) electronic states coupled via intermediate high-lying states that are only virtually populated, by real participation of such intermediate states in a coherent way (when thermal relaxation and dephasing are slow), or by sequential transfer through such states. This issue was extensively discussed[251-253] for three state models of electron transfer that were recently used to describe primary charge separation in bacterial photosynthesis. The possibility to observe similar effects in STM studies of molecules adsorbed at electrochemical interfaces was discussed by Schmickler.[254, 255]

Closely related to this phenomenology is the process of light scattering from molecular systems where the donor and acceptor states are replaced by the incoming and outgoing photons. Elastic (Rayleigh) scattering is the analog of the 2-state 'standard' electron transfer process. Inelastic (Raman) scattering is the analog of the process analyzed by by Stuchebrukhov and coworkers.[246], except that our ability to resolve the energy of the scattered photon make it possible to separate the total rate (or flux), the analog of Eq. (85), into its elastic and different inelastic components [256]. Resonance Raman scattering and resonance fluorescence are the processes that take place when excited molecular states are physically, as opposed to virtually, occupied during the light scattering process. The former is a coherent process that take place in the absence of dephasing and thermal relaxation while the latter follows thermal relaxation in the excited molecular state. Re-emitting the photon after dephasing has occurred, but before full thermal relaxation takes place, is the process known as hot luminescence.

3.3. *Thermal interactions in molecular conduction*

Coming back to electron transfer and transmission, the importance of dephasing effects in the operation of microscopic junctions has long been recognized.[108, 219] The Landauer formula for the conduction of a narrow constriction connecting two macroscopic metals, Eq. (25) or (29), is derived by assuming that the transmission is elastic and coherent, i.e. without dephasing and energy changing interactions taking place in the constriction. If the constriction is small relative to the mean free path of the electron in it, these effects may indeed be disregarded. When the constriction becomes macroscopic multiple scattering and dephasing are essential to obtain the limiting Ohm's law behavior. A simple demonstration is obtained([219]p. 63) by considering a conductor of length L as a series of N macroscopic scatterers, each of the type that, by itself, would yield Eq. (25). At each scatterer the electron can be transmitted with probability $\mathcal{T}$, or reflected with probability $\mathcal{R}=1-\mathcal{T}$. Let the the total transmission through $N$ such objects be $\mathcal{T}_N$, so that $\mathcal{T}=\mathcal{T}_1$. *Provided that the phase of the wavefunction is destroyed after each transmission-reflection event*, so that we can add probabilities, the transmission through an $N$ scatterers system is obtained by considering a connection in series of an $N$-1 scatterers system with an additional scatterer, and summing over all multiple scattering paths

$$\mathcal{T}_N = \mathcal{T}_{N-1}\left(1 + \mathcal{R}\mathcal{R}_{N-1} + (\mathcal{R}\mathcal{R}_{N-1})^2 + ...\right) = \frac{\mathcal{T}\mathcal{T}_{N-1}}{1-\mathcal{R}\mathcal{R}_{N-1}} \quad (88)$$

with $\mathcal{R}=1-\mathcal{T}$ and $\mathcal{R}_N=1-\mathcal{T}_N$. This implies

$$\frac{1-\mathcal{T}_N}{\mathcal{T}_N} = \frac{1-\mathcal{T}_{N-1}}{\mathcal{T}_{N-1}} + \frac{1-\mathcal{T}}{\mathcal{T}} = N\frac{1-\mathcal{T}}{\mathcal{T}} \quad (89)$$

so that

$$\mathcal{T}_N = \frac{\mathcal{T}}{N(1-\mathcal{T})+\mathcal{T}} = \frac{L_0}{L+L_0} \tag{90}$$

where $L_0 = \mathcal{T}/\nu(1-\mathcal{T})$ and $\nu = N/L$ is the scatterers density. Using this in Eq. (25) yields

$$g(E) = \frac{e^2}{\pi\hbar}\frac{L_0}{L+L_0} \tag{91}$$

that gives the inverse length dependence characteristic of Ohm's law as $L \to \infty$. (See however [108], p. 107).

A more detailed treatment of the role played by dephasing in quantum charge transport in microcopic junction was given by Büttiker.[257] He has introduced phase destruction processes by conceptually attaching an electron reservoir onto the constriction (Fig. 10), under the condition that, while charge carriers are exchanged between the current-carrying system and the reservoir, no net averaged current is flowing into this reservoir. Büttiker has observed that such a contact, essentially a voltage probe, acts as a phase breaking scatterer. By adjusting the coupling strength between this device and the system, a controlled amount of incoherent current can be made to be carried through the system. This approach has been very useful in analyzing conduction properties of multi-gate junctions and connected nano-resistors.

In molecular systems, a very different approach to dephasing was considered by Bixon and Jortner,[258, 259] who pointed out that the irregular nature of Franck Condon overlaps between intramolecular vibrational states associated with different electronic centers can lead to phase erosion in resonant electron transfer. Consequently, bridge assisted electron transfer, which proceeds via the superexchange mechanism in off resonance processes, will become sequential in resonance situations. For a finite temperature system with an electronic energy gap between donor and bridge that is not too large relative to $k_B\Theta$, the thermally averaged rate from a canonical distribution of donor states results in a superposition of both superexchange and sequential mechanisms.

While coupling to the thermal environment is implicit in the models described above, using molecular bridges embedded in condensed environments as conductors immediately suggests the need to consider the coupling to intramolecular and environmental nuclear motions explicitly as in the Hamiltonian (78)-(81). The models of Figures 2 and 3, where transition between the two electron reservoirs or between the donor and acceptor species is mediated by a bridge represented by the group of states {n} is again the starting point of our discussion. Several workers have recently addressed the theoretical problem of electron migration in such models, where the electron is coupled to a zero temperature phonon bath. Bonča and Trugman[260, 261] have provided an exact numerical solution for such a problem. Their model is similar to that described by Eqs. (78)-(81), except that the metal leads connected to the molecular target are represented by 1-dimensional semi-infinite tight binding Hamiltonians:

$$H = H_{el} + H_{ph} + H_{el-ph} \tag{92}$$

$$H_{el} = \sum_n E_n c_n c_n + \sum_k E_k c_k c_k + \sum_{n,n'} V_{n,n'} c_n c_{n'} + \sum_{k,k'} V_{k,k'} c_k c_{k'} + \left( \sum_{n,k} V_{n,k} c_n c_k + h.c \right) \tag{93}$$

$$H_{ph} = \sum_v \hbar \omega_v b_v b_v \tag{94}$$

$$H_{el-ph} = \sum_n \sum_v \lambda_{nv} c_n c_n \left( b_v + b_v \right) \tag{95}$$

Here, $H_{el}$ desribes both the metal leads (represented by the manifold(s) of states {k}) and the molecular target (with states {n}). The coupling to the phonon field is assumed to vanish on the metal sites. The electron transport problem is treated as a 1-particle multichannel scattering problem, where each of the (one incoming, many outgoing) channels corresponds to a given vibrational state of the target. A finite basis is employed by using a finite number of phonon modes and limiting the number of phonons quanta associated with each site, and by projecting out leads that carry only outgoing states, however the size of this basis can be increased until convergence is achieved. Yu et al[262, 263] have studied the same 1-dimensional electronic model with a different electron-phonon interaction: instead of the Holstein type interaction taken in Eqs. (81) and (95), they use a model similar to the Su-Schrieffer-Heeger (SSH) Hamiltonian,[264] where Eqs. (93)-(95) are replaced by

$$H_{el} + H_{el-ph} = \sum_n E_n c_n c_n + \sum_n \left\{ \left[ V_{n,n+1} - \alpha_{n,n+1} \left( u_{n+1} - u_n \right) \right] c_n c_{n+1} + h.c. \right\} \tag{96}$$

$$H_{ph} = \frac{1}{2} K \sum_{n=1}^{N-1} \left( u_{n+1} - u_n \right)^2 + \frac{1}{2} \sum_{n=1}^{N} m_n \dot{u}_n^2 \tag{97}$$

where $u_n$ (n = 1,...,N) are displacements of the target atoms. The segment of the lattice between $n=1$ and $n=N$ represents an organic oligomer, connecting between two metals, and the model for the Oligomer is the same as that used in the SSH theory of conducting conjugate polymers, with the nuclear degrees of freedom treated classically. The electron-phonon coupling is again assumed to vanish outside the bridge, i.e., in Eq. (96) $\alpha_{n.n+1}$ is taken zero unless $n=1,2,...N-1$. A special feature (in the context of this review) of this calculation is that it is done using the exact many electron ground state of the metal-oligomer-metal system, which takes into account the Peierl's distortion[265] that leads to a dimerization in the Oligomer's structure.[264] However, the energy of the transmitted electron is taken far above the Fermi energy and electron-electron

43interactions are neglected, so issues associated with Pauli exclusion can be disregarded. The model is used to study the time evolution of an excess electron wavepacket that starts in the metal lead in the direction of the oligomer segment. This time evolution is computed using the quantum-classical time dependent self consistent field (TDSCF) approximation, whereupon the electron wavefunction is propagated under the instantaneous nuclear configuration, while the latter is evolved classically using the expectation value of the Hamiltonian with the instantaneous electronic wavefunction.[s] This approximation for the time evolution conserves the total system energy, so energy exchange between the electronic and nuclear subsystem can be studied as a function of time in addition to the total transmission and reflection probabilities. It is found that lattice dynamics can be quite important at an intermediate window of electron energies, where the electronic and nuclear timescales are comparable. Of particular interest is the energy left in the nuclear subsystem after the electron has traversed the oligomer.

A fully quantum analog of this model was studied by Ness and Fisher.[266] Their Hamiltonian is

$$H_{el} = \sum_n E_n c_n c_n + \sum_\nu \hbar\omega_\nu b_\nu b_\nu + \sum_{\nu,n,m} \gamma_{\nu,n,m}\left(b_\nu + b_\nu\right) c_n c_m \qquad (98)$$

where, again, the distinction between the metal leads and the molecular system enters through the values of the site energies $E_n$ and through the fact that coupling to phonons exists only at the oligomer sites. The ground state of the neutral N electron dimerized chain is the reference system. Electron-electron interaction is disregarded and the time evolution in the corresponding N+1 or N-1 electron system is studied at zero temperature using the multichannel time independent scattering theory approach of Bonča and Trugman.[260, 261] The result of this calculation is a considerable increase in the tunneling current when the electron-phonon interaction is switched on, in particular for long chains. The origin of this behavior seems to be the existence of a polaron state below the conduction band edge of the molecular segment that effectively lower the barrier energy experienced by the tunneling electron. Close to resonance however, the effect of electron-phonon coupling may be reversed, leading to a smaller total overall conduction.[267]

---

[s] An open issue in this calculation is the validity of the TDSCF approximation. This approximation is known to be problematic in tunneling and scattering calculations where the quantum wavefunction splits to several distinct components.

The Bonča Trugman approach[260, 261] has also been used recently by Emberly and Kirczenow,[221] also for a 1-dimensional tight binding model described by the SSH Hamiltonian. These authors attempt to take into account the Pauli exclusion principle in calculating the inelastic contributions to electron transmission and reflection. While the formalism can in principle be applied to finite temperature processes, the implementation is done for a low temperature system. The result again indicates that inelastic processes can substantially modify electron transport for long molecular chains and large potential drops.

3.4. *Reduced density matrix approaches*

The works described above use models for quantum transport that yield practically exact numerical solutions at the cost of model simplicity: 1-dimensional tight binding transport model, only a few harmonic oscillators and essentially zero temperature systems. An alternative approach uses the machinery of non-equilibrium statistical mechanics, starting from an Hamiltonian such as (92) and projecting out the thermal bath part. The resulting reduced equations of motion for the electronic subsystem contain dephasing and energy relaxation rates that are related explicitly to properties of the thermal bath and the system-bath coupling.

Such approaches to bridge mediated electron transport were made by several workers.[119, 120, 268-273]. For simplicity we limit ourselves to the tight binding super-exchange model for bridge mediated electron transfer (see Section 2.1). Also, for simplicity of notation we consider N bridge states between the two electrodes, without assigning special status to 'donor' and 'acceptor' states as in Fig. 3b. (It should be obvious that this makes only a notational difference). The Hamiltonian for the athermal system is

$$H = H_0 + V \tag{99}$$

$$H_0 = \sum_{n=1}^{N} E_n |n\rangle\langle n| + \sum_l E_l |l\rangle\langle l| + \sum_r E_r |r\rangle\langle r| \tag{100}$$

$$V = \sum_l \left(V_{l,1}|l\rangle\langle 1| + V_{1,l}|1\rangle\langle l|\right) + \sum_{n=1}^{N-1} \left(V_{n,n+1}|n\rangle\langle n+1| + V_{n+1,n}|n+1\rangle\langle n|\right) \\ + \sum_r \left(V_{r,N}|r\rangle\langle N| + V_{N,r}|N\rangle\langle r|\right) \tag{101}$$

where $\{l\}$ and $\{r\}$ are again continuous manifolds corresponding to the 'left' and 'right' metal leads and $\{n\}$ is a set of bridge states connecting these leads in the way specified by the corresponding elements of the coupling $V$. In the absence of thermal interactions, and when the left and right electrodes are coupled only to levels 1 and N of the bridge, respectively, transport in this system is descibed by the conduction function (c.f. Eqs. (31) and (40))

$$g(E) = \frac{e^2}{\pi \hbar} |G_{1N}(E)|^2 \, \Gamma_1^{(L)}(E) \Gamma_N^{(R)}(E) \tag{102}$$

with

$$\Gamma_1^{(L)}(E) = 2\pi \sum_l |V_{l1}|^2 \, \delta(E_1 - E) \quad ; \quad \Gamma_N^{(R)}(E) = 2\pi \sum_r |V_{Nr}|^2 \, \delta(E_N - E) \tag{103}$$

In general $G(E)$ is evaluated numerically by inverting the corresponding Hamiltonian matrix. For $E_n = E_B$ and $V_{n,n+1} = V_B$, identical for all bridge levels and for all mutual couplings, respectively, and in the superexchange limit, $|V_B| \ll |E_B - E|$, the Green's function element is $V_B^{N-1}/\Delta E_B^N$ (c.f. Eq. (10)), with $\Delta E_B = E - E_B$. In this case $g$ depends exponentially on the bridge length N according to $g \sim \exp[-\beta' N]$ with $\beta' = 2\ln(|\Delta E_B / V_B|)$ (c.f. Eq. (13)).

*Weak thermal coupling.* To see how this dynamics is modified by thermal relaxation and dephasing effects, we follow the formulation of Ref.[119] The Hamiltonian $H$ is supplemented by terms describing a thermal bath and a system-bath interaction

$$\mathcal{H} = H + H_\Theta + F \tag{104}$$

where $H_\Theta$ is the Hamiltonian for the thermal environment or bath, and where the system-bath interaction $F$ is assumed weak. In this case thermal coupling between different bridge levels is neglected relative to the internal coupling $V$ between them, so

$$F = \sum_{n=1}^{N} F_n \, |n\rangle\langle n| \tag{105}$$

where $F_n$ are operators in the bath degrees of freedom that satisfy $\langle F_n \rangle \equiv Tr_\Theta \left( e^{-\beta H_\Theta} F_n \right) = 0$ ($Tr_\Theta$ is a trace over all thermal bath states). $F$ is characterized by its time correlation function. As a simple model we postulate

$$\langle F_n(t) F_{n'}(0) \rangle = f(t) \delta_{n,n'}, \tag{106}$$

The Fourier transform of the remaining correlation functions satisfies the detailed balance condition

$$\int_{-\infty}^{\infty} dt \, e^{i\omega t} \langle F_n(t) F_n(0) \rangle = e^{\beta \hbar \omega} \int_{-\infty}^{\infty} dt \, e^{i\omega t} \langle F_n(0) F_n(t) \rangle \quad ; \quad \beta = (k_B \Theta)^{-1} \tag{107}$$

where $\Theta$ is the temperature and $\beta$ – the Boltzmann constant. For specificity we sometimes use

$$f(t) = \frac{\kappa}{2\tau_c} \exp(-|t|/\tau_c) \tag{108}$$

which becomes $\kappa \delta(t)$ in the Markovian, $\tau_c \to 0$, limit. Note that (105) is a particular model for the thermal interactions, sufficient to show their general consequences, but by no means adequate for quantitative predictions. In particular, the assumption (106) will be replaced by a more realistic model below.

Galperin et al[111] have shown that the conduction properties of a system like that described by the Hamiltonian (99)-(104) can be obtained by studying a steady state in which the amplitude of one state |0> in the initial {l} manifold remains constant and the amplitudes of other states evolve under this restriction. Segal et al[119] have generalized this approach to thermal systems of the kind described by the Hamiltonian (104), using, in the weak thermal coupling limit, the Redfield approximation.[269, 274, 275] This approximation combines two steps that rest on the weak coupling limit: an expansion up to second order in the coupling $F$ and the assumption that the thermal bath is not affected by its coupling to the molecular system. In this approach one starts from the set of states |0>, |1>,..., |n>, {|l>}, {|r>}, where |0> is the incoming state in the {l} manifold, and projects out the continuous manifolds {l} (except |0>) and {r}). This amounts to replacing $H$ of Eqs. (99)-(104) by an effective Hamiltonian, $H^{eff}$, in the space spanned by states |0>, |1>,..., |n> in which the energies $E_1$ and $E_N$ are modified by adding self energy terms whose imaginary parts are respectively $\Gamma_1^{(L)}/2$ and $\Gamma_N^{(R)}/2$. This effective Hamiltonian of order $N+1$ is then diagonalized and the resulting set of $N+1$ states (originating from $N$ bridge states and one incoming state) is used to represent the Liouville equation for the density operator $\rho$ of the overall electrode-bridge-bath system, $\dot{\rho} = -i[\mathcal{H}, \rho]$. This Liouville equation is expaned to second order in $F$ and traced over bath degrees of freedom using the approximation $\rho(t) = \rho_\Theta \sigma(t)$ with $\rho_\Theta = e^{-\beta H_\Theta}$ and $\sigma(t) = Tr_\Theta \rho(t)$. This leads to an equation of motion for the reduced density matrix $\sigma(t)$ for the electrode-bridge system that takes the form

$$\dot{\sigma}_{jk} = -iE_{jk}\sigma_{jk} - \Gamma_{jk}\sigma_{jk} - \int_0^t dt' \sum_{lm} \{ \langle \tilde{F}_{jl}(t-t')\tilde{F}_{lm}(0) \rangle e^{-iE_{lk}(t-t')} \sigma_{mk}(t')$$
$$- \langle \tilde{F}_{mk}(0)\tilde{F}_{jl}(t-t') \rangle e^{-iE_{lk}(t-t')} \sigma_{lm}(t') \quad (109)$$
$$- \langle \tilde{F}_{mk}(t-t')\tilde{F}_{jl}(0) \rangle e^{-iE_{jm}(t-t')} \sigma_{lm}(t')$$
$$+ \langle \tilde{F}_{ml}(0)\tilde{F}_{lk}(t-t') \rangle e^{-iE_{jl}(t-t')} \sigma_{jm}(t') \}$$

where $E_{jl}=E_j-E_l$ and $\tilde{F}(t) = e^{iH_\Theta t} F e^{-iH_\Theta t}$. Here the indices $j,k,l,m$ refer to molecular states that diagonalize the effective Hamiltonian $H^{eff}$. The damping terms $\Gamma$ originate from the decay of states |1> and |N> distributed into these eigenstates. At steady state all $\sigma$ elements are constant and Eq. (109) become

$$0 = -iE_{jk}\sigma_{jk} - \Gamma_{jk}\sigma_{jk} + \sum_{lm}\{$$

$$\sigma_{lm}\int_0^\infty d\tau \left(\left\langle \tilde{F}_{mk}(0)\tilde{F}_{jl}(\tau)\right\rangle e^{-iE_{lk}\tau} + \left\langle \tilde{F}_{mk}(\tau)\tilde{F}_{jl}(0)\right\rangle e^{-iE_{jm}\tau}\right)$$

$$-\sigma_{mk}\int_0^\infty d\tau \left\langle \tilde{F}_{jl}(\tau)\tilde{F}_{lm}(0)\right\rangle e^{-iE_{lk}\tau} - \sigma_{jm}\int_0^\infty d\tau \left\langle \tilde{F}_{ml}(0)\tilde{F}_{lk}(\tau)\right\rangle e^{-iE_{jl}\tau}\} \quad (110)$$

Transforming (110) back to the local bridge representation {0, n=1,...N} leads to a a set (N+1)(N+1) equations of the form

$$-iE_{nn'}\sigma_{nn'} - i[V,\sigma]_{nn'} + \sum_{n_1}\sum_{n_2} R_{nn'n_1n_2}\sigma_{n_1n_2} = \frac{1}{2}(\Gamma_n + \Gamma_{n'})\sigma_{nn'} \; ; \; n,n' = 0,...,N \quad (111)$$

where the elements of $R$ are linear combinations of the integrals appearing in Eq. (110) and where $\Gamma_n = \Gamma_N^{(R)}\delta_{n,N} + \Gamma_1^{(L)}\delta_{n,1}$. Again, at steady state the first ($n=n'=0$) equation is replaced by the boundary condition $\sigma_{00}=constant$. The remaning (N+1)(N+1)-1 equations constitute a set of linear non-homogeneous algebraic equations in which the terms containing $\sigma_{00}$ constitute source terms. Thus, all elements $\sigma_{nn'}$, and in particular $\sigma_{NN}$, can be obtained in the form $\sigma_{nn'} = U_{nn'}\sigma_{00}$, in terms of the fixed population $\sigma_{00}$ in the incoming state |0> of the {$l$} manifold, where the coefficients $U_{nn'}$ are related to the inverse of the (N+1)(N+1)-1 order matrix of thermal rates. The steady state flux into the {$r$} manifold is $\Gamma_N^{(R)}\sigma_{NN}$, and the corresponding rate is

$$k_{0\to R} = \Gamma_N^{(R)}\sigma_{NN}/\sigma_{00} = \Gamma_N^{(R)}U_{NN} \quad (112)$$

While the general expression for $U_{NN}$ is very cumbersome, involving the inverse of an (N+1)(N+1)-1 order matrix, numerical evaluation of the resulting rate and its dependence on coupling parameters, bridge length and temperature is an easy numerical task for reasonable bridge lengths. A final technical point stems from the observation that the resulting $k_{0\to R}$ must be proportional to $|V_{10}|^2$, the squared coupling between the first bridge level and the left continuous manifold. We therefore rewrite Eq. (112) in terms of new variables $k'_{0\to R}$ and $U'_{NN}$, defined by

$$k_{0\to R} = k'_{0\to R}|V_{10}|^2 = \Gamma_N^{(R)}U'_{NN}|V_{10}|^2 \quad (113)$$

We can make contact with results obtained in the athermal case by writing $|0>=|k_\parallel,k_x>$ where x is the direction of transmision, $k_\parallel$ is the momentum in the yz plane and $(\hbar^2/2m)\left(k_\parallel^2 + k_x^2\right) = E_\parallel + E_x = E_0$. The transmission coefficient $\mathcal{T}(E_0,k_\parallel)$ for electron incident from the left electrode with total energy $E_0$ in channel $k_\parallel$ is related to $k_{0\to R}$ by

$$k_{0 \to R} = \frac{k_x}{mL} \mathcal{T}(E_0, k_\parallel) = \left(2\pi\rho(E_x)\right)^{-1} \mathcal{T}(E_0, k_\parallel) \tag{114}$$

where $\rho(E_x)$ is the 1-dimensional density of states for the motion in the $x$ direction. Therefore

$$\mathcal{T}(E_0, k_\parallel) = 2\pi\rho(E_x) k_{0 \to R} = \Gamma_{1,k_\parallel}^{(L)} k'_{0 \to R} \tag{115}$$

and the all-to-all transmission at energy $E_0$ is the sum over all channels with energy $E_\parallel < E_0$

$$\mathcal{T}(E_0) = \Gamma_1^{(L)} k'_{0 \to R} = \Gamma_1^{(L)} \Gamma_N^{(R)} U'_{NN} \tag{116}$$

Comparing to Eq. (102), we see that Eq. (116) is the analog of Eq. (40), where, in the thermal case, $U'_{NN}$ has replaced $|G_{1N}|^2$.

In the athermal case the conduction of a junction characterized by a given transmission coefficient is obtained from the Landauer formula (29). Here the issue is more complex since, while $\mathcal{T}(E_0)$ is the probability that an incident electron with energy $E_0$ will be transmitted through the molecular barrier, it is obvious that the transmitted electron can carry energy different from $E_0$. As an example consider the case where the bridge has only one intermediate state, i.e. $N=1$. Within the same model and approximations as outlined above it is possible[t] to obtain the *energy resolved transmission*. In the Markovian limit ($\tau_c \to 0$ in Eq. (108)) the result is

$$\mathcal{T}'(E_0, E) = \mathcal{T}_0(E_0)\left[\delta(E_0 - E) + \frac{(\kappa/2\pi)e^{-\beta(E_1 - E_0)}}{(E_1 - E)^2 + (\Gamma_1/2)^2}\right] \tag{117}$$

(we use $\mathcal{T}'$ to denote the differential (per unit energy range) transmission coefficient) where $\Gamma_1 = \Gamma_1^{(L)} + \Gamma_1^{(R)}$ and $\mathcal{T}_0$ is eleastic transmission coefficient

$$\mathcal{T}_0(E_0) = \frac{\Gamma_1^{(L)}\Gamma_1^{(R)}}{(E_1 - E_0)^2 + (\Gamma_1/2)^2}$$

The total transmission coefficient, including inelastic contribution is given by

$$\mathcal{T}(E_0) = \int dE\, \mathcal{T}'(E_0, E) = \mathcal{T}_0(E_0)\left[1 + \frac{\kappa}{\Gamma_1} e^{-\beta(E_1 - E_0)}\right] \tag{118}$$

In the absence of thermal interactions ($\kappa = 0$ in Eq. (108)) $\mathcal{T}$ is reduced to $\mathcal{T}_0$, and the electron is transmitted with $E=E_0$. For a finite $\kappa$ we get an additional, thermally activated, component peaked about the energy $E_1$ of the bridge level.

---

[t] D. Segal and A. Nitzan, Chem. Phys., in press.

How will this affect the conduction? It has been argued (see[219] chapter 2.6) that simple expressions based on the Pauli principle (e.g. Eqs. (21), (35)) are not valid in the presence of inelastic processes including thermal relaxation. It may still be used however in the weak metal-bridge coupling limit (see discussion in Section 2.12). Proceeding along this line, an equation equivalent to (35) can be written

$$I = \frac{e}{\pi\hbar}\int_0^\infty dE_0 \int_0^\infty dE\, \mathcal{T}'(E_0,E)\left[ f(E_0)(1-f(E+e\Phi)) - f(E_0+e\Phi)(1-f(E)) \right] \qquad (119)$$

For small bias and low enough temperature (so that $f(E+e\Phi) \sim f(E) - e\Phi\delta(E-E_F)$) this leads to[t]

$$g(E_0) = \frac{I}{\Phi} = \frac{e^2}{\pi\hbar}\mathcal{T}_0(E_0)\left(1 + (1-f(E_1))\frac{\kappa}{\Gamma_1}e^{-\beta(E_1-E_0)}\right) \qquad (120)$$

The equivalent result for electron transfer rates is familiar: at zero temperature the rate is determined by a tunneling probability, and at higher temperature an activated component takes over. For an experimental manisfestation of this behavior see, e.g. [276]. It is also interesting to examine the bridge length dependence of the transfer rate and the associated conduction. Here analytical results are combersome but numerical evaluation of the rate, Eq. (112), and the transmission coefficients (115) and (116) in terms of the system parameters (Hamiltonian couplings and the parameters $\kappa$ and $\tau_c$ of Eq. (108)) is straighforward.[119] Figure 11 shows the conduction (in units of $e^2/\pi\hbar$) obtained from such a model calculation using $V_B$=0.05eV, $\Delta E_B = E_B - E_F$ = 0.2eV, $\Gamma_1^{(L)} = \Gamma_N^{(R)} = 0.1 eV$, $\tau_c$=0, 0.1eV and $T$=300K and 500K, plotted against the number of bridge segments $N$ for two different temperatures. An exponential dependence on $N$, characteristic of the superexchange model, is seen to give way to a weak bridge length dependence at some cross-over value of $N$. Further analysis of this results[119, 120] reveals that the dependence on bridge length beyond the cross-over may be written in the form $\left(k_{up}^{-1} + k_{diff}^{-1}N\right)^{-1}$, where $k_{up}$ is the rate associated with the thermal activated rate from the Fermi-level into the bridge, while $k_{diff}$ corresponds to hopping (diffusion) between bridge sites. As $N$ increases, the conduction behaves as $N^{-1}$, indicating Ohmic behavior. This inverse length dependence should be contrasted with non-directional diffusion, where the rate to reach a distance $N$ from the starting position behaves like $N^{-2}$. Furthermore, if other loss channels exist, so carriers may be redirected or absorbed with a rate $\Gamma_B$ once they populate the bridge, the bridge length dependence again becomes exponential and may be written $g \sim \left(k_{up}^{-1} + k_{diff}^{-1}N\right)^{-1}e^{-\alpha N}$, where $\alpha$ is related to this loss rate.[49, 278-281] Table 1[119] summarizes these results for the Markovian limit of the thermal relaxation process

Table 1 Bridge length dependence of the transmission rate[119]

| Physical Process | Bridge length ($N$) dependence | |
|---|---|---|
| Super exchange (*small N, large $\Delta E_B/V_B$, large $\Delta E_B/k_B\Theta$*) | $e^{-\beta' N}$ | $\beta' = 2\ln(V_B/\Delta E_B)$ |
| Steady state hopping (*large N, small $\Delta E_B/V_B$, small $\Delta E_B/k_B\Theta$*) | $N^{-1}$ | |
| Non-directional hopping (*large N, small $\Delta E_B/V_B$, small $\Delta E_B/k_B\Theta$*) | $N^{-2}$ | |
| Intermediate range (*intermediate N, small $\Delta E_B/V_B$*) | $\left(k_{up}^{-1} + k_{diff}^{-1} N\right)^{-1}$ | $k_{up} \sim \left(V_B^2 \kappa/\Delta E^2\right) e^{-\Delta E_B/k_B\Theta}$ $k_{diff} \sim \left(4V_B^2/\kappa\right) e^{-\Delta E_B/k_B\Theta}$ |
| Steady state hopping + competing loss at every bridge site | $e^{-\alpha N}$ | $\alpha = \sqrt{\Gamma_B(\Gamma_B+\kappa)}/2V_B$ |

Observing the behaviors indicated by this Table experimentally is not easy since it is usually not possible to change the length of a molecular bridge without affecting its other properties, e.g. the positions of molecular HOMOs and LUMOs relative to donor and acceptor energies or an electrode Fermi energy.[282] A nice example of a cross-over behavior observed in a LEET experiment (see section 6) as a function of thickness of an absorbed molecular layer is seen in Fig. 12. Here electrons are injected into N-hexane films adsorbed on a polycrystaline Pt foil at energies below the bottom of the conduction band (~0.8eV). The role of bridge states is here assumed by impurity states in the hydrocarbon band gap. Since the energy and localization position of these states is not known, the observed results cannot be quantitatively analyzed with the model described above. However a crossover from tunneling to hopping behavior is clearly seen.

*Strong thermal coupling.* The weak system-thermal bath coupling model discussed above rests on two approximations: (a) The system-bath interaction can be considered in low order, and (b) the bath degrees of freedom are essentially unaffected by the electronic process. Using these assumptions has enabled us to obtain the general charactersitics of electron transmission through molecular barriers in the presence of barrier-localized thermal interactions.

When the interaction between the electronic system and the underlying bath is stronger these assumptions break down, and distortions in the bath configuration induced by the electronic process can play an important role. One example is the analysis of Ness and Fischer[266] discussed below Eq. (98), where coupling to phonons increases the overall transmission because of the existance of polaron state below the conduction band edge of the electronic system. However, because the overall transmission efficiency depends both on energetics (the polaron state lowers the effective barrier height) and coupling strength (small nuclear overlaps between distorted and undistorted nuclear configurations decreases the effective coupling) the issue is more involved and, depending on details of coupling and frequencies, both enhancement or reduction of transmission probabilities can occur. Similarly, at finite temperatures, the relative importance of the two transmission routes, tunneling and the activated hopping, is sensitive to these details. Relatively simple results are obtained in the particular limit where the thermal coupling is strong while the bare electronic coupling $V_B$ is weak.[t] In this case it may still be assumed that the bath degrees of freedom remain in thermal equilibrium throughout the process. Taking the bath to be a system of harmonic oscillators, $H_B = \sum_\alpha \left[ \left( p_\alpha^2 / 2m_\alpha \right) + \left( m_\alpha \omega_\alpha^2 / 2 \right) x_\alpha^2 \right]$ and taking $F_n$ in Eq. (105) to be linear in the coordinates $x_\alpha$

$$F_n = (1/2) \sum_\alpha C_{n\alpha} x_\alpha \qquad (121)$$

(so that the Hamiltonian (104) is similar to the polaron-type Hamiltonian used in Eqs. (78)-(81) and (92)-(95)), a small polaron transformation is applied in the form

$$\begin{aligned} \mathcal{H}' &= U \mathcal{H} U^{-1} \\ U &= U_1 U_2 ... U_N \\ U_n &= \exp\left( -|n><n| \Omega_n \right) \\ \Omega_n &= \sum_\alpha \Omega_{n\alpha} \; ; \quad \Omega_{n\alpha} = \frac{C_{n\alpha} p_\alpha}{2 m_\alpha \omega_\alpha^2} \end{aligned} \qquad (122)$$

leading to the transformed Hamiltonian

$$\begin{aligned} \mathcal{H}' &= H + H_B + F' + E_{shift} \\ F' &= V_B \sum_{n=1}^{N-1} \left( |n><n+1| e^{i(\Omega_{n+1} - \Omega_n)} + |n+1><n| e^{-i(\Omega_{n+1} - \Omega_n)} \right) \\ E_{shift} &= -\frac{1}{8} \sum_n \sum_\alpha \frac{C_{n\alpha}^2}{m_\alpha \omega_\alpha^2} |n><n| \end{aligned} \qquad (123)$$

where $H$ is given by Eqs. (99)-(101). If $V_B$ is small the procedure based on the Redfield approximation, that lead to Eq. (111), can be repeated. Note that keeping only terms up to second order in F' still includes terms of arbitrary order in the system-bath coupling. This procedure leads to[t]

$$\dot{\sigma}_{jk} = -i\omega_{jk}\sigma_{jk} - iV_B \sum_m \left( <F'_{jm}> \sigma_{mk} - <F'_{mk}> \sigma_{jm} \right)$$

$$+ V_B^2 \sum_{l,m} \left\{ \sigma_{lm} \int_0^\infty d\tau \left( \left\langle \tilde{F}_{mk}(0)\tilde{F}_{jl}(\tau) \right\rangle e^{-iE_{lk}\tau} + \left\langle \tilde{F}_{mk}(\tau)\tilde{F}_{jl}(0) \right\rangle e^{-iE_{jm}\tau} \right) \right. \quad (124)$$

$$\left. - \sigma_{mk} \int_0^\infty d\tau \left\langle \tilde{F}_{jl}(\tau)\tilde{F}_{lm}(0) \right\rangle e^{-iE_{lk}\tau} - \sigma_{jm} \int_0^\infty d\tau \left\langle \tilde{F}_{ml}(0)\tilde{F}_{lk}(\tau) \right\rangle e^{-iE_{jl}\tau} \right\}$$

where $\tilde{F} = F' - <F'>$. The terms in the first line of Eq. (124) account for coherent motion with a modified coupling operator, while the terms proportional to $V_B^2$ describe incoerent hopping between bridge sites. An important new element in this formulation is the themperature dependent renormalization of the coupling responsible for the coherent transmission. Using Eq. (123) results in

$$<F'> = \exp(-S_T)$$
$$S_T = (1/2) \sum_\alpha d_{n\alpha}^2 (2\bar{n}_\alpha + 1)$$
$$\bar{n}_\alpha = \left( \exp(\omega_\alpha / k_B T) - 1 \right)^{-1} \qquad (125)$$
$$d_{n\alpha}^2 = \frac{(C_{n\alpha} - C_{n+1,\alpha})^2}{8m_\alpha \omega_\alpha^3}$$

so that coherent transfer becomes less important at higher temperatures. This reduction in the coherent hopping rate is associated with the small overlap between bath degrees of freedom accomodating the electron at different sites. In fact <F'> is recognized as the thermally averaged Franck Condon factor associated with the electron transfer between two neighboring bridge sites. In terms of the spectral density

$$J(\omega) = \frac{\pi}{2} \sum_\alpha \frac{(C_{n\alpha} - C_{n+1,\alpha})^2}{m_\alpha \omega_\alpha} \delta(\omega - \omega_\alpha) \qquad (126)$$

(independent of n if the bridge sites are equivalent) we have

$$S_T = \frac{1}{8\pi} \int_0^\infty \frac{J(\omega) \coth(\omega / 2k_B T)}{\omega^2} d\omega \xrightarrow[\text{finite } T]{\omega \to 0} \frac{k_B T}{4\pi} \int_0^\infty \frac{J(\omega)}{\omega^3} d\omega \qquad (127)$$

Depending on the spectral density this integral may diverge. More specifically, if $J(\omega) \sim \omega^s$ with s<2, $S_T$ diverge at any finite temperature and the coherent route is blocked. In other cases the coherent route quickly becomes insignificant with increasing temperature.

We have gone with some length into this discussion of thermal relaxation and dephasing effects in bridge assisted electron transport both because these effects are inherently important in determining transport and conduction properties of molecular junctions, and because the issue of heat generation in these current carrying nano-structures is intimately related to these relaxation phenomena. As we have seen this problem is far from being solved and more research along these lines should be expected.

## 4. Electron tunneling through water

Electron tunneling through water is obviously an important element in all electron transfer processes involving hydrated solutes, and in many processes that occur in water based electrochemistry. Still, only a few systematic experimental studies of the effect of the water structure on electron transfer processes have been done.[73, 76, 80, 283-290] Porter and Zinn[80] have found, for a tunnel junction made of a water film confined between two mercury droplets, that at low (<1nm) film thickness conduction reflects the discrete nature of the water structure. Nagy[76, 288, 289] have studied STM current through adsorbed water layers and has found that the distance dependence of the tunneling current depends on the nature of the substrate and possibly indicates the existing of resonance states of the excess electron in the water layer. Vaught et al[287] have seen a non-exponential dependence on tip-substrate distance of tunneling in water, again indicating that at small distances water structure and possibly resonance states become important in affecting the junction conductance. Several workers have found that the barrier to tunneling through water is significantly lower than in vacuum for the same junction geometry.[73, 283-286, 289-291] The observed barrier is considerably lower than the threshold observed in photoemission into water[292, 293] and, in contrast to tunneling in vacuum, can not be simply explained by image effects.[80]

The present section focuses on attempts[294-302] to correlate these observations with numerical and theoretical studies. In the spirit of most calculations of electron transfer rates (as in Sect. 2) and of earlier dielectric continuum modes that neglect the water structure altogether, we assume at the outset that in films consisting of a few monolayers transmission is dominated by elastic processes. The discussion of Sect. 3 emphasizes the need to justify this assumption. Since we are dealing with negative energy (tunneling) processes, electronic excitations of water molecules by the transmitting electron can be ruled out. In addition, photoemission through thin water films adsorbed on metals indicates that inelastic processes associated with the water nuclear motion contributes relatively weakly at such energies.[303, 304] Numerical simulations of sub-excitation electron transmission through 1-4 water monolayers adsorbed on Pt(1,1,1)[305] are in agreement with this observation.[u] Theoretical calculations of inelastic tunneling[309] similarly show that sufficiently far from resonance the overall transmission is only weakly affected by inelastic processes. In both cases this can be rationalized by the short interaction times (see Ref [305] and Section 3.1). In such cases a *static medium assumption* appears to provide a reasonable starting point for discussing the overall transmission, i.e. we assume that the transmission event is completed before substantial nuclear motion takes place. The computation of the transmission probability can therefore be done for individual static water configurations sampled from an equilibrium ensemble, and the results averaged over this ensemble. This assumption is critically examined below. It should be emphasized that while solvent nuclear motion is slow relative to the transmission timescale, solvent electronic response (electronic polarizability) is not. We return to this issue also below.

In section 2 we have summarized theoretical and computational approaches available for studying electron transfer and electron transmission. The following account (see also [301]) summarizes recent computational work on electron transmission through water that use the pseudo-potential method. [294-300, 302] Here the detailed information about the electronic structure of the molecular spacer is disregarded, and replaced by the assumption that the underlying electron scattering or tunneling can be described by a one electron potential surface. This potential is taken to be a superposition of the vacuum potential experienced by the electron and the interaction potential between an excess electron and the molecular spacer. The latter is written as a sum of terms representing the interaction between the electron and the different atomic (and sometimes other suitably chosen) centers. The applicability of this method depends on our ability to construct reliable pseudopotentials of this type. In the work described below we use the electron-water pseudopotential derived and tested in studies of electron hydration,[310] and a modified pseudopotential that includes the many-body interaction associated with the water electronic polarizability. Other electron-solvent pseudopotentials have been used for water,[311] ammonia[312] methanol,[313] rare gases[314] and hydrocarbons. [315]

With such a potential given, the problem is reduced to evaluating the transmission probability of an electron when it is incident on the molecular layer from one side, say the left.

---

[u] It should be kept in mind that energy transfer from the transmitting electron to water nuclear degrees of freedom, the mechanism responsible for capturing and localizing the electron as a solvated species must play an important role for thicker layers.[306-308].

In recent years various time dependent and time independent numerical grid techniques were developed for such calculations. In the time dependent mode an electron wavepacket is sent towards the molecular barrier, and propagated on the grid using a numerical solver for the time dependent Schrödinger equation. This propagation continues until such time $t_f$ at which the 'collision' with the barrier has ended, i.e. until the probability that the electron is in the barrier region, $\int_{barrier} |\psi(\mathbf{r},t)|^2 d\mathbf{r}$, has fallen below a predetermined margin. Since only the result at the end of the time evolution is needed, a propagation method based on the Chebychev polynomial expansion of the time evolution operator[316, 317] is particularly useful.

In the time independent mode, Nitzan and coworkers [295-299, 302, 318, 319] have applied the spatial grid based absorption boundary condition Green's function (ABCGF) technique described in Section 2.7 (Eqs. (63) and (66)). Taking $x$ be the tunneling direction, periodic boundary conditions are used in the $y$-$z$ plane parallel to the molecular layer, and the absorption function, $\varepsilon(r)=\varepsilon(x)$, is taken to be different from zero near the grid boundaries in the z direction, far enough from the interaction region (i.e. the tunneling barrier), and gradually diminishing to zero as the interaction region is approached from the outside. The stability of the computed transmission to moderate variations of this function provides one confidence test for this numerical procedure. The cumulative microcanonical transition probability and the one-to-all transition rates are calculated as outlined in Sect. 2.7. In addition, exact outgoing and incoming wavefunctions $\Psi_i^+$ and $\Psi_f^-$ which correspond to initial and final states (eigenfunctions of $H_0$ with energy E) $\phi_i$ and $\phi_f$, respectively, can be computed from

$$\psi_i^+ = \frac{1}{E-H+i\varepsilon} i\varepsilon\phi_i$$
$$\psi_f^- = \frac{1}{E-H-i\varepsilon} (-i\varepsilon)\phi_f$$
(128)

and provide a route for evaluating state selected transition probabilities, $S_{if}= \langle\psi_f^-|\psi_i^+\rangle$. The evaluation of these expressions requires (a) evaluating the Hamiltonian matrix on the grid, and (b) evaluating the operation of the corresponding Green's operator on a known vector. In the implementation of Refs [296-299] 7$^{th}$ order finite-differencing representation is used to evaluate the kinetic energy operator on the grid. As in most implementations of grid Hamiltonians the resulting matrix is extremely sparse, suggesting the applicability of Krylov space based iterative methods such as the Generalized Minimum Residual method (GMRES),[320] or Quasi Minimal Residual method (QMR). [321]

While considerable sensitivity to the water structure is found in these studies, water layers prepared with different reasonable water-water interaction models had similar transmission properties.[297, 298] On the other hand the results are extremely sensitive to the choice of the electron-water pseudopotential. Most previous studies of electron solvation in water represent the electron-water pseudopotential as a sum of two-body interactions. Studies of electron hydration and hydrated electron spectroscopy show that the potential developed by Barnett et al[310] as well as that developed by Schnitker and Rossky[311] could account semi-quantitatively for the general features of electron solvation structure and energetics and in water and water clusters. Taking into account the many-body aspects of the electronic polarizability contributions to the electron-water pseudopotential[322] have lead to improved energy values that were typically different by 10-20% from the original results. In contrast, including these many-body interactions in the tunneling calculation is found (see below) to make a profound effect, an increase of ~ 2 orders of magnitudes in the transmission probability of electron through water in the deep tunneling regime. There are two reasons for this: First, as already noted, tunneling processes are fast relative to characteristic nuclear relaxation times. The latter is disregarded, leaving the electronic polarizability as the only solvent response in the present treatment. Secondly, variations of the interaction potentials enter exponentially into the tunneling probability, making their effects far larger than the corresponding effect on solvation. It should be kept in mind that including the solvent electronic polarizability in simulations of quantum mechanical processes in solution raises some conceptual difficulties. The simulation results described below are based on the approach to this problem described in references [295] and [298]. In what follows model B refers to the the corrected electron-water pseudo-potential used in these papers while model A refers to the original pseudopotential of Barnett et al.[310] (see the original publications[295-299, 301] for details of the water-water and water-metal potentials used in these calculations.

The results described below illustrate the principal factors affecting the transmission process: (a) the dimensionality of the process, (b) the effect of layer structure and order, (c) effect of resonances in the barrier and (d) signature of band motion. The simulations consist of first preparing water layer structures on (or between) the desired substrates using classical MD simulations; secondly, setting the Schrödinger equation for the electron transmission problem on a suitable grid and, finally, computing the transmission probabilities.

Figure 13 shows results of such calculations for the transmission probability as a function of the incident electron energy. The results for the polarizable model (B) are seen to be in remarkable agreement with the expectation based on lowering of the effective rectangular barrier by 1.2eV, while those obtained using model A, which does not take into account the many-body nature of the interaction associated with the water electronic polarizability, strongly underestimates the transmission probability. In fact, model A predicts transmission probability in water to be lower than in vacuum, in qualitative contrast to observations.

Next consider the effect of orientational ordering of water dipoles on the metal walls. Water adsorbs with its oxygen on the metal surface and the hydrogen atoms pointing away from it, leading to net surface dipole density directed away from the wall. Simulations yield ~$5 \cdot 10^{-11}$ Coul/m for this density.[v] This is an important factor in the reduction of the surface work function of many metals due to water adsorption.[292, 323, 324] Fig. 14 compares, for Model A, the transmission probabilities computed with two water configurations (sampled as described in Fig. 13). One is the same as the model A result shown in Fig. 13 and the other is obtained from a similar model in which the attractive oxygen-metal wall interaction, therefore the preferred orientational ordering, was eliminated.[295] We see that the existence of surface dipole in the direction that reduces the work function is associated with a larger transmission probability as expected.

Traditional approaches to electron transfer are based on continuum dielectric picture of the solvent, where the issue of tunneling path rarely arises. Barring other considerations, the exponential dependence of tunneling probabilities on the path length suggests that the tunneling process will be dominated by the shortest possible, i.e. 1-dimensional, route. A closer look reveals that electron tunneling through water is inherently 3-dimensional (see e.g. Fig. 7 of Ref. [295]). An interesting demonstration of the importance of the 3-d structure of the water layer in determining the outcome of the tunneling process is shown in Fig. 15. This Figure compares, using the configuration of Fig. 13 and model B at room temperature, tunneling through the given water layer and tunneling through another water configuration that was prepared in the presence of a strong electric field pointing along the tunneling (x) axis. In the resulting layer structure the water dipoles point on the average along this axis. This structure is frozen and the electric field used to generate it is removed during the tunneling calculation. The computed one-to-all transmission for electrons incident in the x direction shows several orders of magnitude difference between the probabilities calculated for electron incident in the direction of the induced polarization and against this direction. Microscopic reversibility implies that the corresponding 1-dimensional process should not depend on the tunneling direction, positive or negative, along the x axis. The observed behaviour is therefore associated with the 3-dimensional nature of the process. it shows that the angular distribution associated with the transmission through such layer depends strongly on the transmission direction, and suggests that asymmetry in current-voltage dependence of transmission current should exist beyond the linear regime.

---

[v] I. Benjamin and A. Nitzan, unpublished results

Next consider the possibility of resonance assisted tunneling. Fig. 16 shows such resonances in a range of ~1eV below the 5eV vacuum barrier. The existence of such resonances correlates with the observation of weakly bound states of an electron in neutral configurations of bulk water. Mosyak et al[298] have found that such states appear in neutral water configurations in both models A and B, however only model B shows such states at negative energies. Moreover, these states are considerably more extended in systems described by model B compared with the corresponding states of model A.[298] The possible effect of bound electron states in water on electron transmission probability through water was raised by several workers in the past.[325-327] Peskin et al[302] have recently identified the source of the resonances seen in our simulations as transient vacancies in the water structure. We emphasize again that because these results were obtained for static water configurations, their actual role in electron transmission through water is yet to be clarified.

The effective barrier to electron tunneling in water has been subject to many discussions in the STM literature.[73, 286, 290, 328] While the absolute numbers obtained vary considerably depending on the systems studied and on experimental setups and conditions, three observations can be made: (a) Tunneling is observed at large tip-surface distances, sometimes exceeding 20Å.[73, 290, 328] (b) The barrier, estimated using a 1-dimensional model from the distance dependence of the observed current, is unusually low, of the order of 1eV in systems involving metals with work-functions of 4-5eV. (c) The numbers obtained scatter strongly: the estimated barrier height may be stated to be 11eV. (d) The apparent barrier height appears to depend on the polarity of the bias potential.

It should be kept in mind that even in vacuum STM the barrier to tunneling is expected to be lower than the workfunctions of the metals involved because of image effects associated with the fast electronic response of the electrodes.[91] Nevertheless, the reduction of barrier height in the aqueous phase seems to be considerably larger. Taking the vacuum barrier as input in our discussion lets consider the possible role of the solvent. These can arise from the following factors: (1) The position, on the energy scale, of the "conduction band" of the pure solvent. By "conduction band" we mean extended electronic states of an *excess* electron in the neutral solvent configuration. (2) The effect of the solvent on the electrode workfunction. (3) The hard cores of the atomic constituents, in the present case the water oxygens, which make a substantial part of the physical space between the electrodes inaccessible to the electron. (4) The possibility that the tunneling is assisted by resonance states supported by the solvent. Such resonances can be associated with available molecular orbitals - this does not appear to be the case in water- or with particular transient structures in the solvent configurations as discussed above.

Factors (2)-(4) are usually disregarded in theories of electron transfer, while a common practice is to account for the first factor by setting the potential barrier height at a value, below the vacuum level, determined by the contribution of the solvent electronic polarizability. This value can be estimated as the Born energy of a point charge in a cavity of intermolecular dimensions, say a radius of a~5au, in a continuum with the proper dielectric constant, here the optical dielectric constant of water, $\varepsilon_\infty=1.88$. This yields $e^2(2a)^{-1}[\varepsilon_\infty^{-1}-1]\sim-1.3eV$, same order as the result of a more rigorous calculation by Schmickler and Henderson,[329] and in agreement with experimental results on photoemission into water.[292, 324] It should be noted that this number was obtained for an infinite bulk of water, and should be regarded as an upper limit for the present problem.

The simulations described above shed some light on the roles played by the other factors listed above. First, we find that lowering the metal workfunction by the orientational ordering of water dipoles at the metal surface does affect the tunneling probability, see Fig. 14. Secondly, the occupation of much of the physical space between the electrodes by the impenetrable oxygen cores strongly reduces the tunneling probability. In fact, if these two factors exist alone, the computed tunneling probability is found to be considerably lower than in the corresponding vacuum process, see Fig 7 of Ref. [295]. Even including the effect of the water electronic polarizability (i.e. attractive $r^{-4}$ terms) in the two-body electron-water pseudopotential (model A) is not sufficient to reverse this trend, as seen in Fig. 13. Taking into account the full many body nature of this interaction was found to be essential for obtaining the correct qualitative effect of water, i.e. barrier lowering relative to vacuum.

The estimate of the magnitude of this lowering effect in our simulations can be done in two ways. One is to fit the absolute magnitude of the computed transmission probability to the result obtained from a 1-dimensional rectangular barrier of width given by the distance s between the electrodes. [297] This is done in Figure 17 for systems with 1-4 monolayers of water (s=3.6, 6.6, 10.0, 13.3Å).[w] The following points should be noted:

(a) The effective barrier to tunneling computed with the fully polarizable model B is reduced by at least 0.5eV (from the bare value of 5eV used in these simulations) once a 'bulk' has been developed in the water layer, i.e. once the number of monolayers is larger than 2.

(b) The equivalent calculation done with model A, in which water polarizability is accounted for only on the 2-body level, yields an effective barrier higher than the vacuum barrier.

(c) For the very thin layers studied, the effective barrier height depends on the layer thickness. This behavior (which support a recent experimental observation by Nagy[289]) is expected to saturate once a well-defined bulk is developed.

Following common practice in STM studies, another way to discuss the effective simulated barrier is to fit the distance dependence of the observed tunneling probability to the analytical result for a rectangular barrier. This practice can yield very low apparent barriers in cases where tunneling is influenced by resonance structures.[301] Moreover, since the existence and energies of these resonances in water depend on local structures that evolve in time, it is possible that the characteristic scatter of data that appears in these measurements[73, 286, 290, 328] may arise not only because of experimental difficulties but also from intrinsic system properties.

The existence in water of transient structures that support excess electron resonaces and the possible implications of these resonances in enhancing the tunneling probability and the apparent barier height raises again the issue of timescales. In particular, the lifetimes of these resonance states is of considerable interest, since they determine the duration of the electron 'capture' by the water film and, as a consequence, the possibility that water dynamics and thermal relaxation become important on this timescale. Peskin et al[302] have determined these lifetimes by a direct evaluation of the complex eigenvalues associated with the corresponding resonance structures, using a filter diagonalization method with the imaginary boundary conditions Hamiltonian. The resulting eigenvalues have imaginary parts of the order ~0.05eV, implying lifetimes of the order $\leq 10 \text{fs}$. An alternative way to probe the dynamics of electron tunneling in water is by evaluating the corresponding traversal times (see Setion 3.1). Here the timescale for possible interaction between the excess electron and barrier motions can be determined both near and away from resonance energies. Galperin et al[232] have applied the internal clock approach of Sect. 3.1 to this problem, starting from the one-to-all transmission probability, Eq. (66), written in the form

$$\sigma = \frac{1}{\pi} <\phi_{in}(E) | \varepsilon_{in}^* G\ \varepsilon_{out} G \varepsilon_{in} | \phi_{in}(E)> \qquad (129)$$

where $\phi_{in}$ denotes an incoming state in the reactant region and $\varepsilon_{in}$ and $\varepsilon_{out}$ are the absorbing boundary functions in the reactant (incoming) and product (outgoing) regions, respectively. In the present application the electron is taken to have two internal states, so that if x is the tunneling direction, $\phi_{in} = \left(e^{ikx}/\sqrt{v}\right)\binom{1}{0}$. The Green's operator is given by $G = \left(E - H_0 + i(\varepsilon_{in} + \varepsilon_{out})\right)^{-1}$ with $H_0$ replaced by

$$H = H_0 \begin{pmatrix} 1 & 0 \\ 0 & 1 \end{pmatrix} + \lambda F(x) \begin{pmatrix} 0 & 1 \\ 1 & 0 \end{pmatrix} \qquad (130)$$

where $\lambda$ is a constant and where $F(x)=1$ in the barrier region and 0 outside it. The approximate scattering wave function,

$$|\psi(E)> = iG(E)\varepsilon_{in} |\phi_{in}(E)> = \begin{pmatrix} \psi_1(E) \\ \psi_2(E) \end{pmatrix} \qquad (131)$$

is evaluated using iterative inversion methods,[320, 321]. The transmission probabilities into the $|1>$ and $|2>$ states are obtained from

$$\mathcal{T}_i(E) = <\psi_i(E) | \varepsilon_{out} | \psi_i(E)> \quad ; \qquad i = 1, 2 \qquad (132)$$

---

[w] It should be emphasized that these results were not statistically averaged over many water configurations, so the absolute numbers obtained should be taken only as examples of a general qualitative behavior.

$\mathcal{T}_i$ are equivalent to $|c_i|^2$ where $c_i$ (i=1,2) are defined above Eq. (77). Accordingly

$$\tau(E) = \lim_{\lambda \to 0} \left( \frac{\hbar}{|\lambda|} \sqrt{\frac{\mathcal{T}_2(E)}{\mathcal{T}_1(E)}} \right) \qquad (133)$$

Figs. 18 and 19[232] display some results of this calculation. Fig. 18 shows calculated traversal times as functions of incident electron energy for an electron transmitted through a layer of three water films between two platinum electrodes (the distance between the electrodes is $d$ =18.9au). Shown is $\tau/\tau_0$ for several configurations of this system, where $\tau_0$ is the tunneling time associated with the bare vacuum barrier (same geometry with no water). The transient nature of the water structures that give rise to the resonance features is seen here. Note that the difference between different configurations practically disappears for energies sufficiently below the resonance regime, where the ratio between the time computed in the water system and in the bare barrier is practically constant, approximately 1.1. Fig. 19 shows, for one of these configurations, the tunneling time and the transmission probability, both as functions of the incident electron energy. We see that the energy dependence of the tunneling time follows this resonance structure closely. In fact, the times (3-15 fs) obtained from the peaks in Figs. 18 are consistent with the resonance lifetimes estimated in Ref. [302].

We conclude this discussion with two more comments: First, in the above analysis the possibility of transient 'contamination' of the tunneling medium by foreign ions has been disregarded. Such ions exist in most systems used in underwater STM studies, and the apearance of even one such ion in the space of 10-20Å between the electrodes can have a profound effect on the tunneling current. This may add another source of scatter in the experimental results. Secondly, as already discussed, changes in the water structure between the electrodes may appear also as bias dependent systematic effects. Thus, the asymmetry in the bias dependence of the barrier height observed in Refs.[286] and[73, 290] may be related to the asymmetric transmission properties of orientationally ordered layers.

## 5. Overbarrier transmission

Our discussion so far has focused on electron transmission processes that at zero temperature can take place only by tunneling. The present section provides a brief overview of transmission processes where an electron incident on a molecular barrier carries a positive (above ground state vacuum) energy. It should be emphasized that this in itself does not mean that transmission can take place classically. If the incident energy is in the bandgap of the molecular spacer, zero temperature transmission is still a tunneling process. Still, this type of phenomena is distinct from those discussed in the other parts of this review for several reasons: first, positive energy transmission (and reflection), essentially scattering processes are amenable to initial state control and to final state resolution that are not possible in negative energy processes. Second, a positive energy electron interacts with a large density of medium states, therefore the probability for resonance or near resonance transfer is considerably larger, implying also a larger cross-section for dephasing and inelastic energy loss. Third, at this range of energies conventional quantum chemistry approaches, as well as pseudopotentials derived from low-energy electronic structure data can be very inaccurate. Finally, at high enough energies elecronic excitations and secondary electron generation become important factors in the transmission mechanism. For the last two reasons the numerical approaches described in Section 2.6-8 are not immediately applicable.

The effect of adsorbates on photoelectrons emitted from surfaces has been studied for almost a century.[330, 331] These experiments were partially motivated by their practical ramifications whereby the surface workfunction was modified by the adsorbate.[332, 333] Recently, the development of tuneable UV light sources has enabled studies of energy resolved photoelectron spectroscopy.[334] This eventaully lead to studies of photoelectron energy distribution for photoelectrons produced from metal surfaces covered with self assembled monolayers (SAMs) of organic molecules, or organized organic thin films (OOTFs).[318, 319, 335-349] These films are prepared either with the Langmuir Blodgett technique[350, 351] or by self assembly from vapor or solution. One of the earlier experiments of this kind was the measurement of transmitted electron energy distribution for photoelectrons produced from a Pt(111) surface covered with several layers of water.[304] It was found that the transmission probability decreases exponentially with increasing number of water layers, however this numebr does not affect the energy distribution of the emitted electrons, indicating that transmission in this system is independent of the electron energy and that inelastic energy loss is small. These results however should be regarded with caution in view of low energy electron transmission (LEET, see below) data[308] that indicate that energy loss from a transmitted electron to water nuclear motion may be quite efficient. The latter observation is supported by estimates[352] of the distance (20-50Å) traversed by electrons photoejected into water at subexcitation energies before their capture to form the precursor of solvated electrons.

Unlike water, the electron affinity $A = -V_0$ of hydrocarbon layers is negative, i.e. their LUMOs, or in the language of solid state physics, the bottom of their electron condition band is above vacuum energy ($V_0=0.8eV$ for bulk hydrocarbons[353]). Indeed a threshold for electron photoemission from silver covered with a monolayer of cadmium stearate $CH_3(CH_2)_{16}COO\ _2Cd^{2+}$ or arachdic acid $CH_3(CH_2)_{16}COOH$ is observed.[341] Above 0.8eV photoemission from these surfaces proceed with efficiency close to 1, turning down again at higher energies. Oscillations in the transmission probability through similar films as function of the initial electron energy were interpreted in terms of the electronic band structure of the film.[318, 340] This interpretation gains further support from the observation of the large sensitivity of the transmission probability to the film structure in the lateral dimension[343, 346] and from the strong effect of film ordering.[346] This does not exclude what is often taken to express a single molecule effect – a strong preference of the phtoemission to be directed along the axis of the molecular adsorbate.[339] The latter manifestation of ordering effect was invoked for the interpretation of the observed dependence of the photoemission yield on substrate temperature.[354] Recently, vibrational structure was observed in the photoemission spectra from gold covered with molecular layers containing benzene, naphthalene and anthracene rings.[348] These structures are usually associated with resonances related to temporary electron capture in the film, however unlike the usual assignment to temporary ion formation (see below), experimental data offer evidence to an interesting collective shape resonance, resulting from a two dimensional quantum well associated with the ordered aromatic rings in the direction parallel to the substrate surface. Finally, using chiral molecular SAMs (L or D polyalanine polypeptides) has revealed that electron transmission of spin-polarized electrons depends, with high degree of selectivity, on the chirality of the layer.[349][x]

Another way to study electron interactions with molecular layers is to send an electron beem from the vacuum side onto a molecular film condensed on a suitable, usually metalic, substarte. In the low energy electron transmission (LEET) spectroscopy developed by Sanche and coworkers the electron transmission spectrum is measured by monitoring the current arriving at the metal substrate as a function of the incident electron energy and direction. Similarly, the reflected electron beam can be analyzed with respect to energy and angular distribution, yielding electron diffraction data, energy loss spectra and energy loss excitation spectra. The same experimental setup can be used to study the effect of electron trapping, electron stimulated desorption and electron induced chemical reactions in the molecular films. For a recent review of these types of studies and references to earlier work see Sanche[355]. Here we focus on observations from LEET experiments that are relevant to our present subject. First, the prominence of the elastic and quasielastic component of the transmitted intensity, observed in most experiments of this kind, is in agreement with the photoemission experiment discussed above. Secondly, a threshold of a few tens of eV (relative to the vacuum level) is seen for transmission through alkane and through rare gas layers, indicating negative electron affinities of these layers and providing an estimate for the position of the bottom of the layers conduction bands. Third, conduction peaks below this threshold are attributed to tunneling assisted by local states inside the gap.[356] This is the analog of the bridge assisted tunneling discussed in Section 2, except that the film constitues a 3-dimensional barrier in which the local states are distributed randomly in position and energy. As discussed in Section 3.4, thermal relaxation and dephasing processes manifests themeselves in a characteristic thickness dependence of the transmission probability as the processes changes from tunneling to hopping dominated with increasing barrier width (see Fig. 12). fourth, the electron transmission spectra closely reflect the band structure of the corresponding layer. This should not be taken as an evidence for ballistic transport, in fact this observation holds for the inelatic components of the emission intensity. Rather, the electron propagation through the molecular environment is viewed as a sequence of scattering events, with cross-sections that are proportional to the density of available states.[357] The resulting averaged mean free path is therefore inversely proprtional to the density of states at an energy that (as long as the absolute energy loss is small) may be approximated by the incident energy. Finally, the transmission can be strongly affected by resonances, i.e. negative ion formation. This in turn may greatly increase the probability for inelastic energy loss.[355] These processes are observed in the high resolution electron energy loss (HREEL) spectroscopy, by monitoring the energy of reflected electrons, but they undoubtly play an equally important part in the transmission process.

---

[x] Recent results, (Carmeli I, Naaman R, Vager Z. To be published) indicate that when the chiral molecules also carry and electric dipole, the effect of chirality depends on the direction in which the electron travels along the helical structure.

As already mentioned, while the theoretical methods discussed in previous sections of these review are general, their applicability to electron transmission in the positive energy regime needs special work because standard quantum chemistry calculations usually address negative energy regimes and bound electronic states, and because pseudopotentials are usually derived from fitting results of such ab-initio calculation to analytical forms based on physical insight. Model calculations that demonstrate some of the concepts discussed above are shown in Figs. 20 and 21.[319] [318] Figure 20 compares the transmission probability ('one to all' with the incident electron perpendicular to the barrier) through a 1-dimensional rectangular barrier of height 3eV and width 1.2 nm as a function of the incident electron energy measured relative to the barrier top, to the transmission through a 3-d slab of 4 Ar layers cut out of an Ar crystal in the (100) direction. The latter results are obtained with a spatial grid technique using the electron-Ar pseudo-potential of Space et al.[314] The oscillations shown in Fig. 20a are interference patterns associated with the finite width of the layers. The full line in Fig. 20b also shows such oscillations, but in addition, a prominent dip above 4eV corresponds to a conduction band gap of this thin ordered layer. The dashed line in Fig 20b shows similar transmission results for disordered layers, obtained from the crystalline layer by a numerical thermal annealing at 400K next to an adsorbing wall using molecular dynamics propagation. The results shown are averaged over four such disordered Ar configurations. The transmission through the disordered layer is considerably less structured (smoother shapes should be obtained with more configurational averaging), in particular, the dip associated with the band gap has largely disappeared. Figure 21 compares the transmission (one-to-all) versus electron energy, for an electron incident in the normal direction on ordered Ar films made of 2, 4 and 6 atomic monolayers ('prepared' by cutting them off an Ar crystal as described above). Already at 6-layer thickness the observed transmission dip is very close to its bulk value, indicating that the band structure is already well developed.

These calculations invetigate transmission through static nuclear structures, and consequently cannot account for thermal relaxation and dephasing effects. In the other extreme limit one uses stochastic models[358, 359] that become accurate when the molecule film is thick enough so that the electron goes through mutiple scattering events before being transmitted through or reflected from the film. Such an approach has been used[306, 307, 360, 361] to describe the energy distributions of electron reflected from molecular films and its relation to the density of excess electron states in the film.

## 6. Conclusions and outlook

This review has described the current status of theoretical approaches to electron transmission and conduction in molecular junctions. In particular, Section 2 consitutes an account of theoretical approaches to this problem for *static junctions*, while Section 3 discusses approaches that focus on dephasing and thermal relaxation effects. It is important to note that even though our methodology follows a stationary, steady state viewpoint of all processes studied, the issue of relative timescales of different processes has played a central role in our analysis.

Current studies of molecular junctions focus on general methodologies on one hand and on detailed studies of specific systems on another. We have described in some details recent computations of electron transmission through water layers and have described other studies on prototypes of molecular wires. Two imporant classes of molecular wires have now become subjects of intense research, even development effort. These are DNA wires[42, 44, 47, 49, 278-281, 362-402] on one hand, and carbon nanotubes[13, 15, 403-445] on the other. While the general principles discussed in the present review apply also to these systems, the scope of recent research on special structure-function properties of these wires merits a separate coverage.

Coming back to theoretical issues, we have outlined some open problems in the methodology of treating these many-body, strongly interacting, non-equilibrium open systems. One additional direction not covered in the present review is the possibility to control the operation of such junctions using external forces (as opposed to control of function by varying the structure). Several recent studies point out the possiility to control transport processes by external fields.[446-461] The specific and selective nature of molecular optical response make molecular junctions strong potential candidates for such applications.

In conclusion, electron transmission and conduction processes in small molecular junctions combines the phenomenology of molecular electron transfer with structural problems associated with design and construction of such junctions on one hand, and with the need to understand their macroscopic transport properties on the other. In addition, the potential technological promise suggests that research in this area will intensify.

**Acknoledgements.** This work was supported by the U.S. Israel Binational Science foundation. I am indebted to many colleagues with whom I have collaborated and or discussed various aspects of this work: I. Benjamin, M. Bixon, A. Burin, B. Davis, D. Evans, M. Galperin, P. Hänggi, J. Jortner, R. Kosloff, A. Mosyak, V. Mujica, R. Naaman, U. Peskin, E. Pollak, M. Ratner, D. Segal, T. Seideman, D. Tannor, M. R. Wasielewski, J. Wilkie and S. Yaliraki. I also thank S. Datta and L. Sanche for allowing me to use Figures from their work.

**Figure Captions**

Fig. 1. Schematic views of typical electron transmission systems: (a) A 'standard' electron transfer system containing a donor, an acceptor and a molecular bridge connecting them (not shown are nuclear motion baths that must be coupled to the donor and acceptor species). (b) A molecular bridge connecting two electronic continua, L and R, representing, e.g., two metal electrodes. (c) Same as (b) with the bridge replaced by a molecular layer.

Fig. 2. A schematic view of the electronic and nuclear states involved in typical electron transmission systems. See text for details

Fig. 3. Simple level structure models for molecular electron transfer (a) and for electron transmission (b). The molecular bridge is represented by a simple set of levels that represent local orbitals of appropriately chosen bridge sites. This set of levels is coupled to the donor and acceptor species (with their corresponding nuclear environments) in (a), and to electronic continua representing metal leads (say) in (b). In the latter case the physical meaning of states 0 and N+1 depends on the particular physical problem: They can denote donor and acceptor states coupled to the continua of environmental states (hence the notation 0=D, N+1=A), surface localized states in a metal-molecule-metal junction, or they can belong to the right and left scattering continua.

Fig. 4. Tunneling gap between two metal electrodes in an unbiased (left) and a biased (right) situations. The bare gap, given by the work function W, is modified by the image interaction – the resulting barriers are represented by the curved lines.

Fig. 5. Measured and computed differential conduction of a single $\alpha,\alpha'$-xylyl dithiol molecule adsorbed between two gold contacts (From Ref.[72]). See text for details.

Fig. 6. Models for electrostatic potential profiles on a molecule connecting two metal leads with different electrochemical potentials ($\mu_i = E_F - e\Phi_i$). See discussion in text above Eq. (71).

Fig. 7. A model for current rectification in a molecular junction: Shown are the chemical potentials $\mu_L$ and $\mu_R$ in the two electrodes, and the HOMO and LUMO levels of the donor, acceptor and bridge. When the right electrode is positively biased (as shown) electrons can hop from left to right as indicated by the dotted arrows. If the opposite bias can be set without affecting too much the electronic structure of the DBA system the reverese current will be blocked.

Fig. 8. A nano-dot between two conductiong leads: A model for Coulomb blockade phenomena

Fig. 9. Parameters used in the expressions for tunneling traversal times. Left: tunneling through a rectangular barrier. Right: bridge mediated transfer, where the grey area denotes the band associated with the tight binding level structure of the bridge.

Fig. 10. The Buttiker dephasing model (see text)

Fig. 11. Finite temperature conduction of a simple tight binding model of a molecular junction as a function of bridge length $N$. See text for details

Fig. 12. (Reproduced from Ref.[356]). Transmitted current in n-hexane films as a function of thickness for various incident energies, showing the transition from tunneling to activation induced transport.

Fig. 13. (Reproduced from Ref.[298]). Electron transmission probability as a function of the incident energy. Shown are one-to-all transmission results with the electron incident in the direction normal to the water layer. These results are averaged over six equilibrium water configurations sampled from an equilibrium trajectory for the water system. This system contains 192 water molecules confined between two walls separated by 10Å, with periodic boundary conditions with period 23.5Å in the directions parallel to the walls, at 300K. These data correspond to three water monolayers between the walls. Thin dashed line: results from model A (see text). Full line: results of model B. Also shown are the corresponding results for tunneling through vacuum, i.e. through a bare rectangular potential barrier of height 5eV (dotted line), and through a similar barrier of height 3.8eV (thick-dashed line), which corresponds to the expected lowering of the effective barrier for tunneling through water.

Fig. 14. (Reproduced from Ref.[295]). Electron tunneling probabilities through water between two electrodes with (full line) and without (dotted line) orientational ordering at the metal wall.

Fig. 15. (Reproduced from Ref.[299]). Electron transmission probabilities between the two electrodes as described in the text. Full line: vacuum tunneling (bare barrier 5eV), dotted line: normal equilibrium water configuration (model B1), dashed and dashed-dotted lines: water oriented by a field 5eV/Å with tunneling direction opposite and identical to the orienting field, respectively.

Fig. 16. (Reproduced from Ref.[302]). Transmission probability vs. electron energy for electron tunneling through a water layer (model B, configuration as in Fig. 13 with bare barrier 5eV), showing tunneling resonances below the vacuum barrier).

Fig. 17. (Reproduced from Ref.[297]). Effective 1-dimensional barrier height for electron transmission through water, displayed as a function of number of water layers. Solid, dotted and dashed lines correspond to models B, A and to the bare (5eV) barrier respectively. See text for further details.

Fig. 18. (Reproduced from Ref.[232]). The ratio $\tau/\tau_0$ (see text) computed for different static configurations of (a) three and (b) four monolayer water films, displayed against the incident electron energy. The inset shows an enlarged vertical scale for the deep tunneling regime.

Fig. 19. (Reproduced from Ref.[232]). The tunneling traversal time (full line; left vertical scale) and the transmission probability (dotted line; right vertical scale) computed as functions of incident electron energy for one static configuration of the 3-monolayer water film.

Fig. 20. (Reproduced from[318, 319]). (a) Transmission probability through 1-d rectangular barrier characterized by height of 3 eV and width of 12Å, as a function of incident electron energy measured relative to the barrier top. (b) Full line: electron transmission through a slab made of 4 Ar layers, cut out of an FCC Ar crystal in the (100) direction. Dashed line: same results obtained for a disordered Ar slab (see Ref. [319] [318] for details.

Fig. 21. (Reproduced from[318, 319]). The computed transmission probabilities, Vs. Electron energy, for an electron incident on slabs cut out of an FCC Ar crystal in the (100) direction. (a) Slabs made of 2 (dashed line) and 4 (full line) monolayers. (b) Slabs with 4 (full line) and 6 (dashed line) monolayers. (The full lines in (a) and (b) are identical).

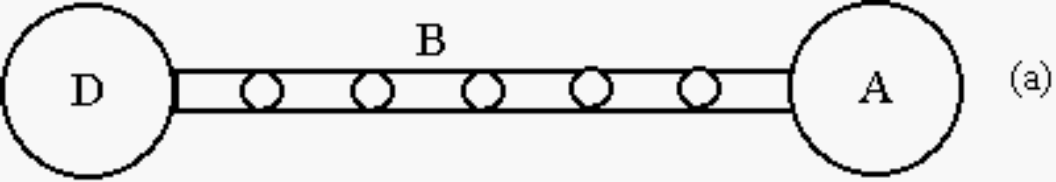

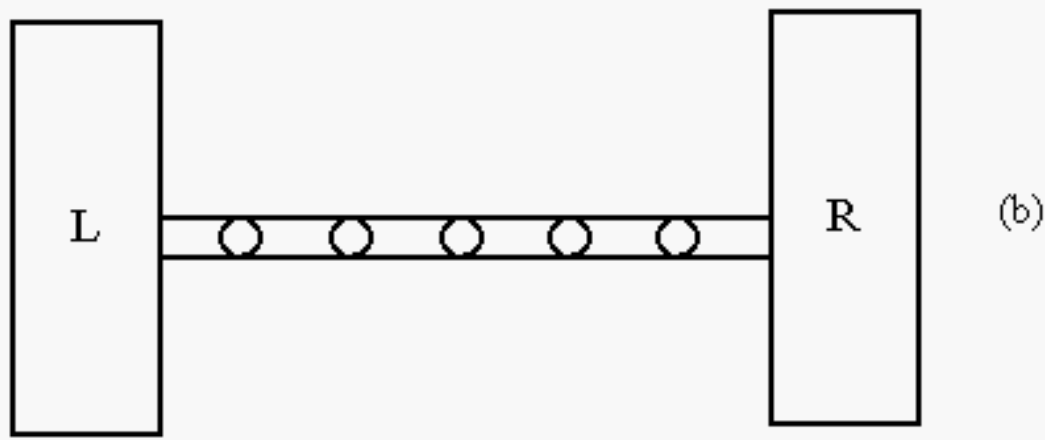

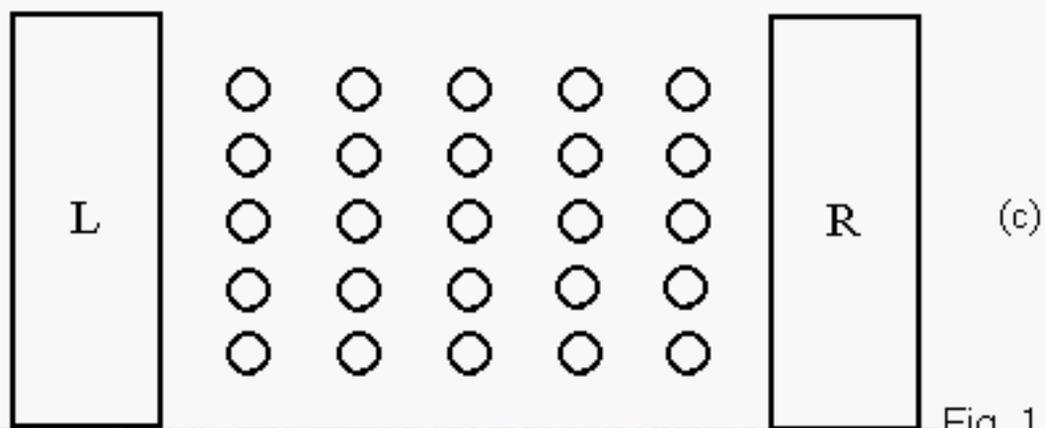

Fig. 1

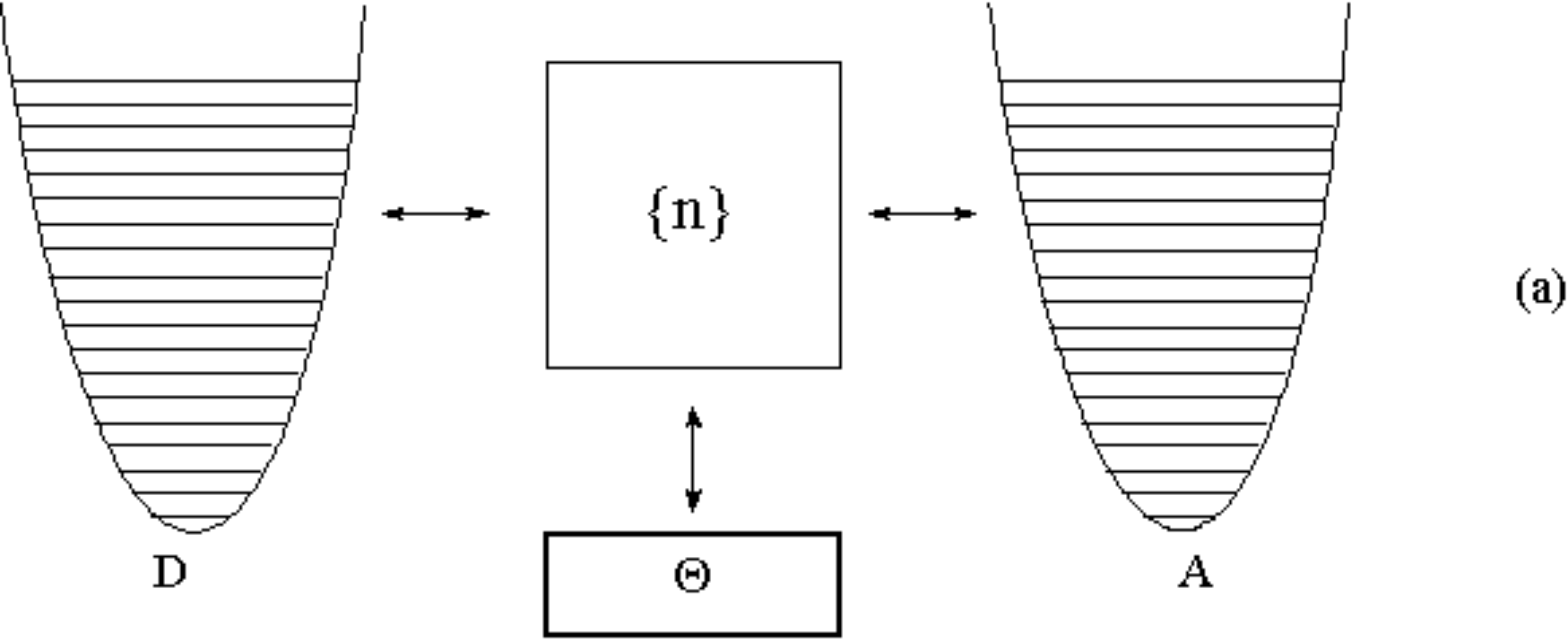

(a)

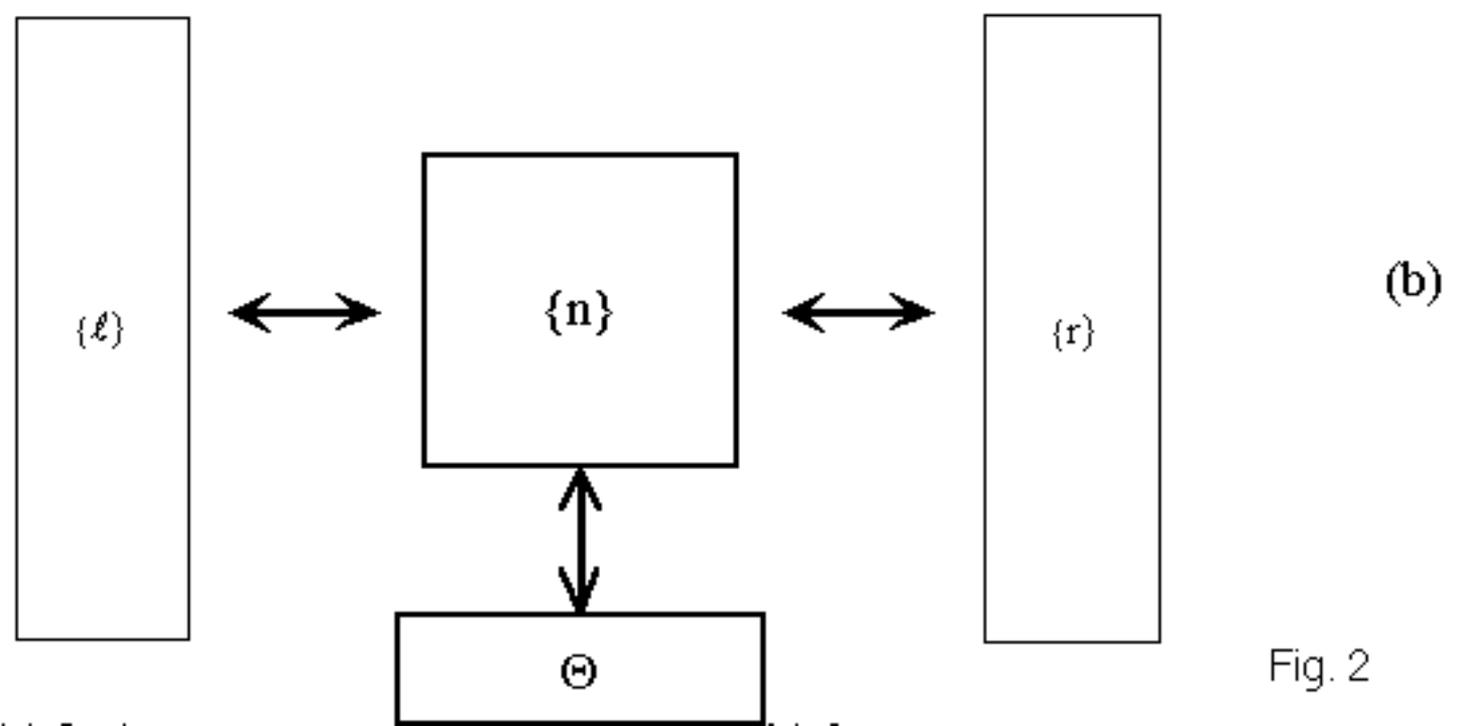

(b)

Fig. 2

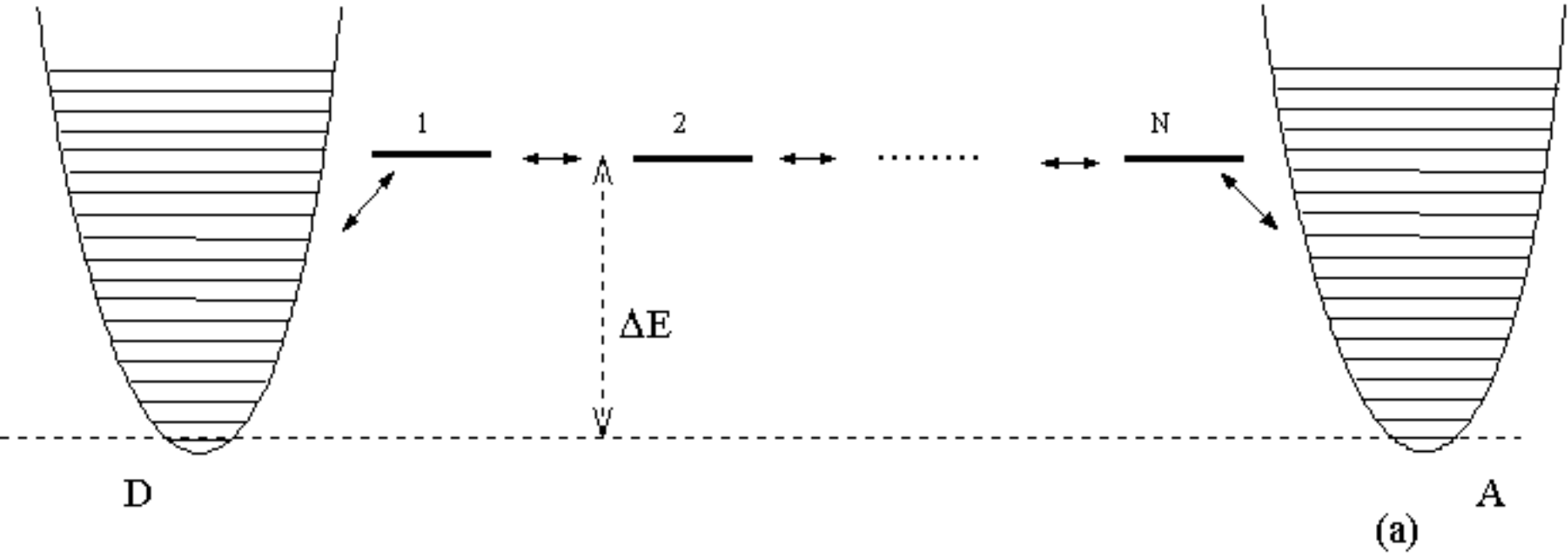

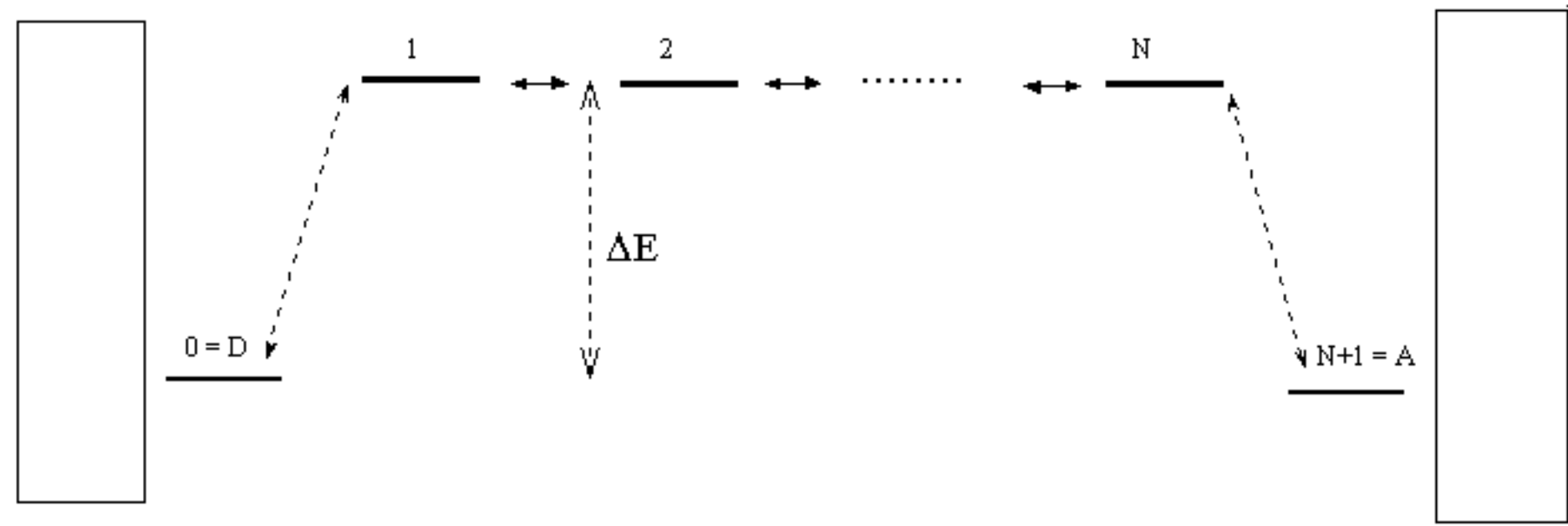

Fig. 3

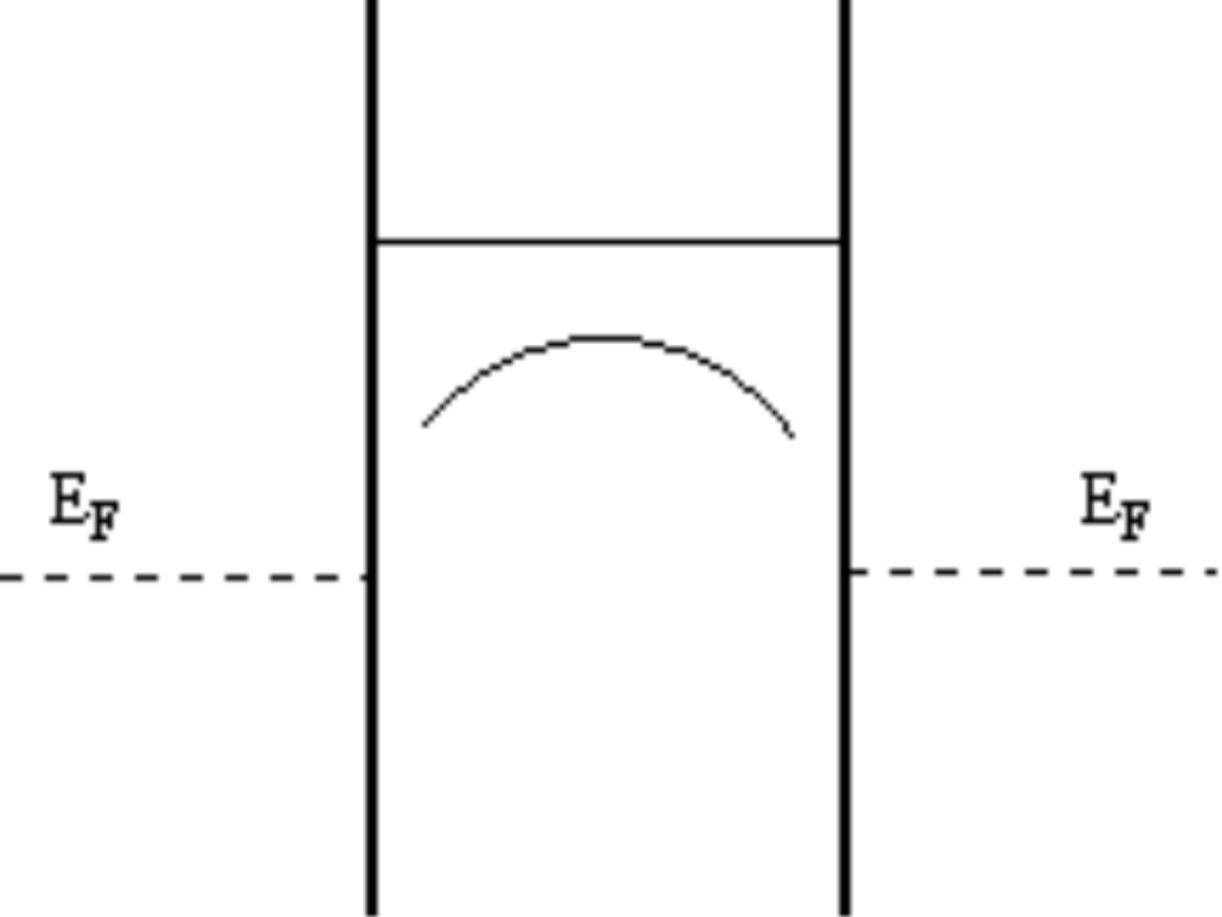 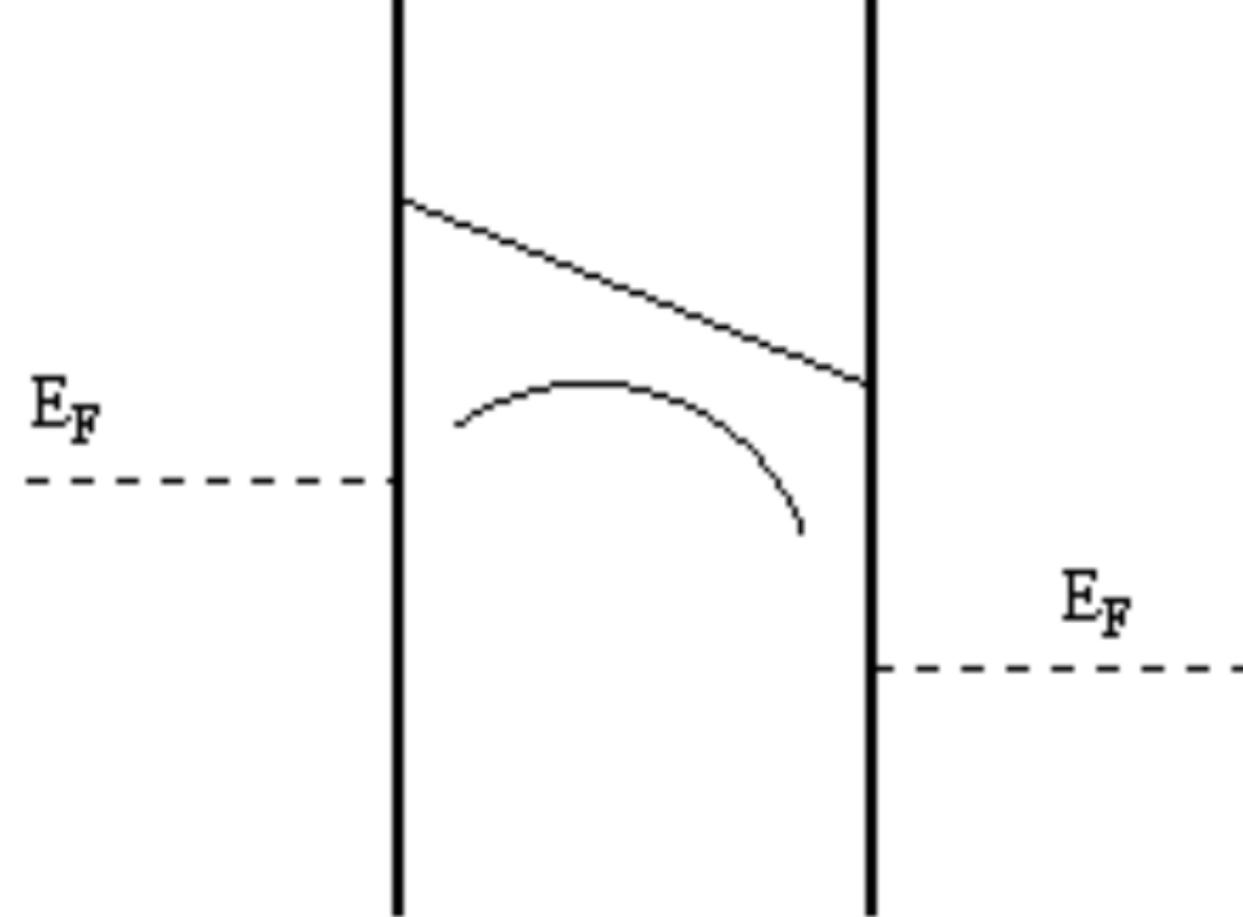

Fig. 4

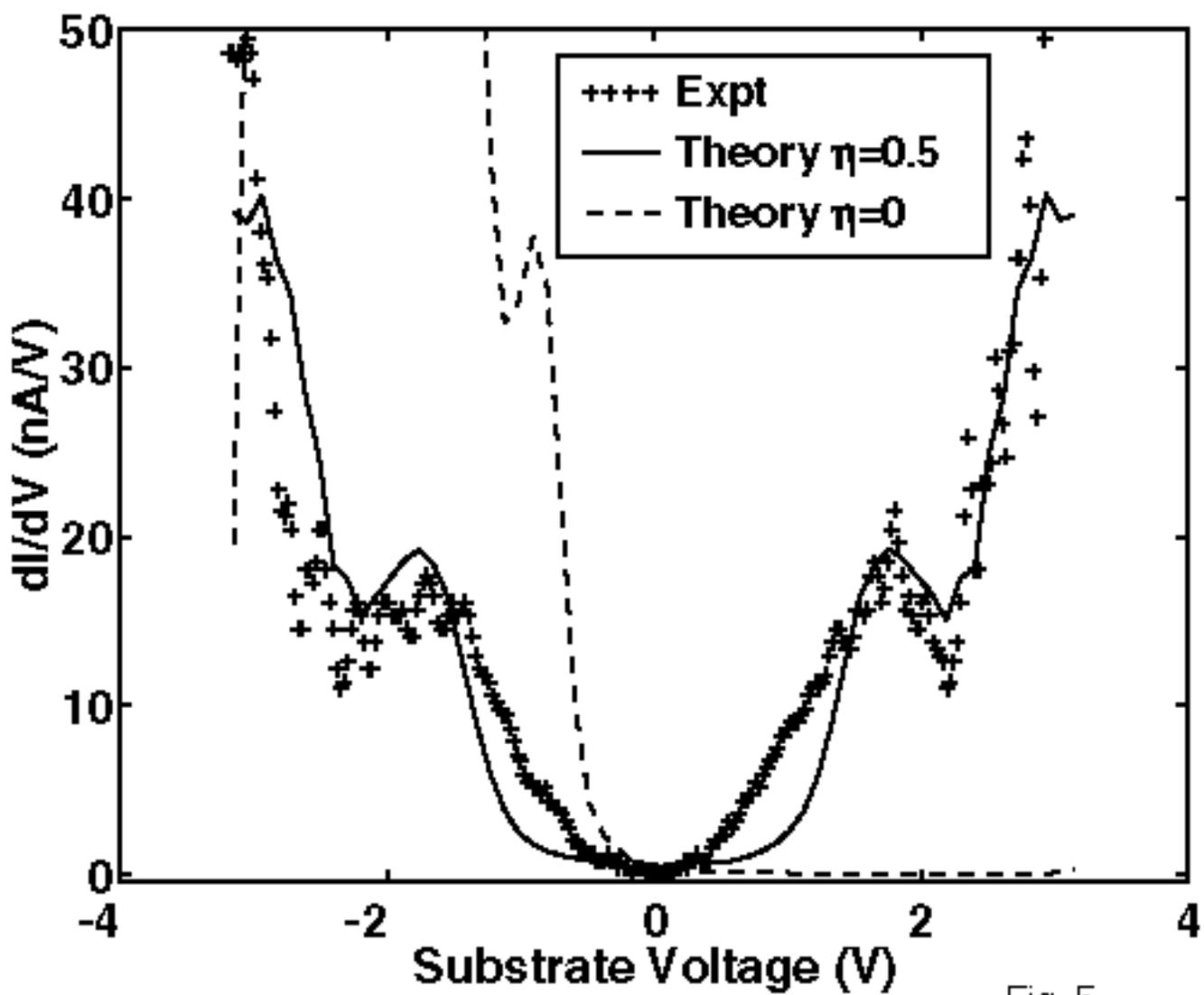

Fig. 5

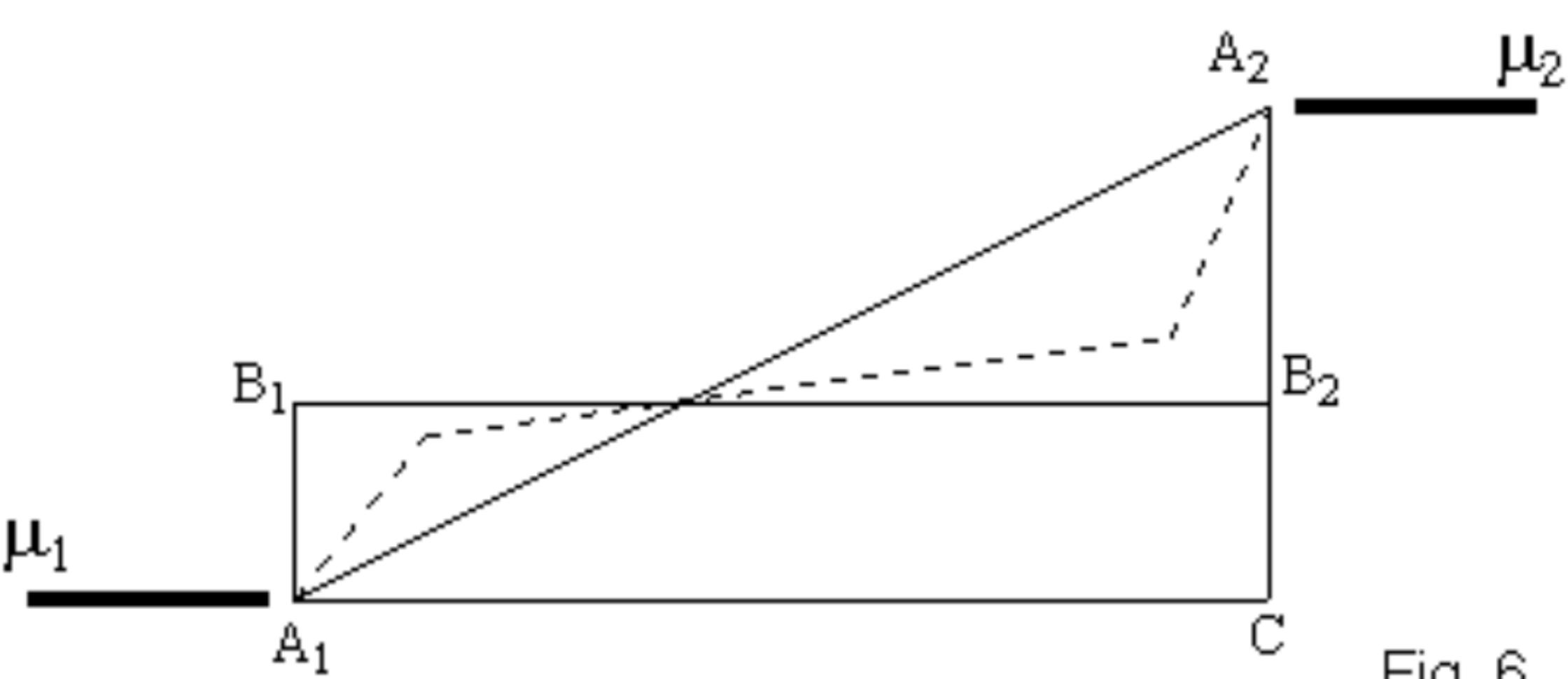

Fig. 6

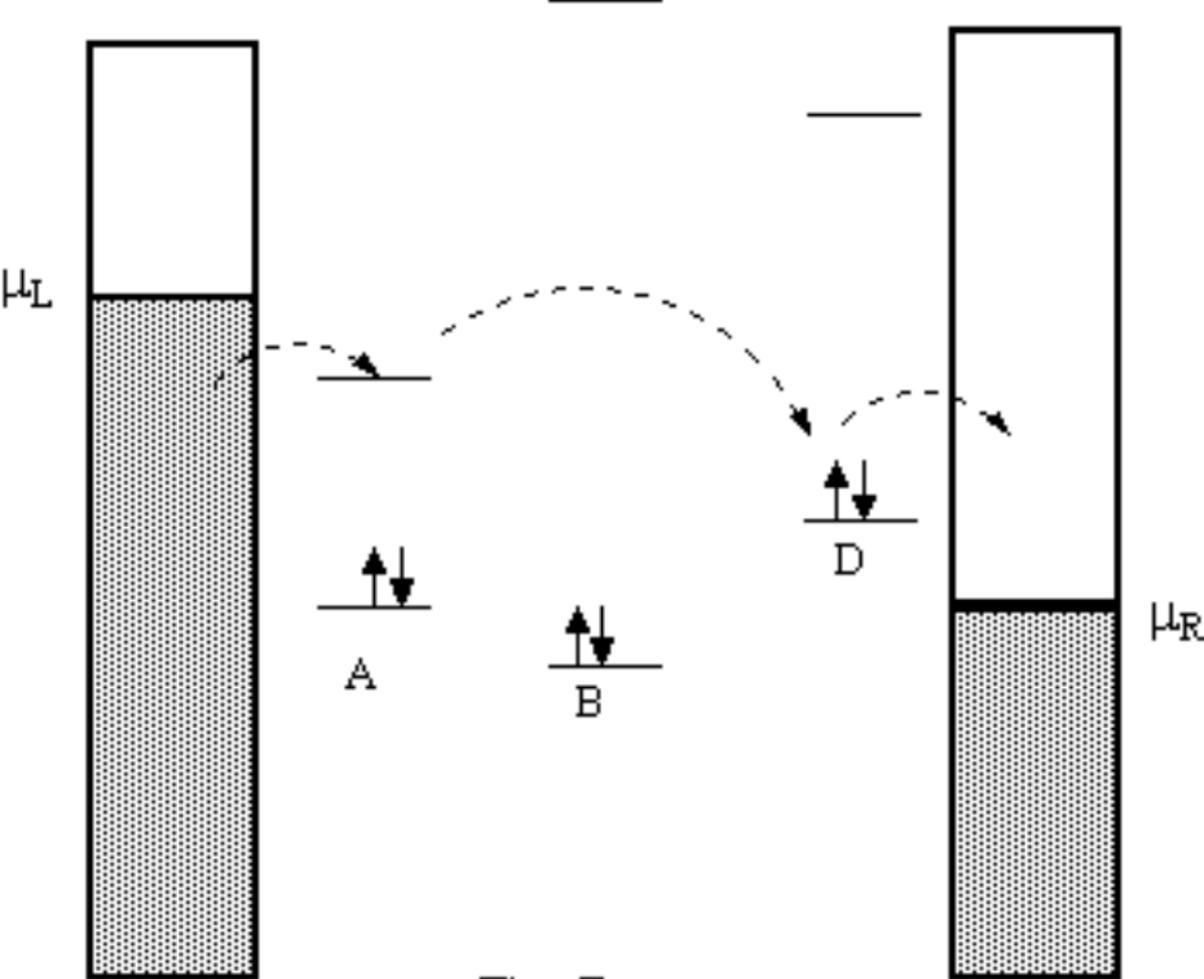

Fig. 7

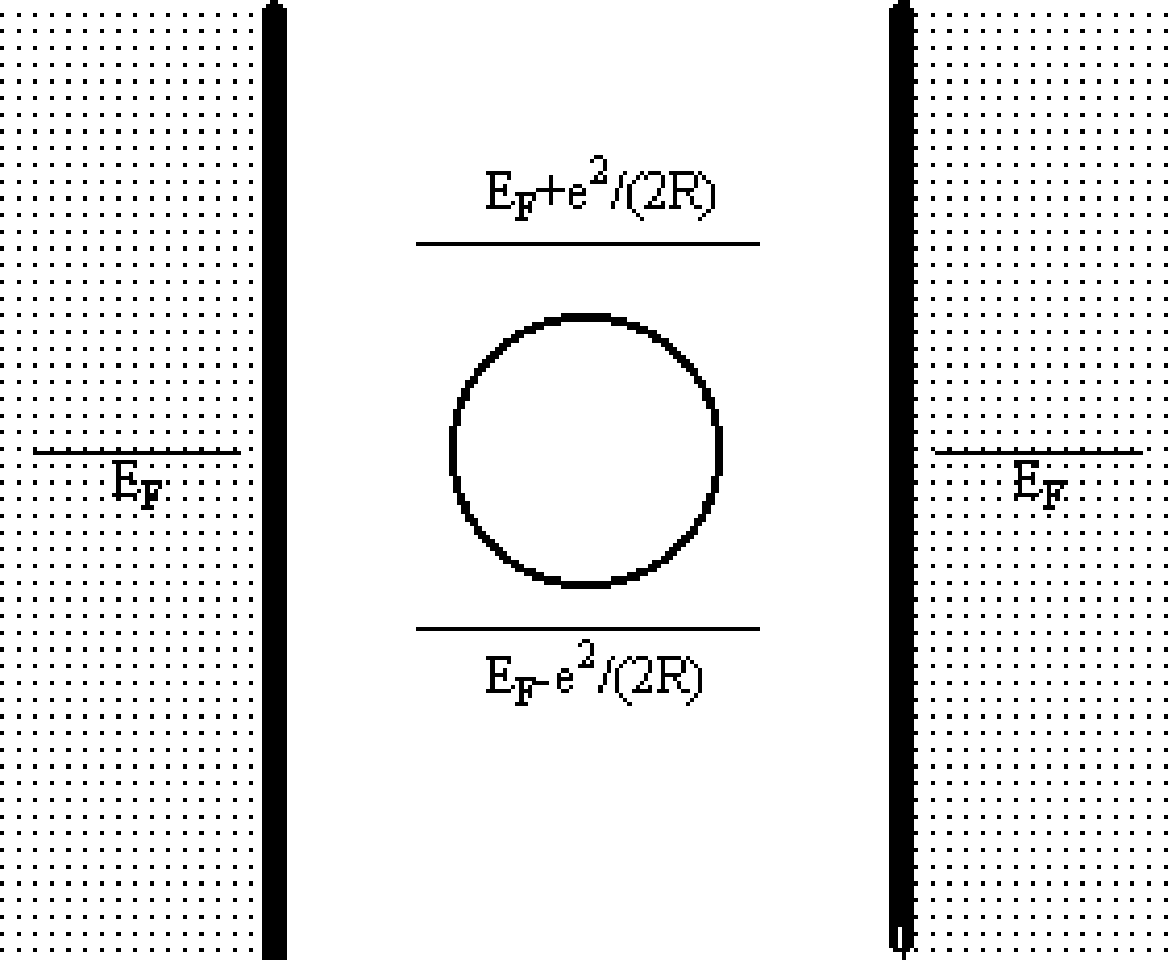

Fig. 8

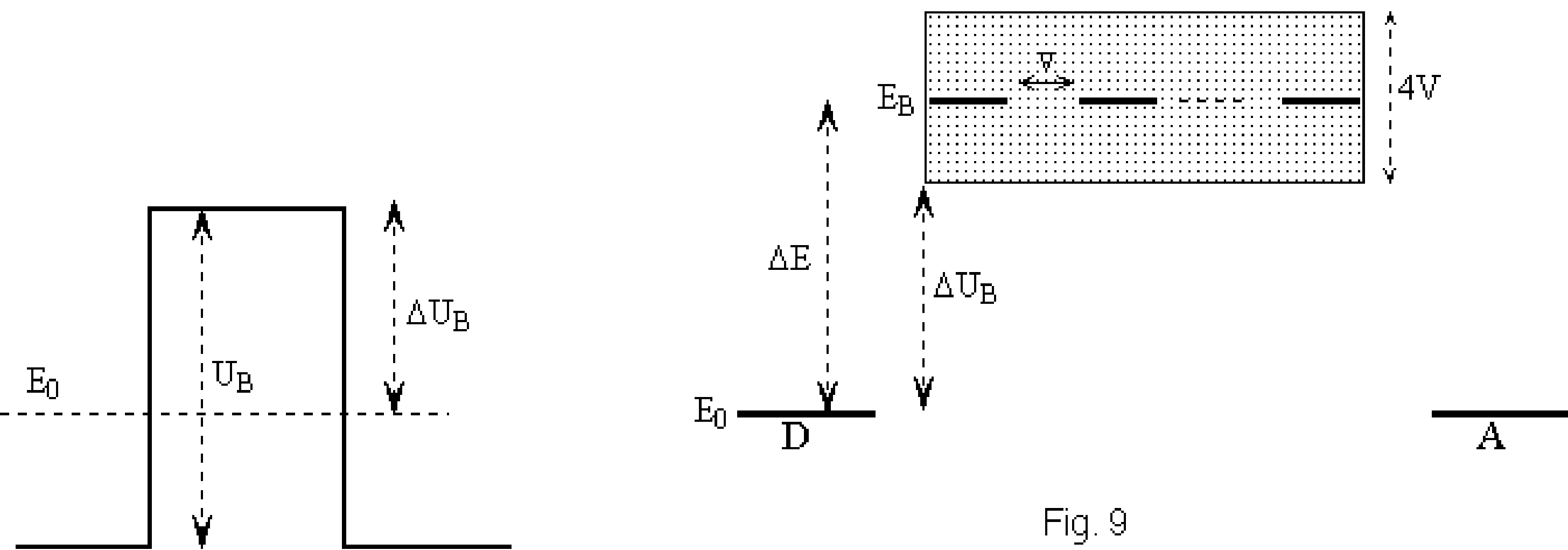

Fig. 9

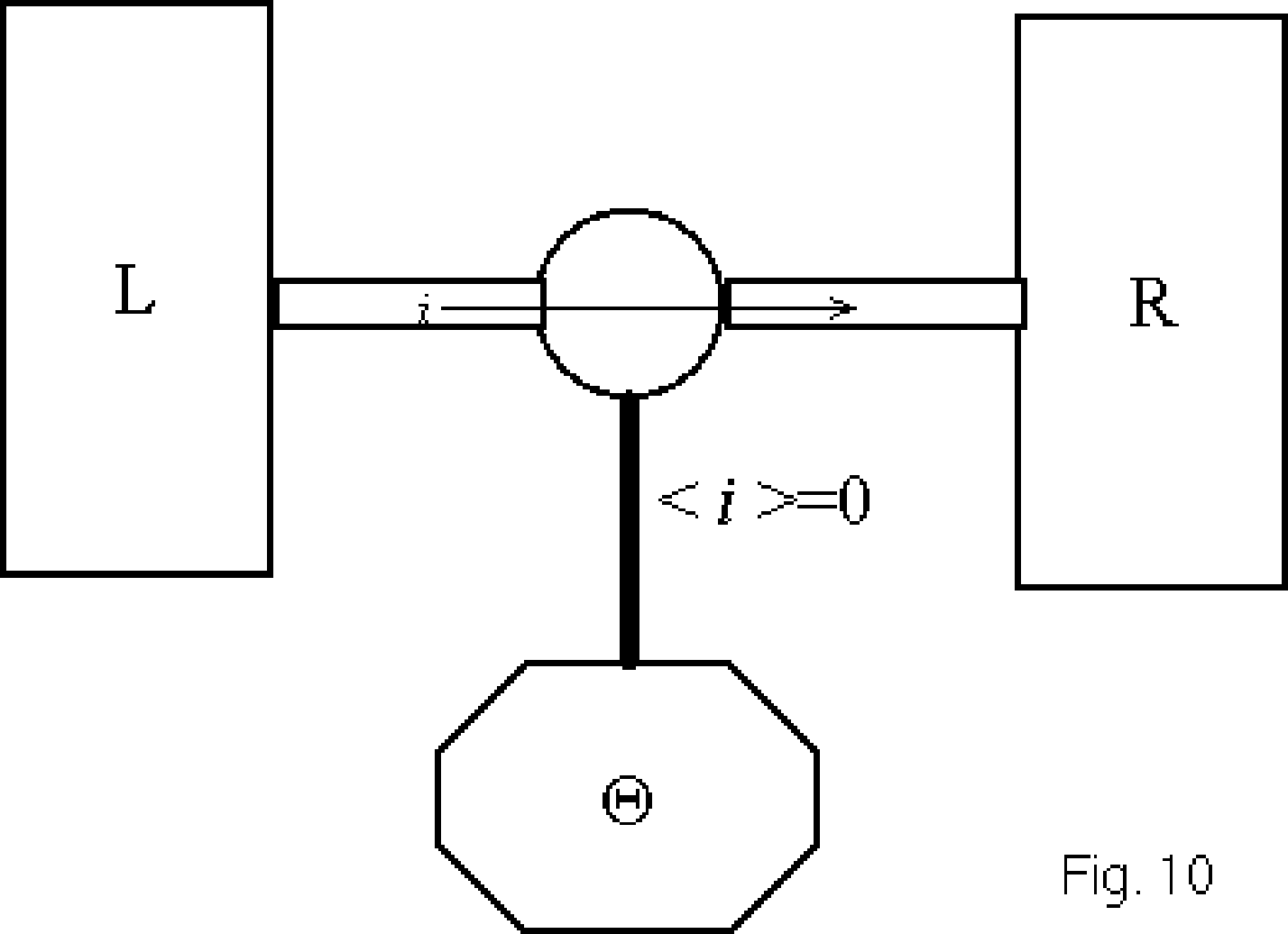

Fig. 10

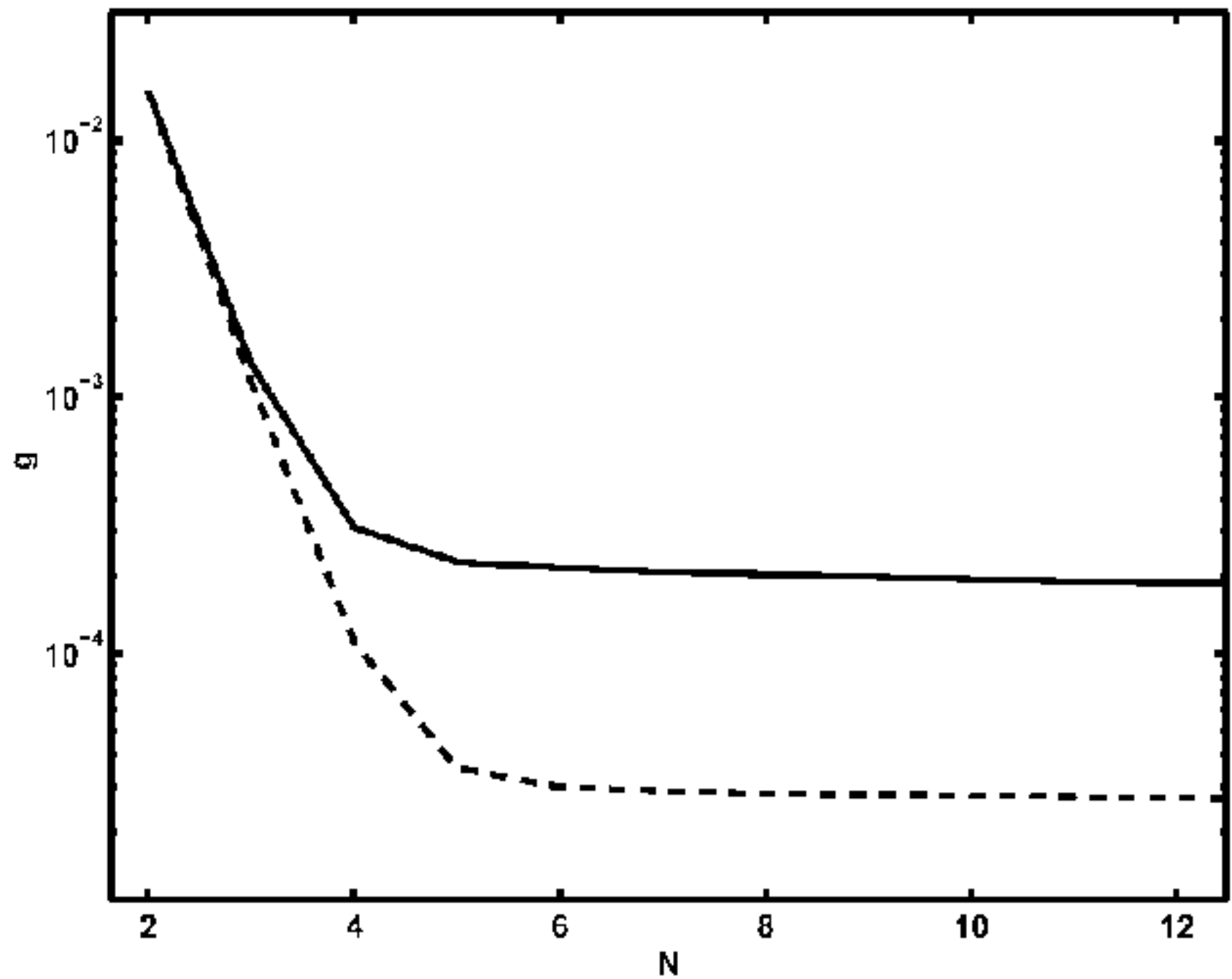

Fig. 11

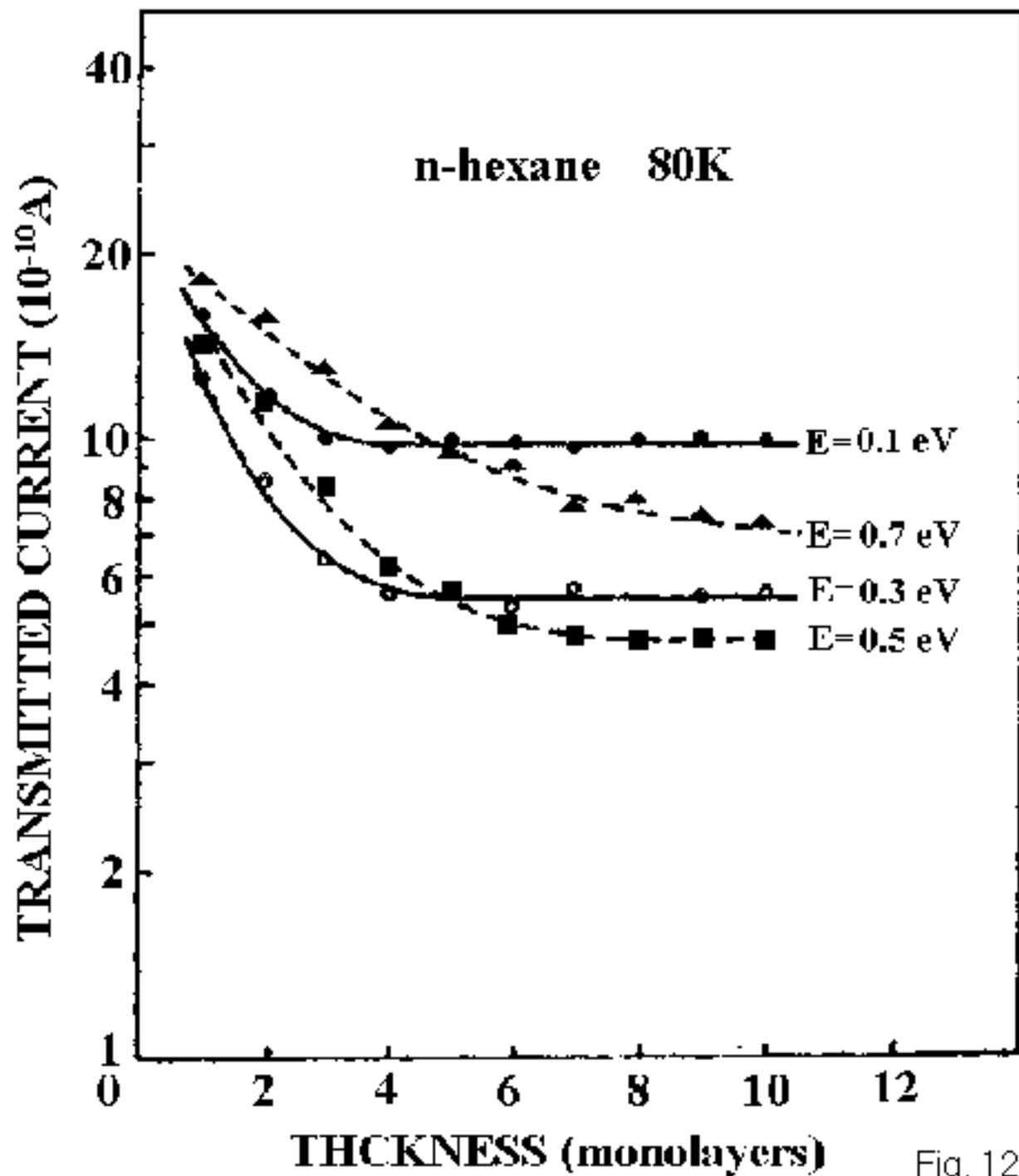

Fig. 12

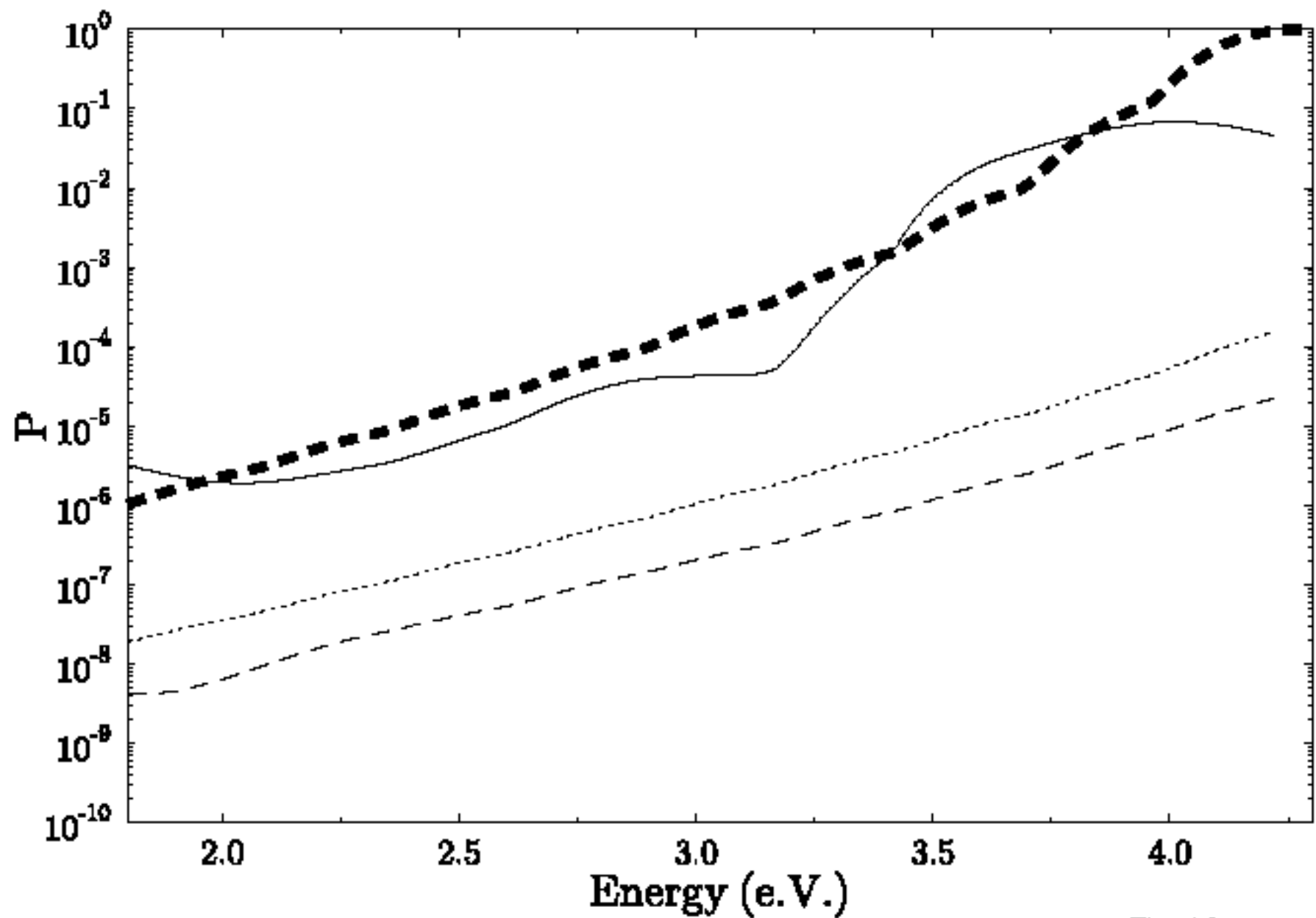

Fig. 13

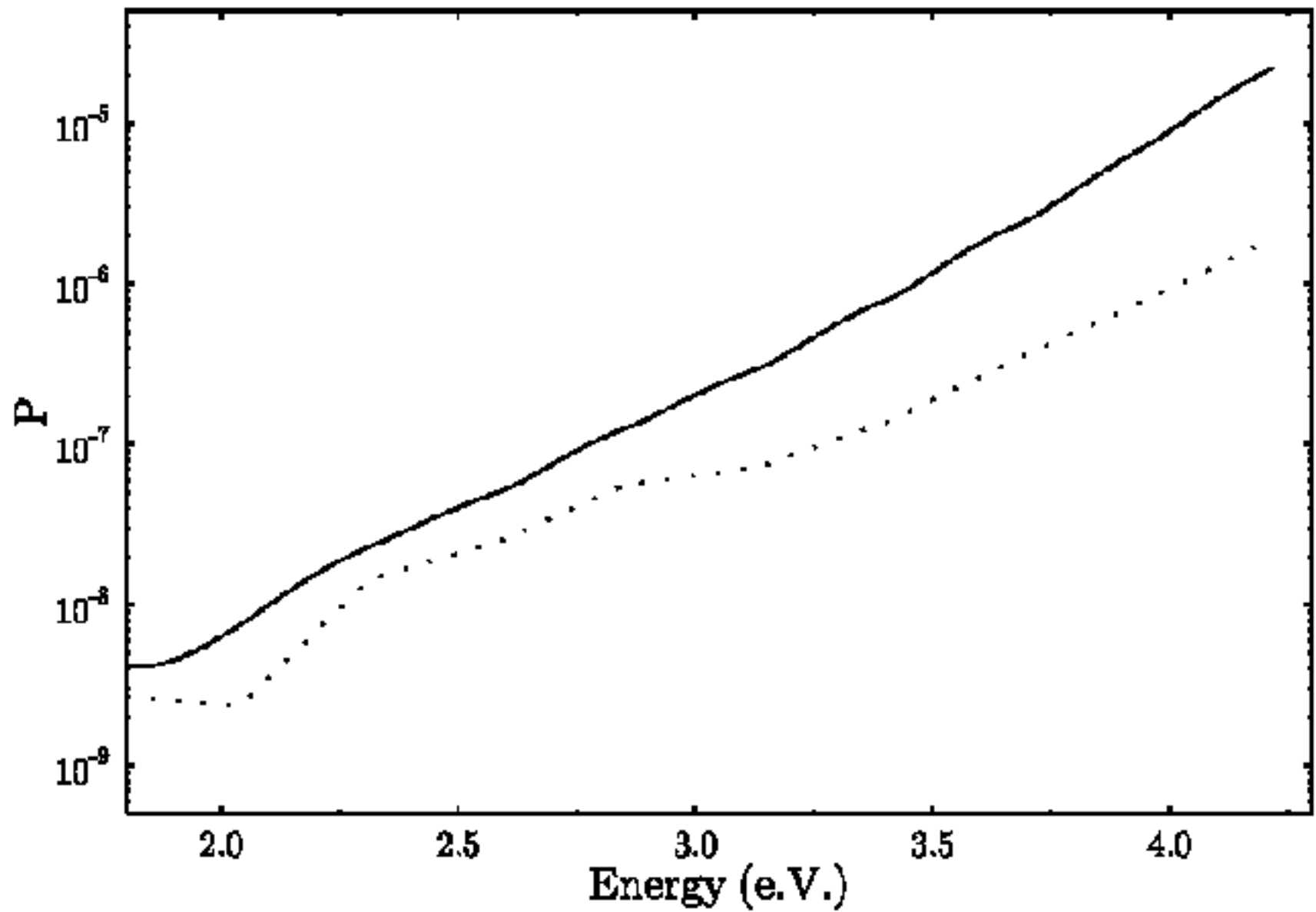

Fig. 14

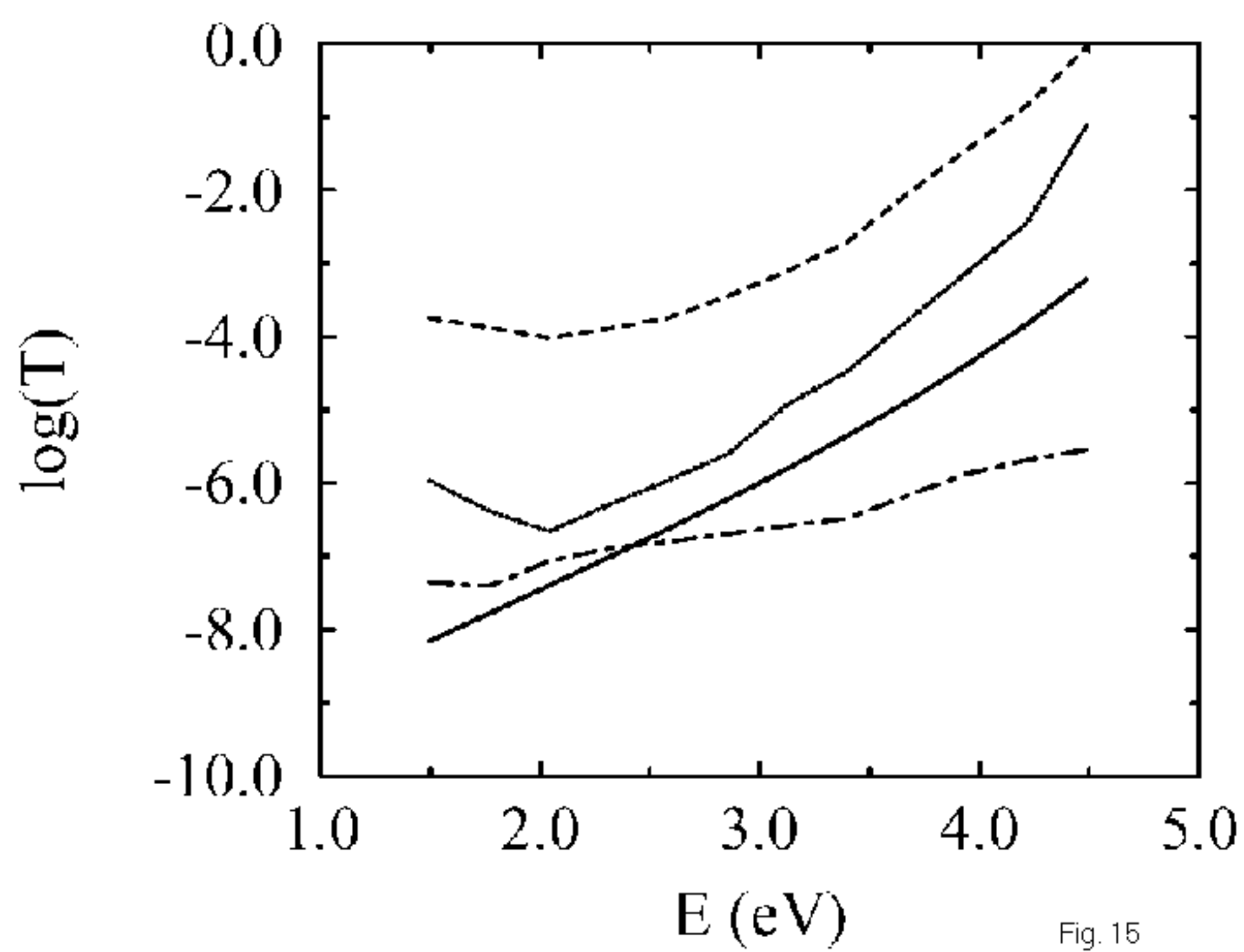

Fig. 15

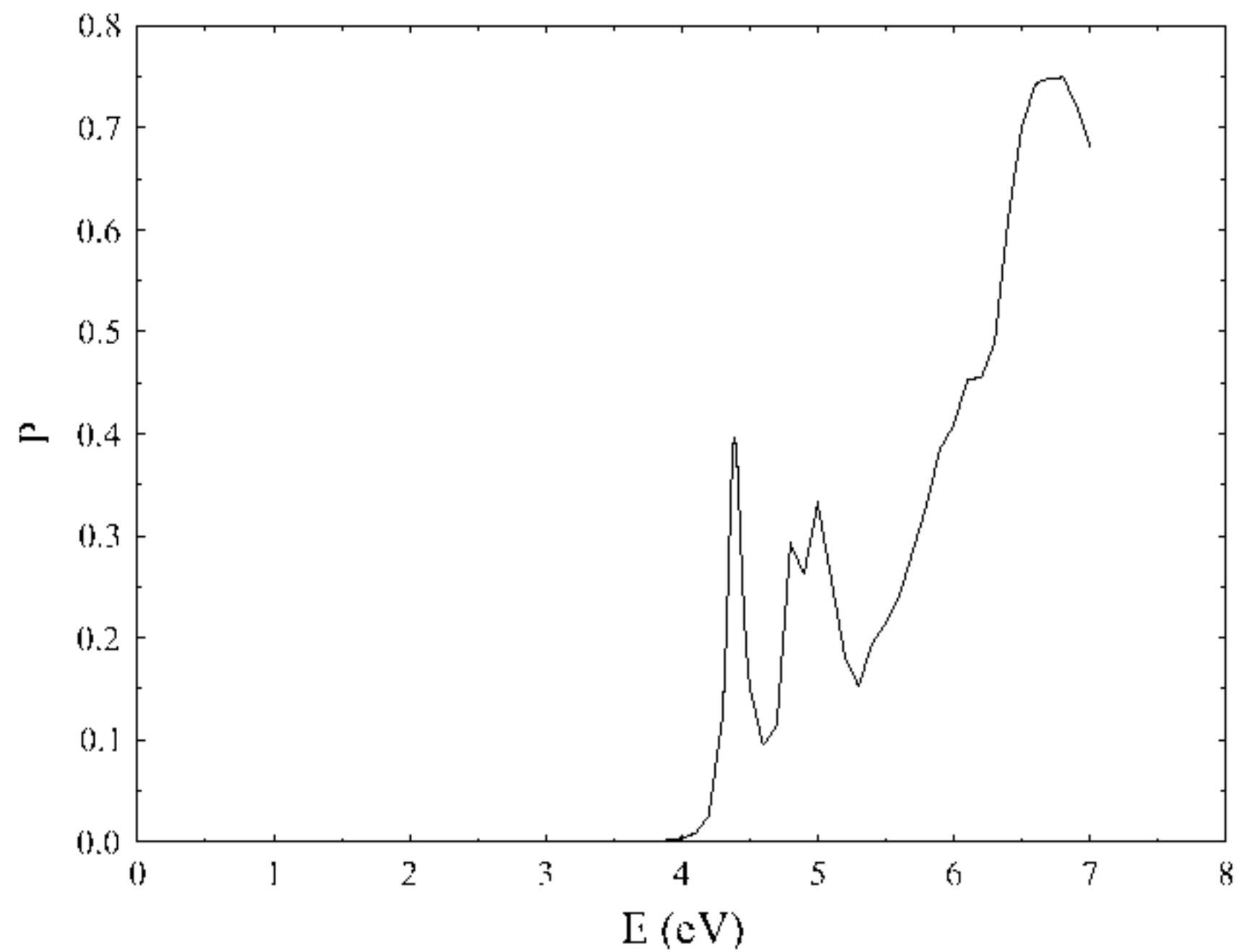

Fig. 16

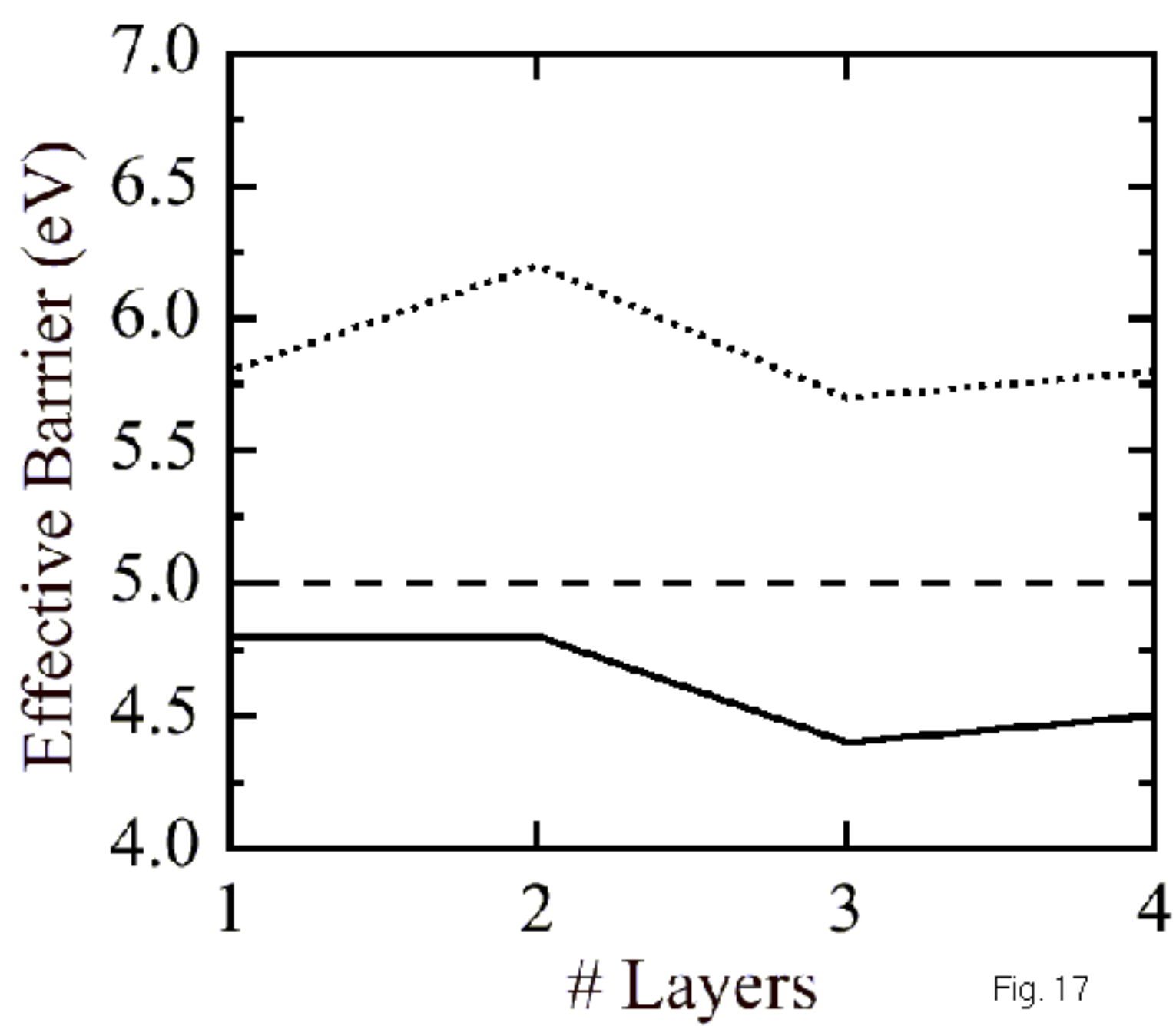

Fig. 17

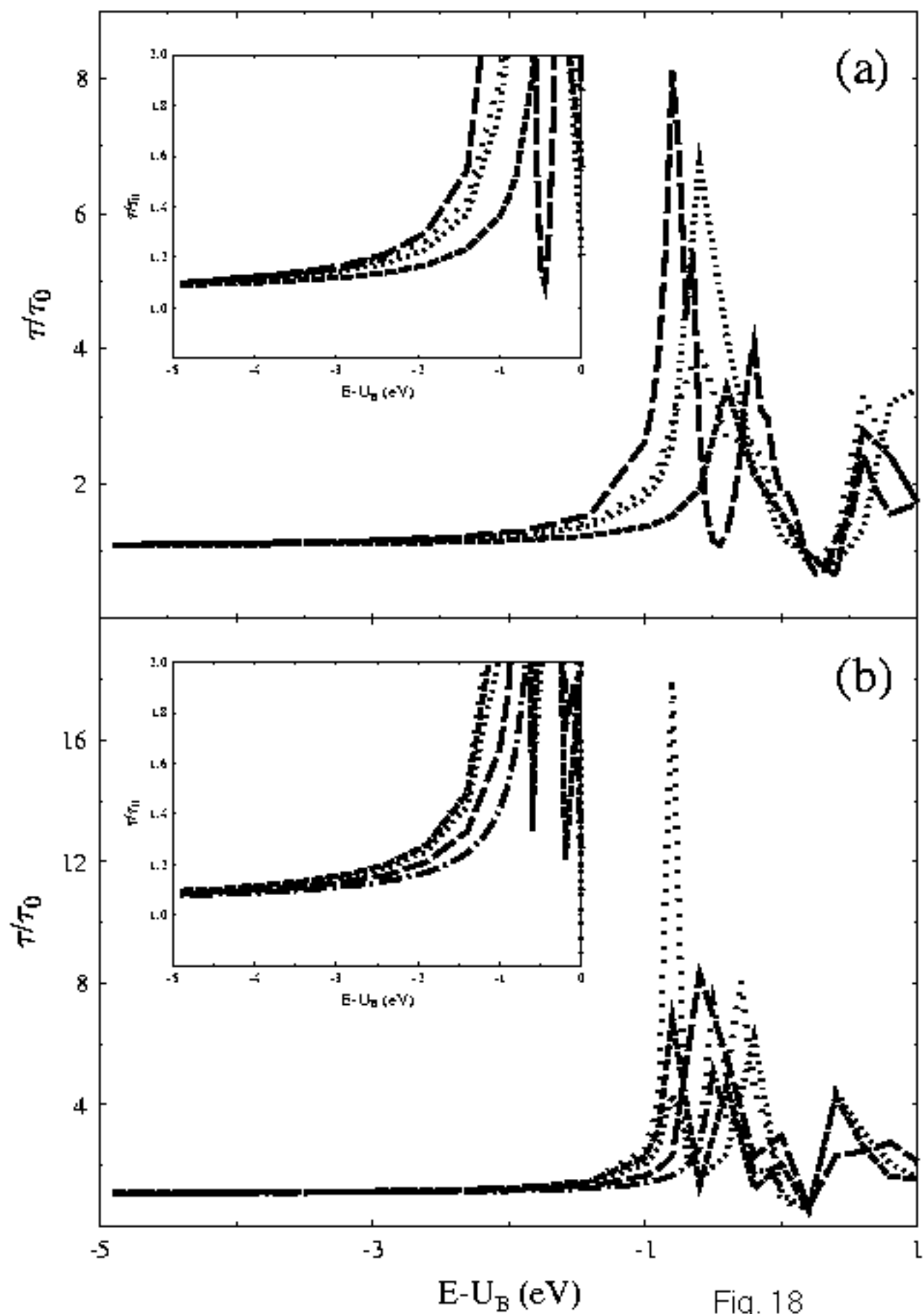

Fig. 18

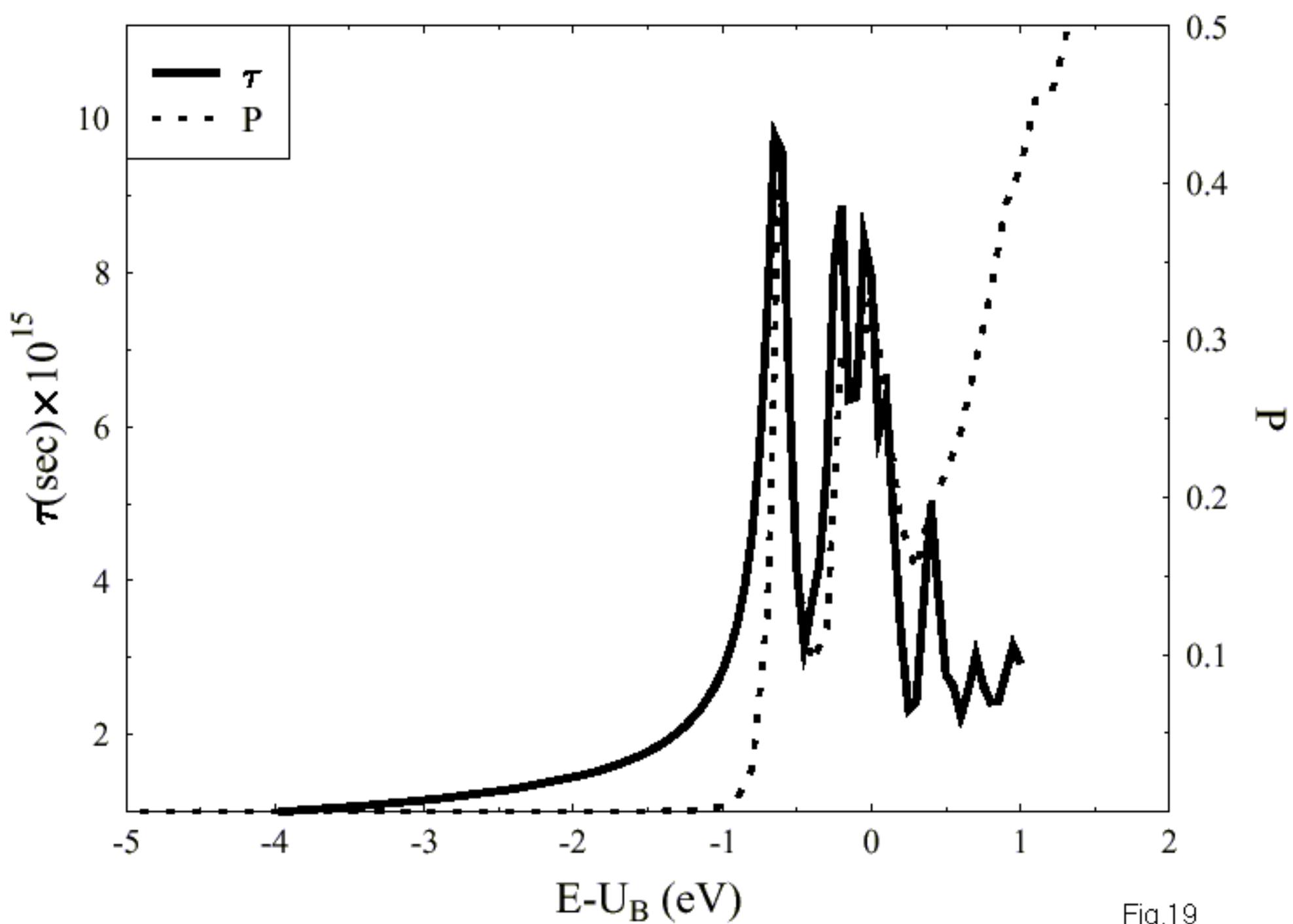

Fig.19

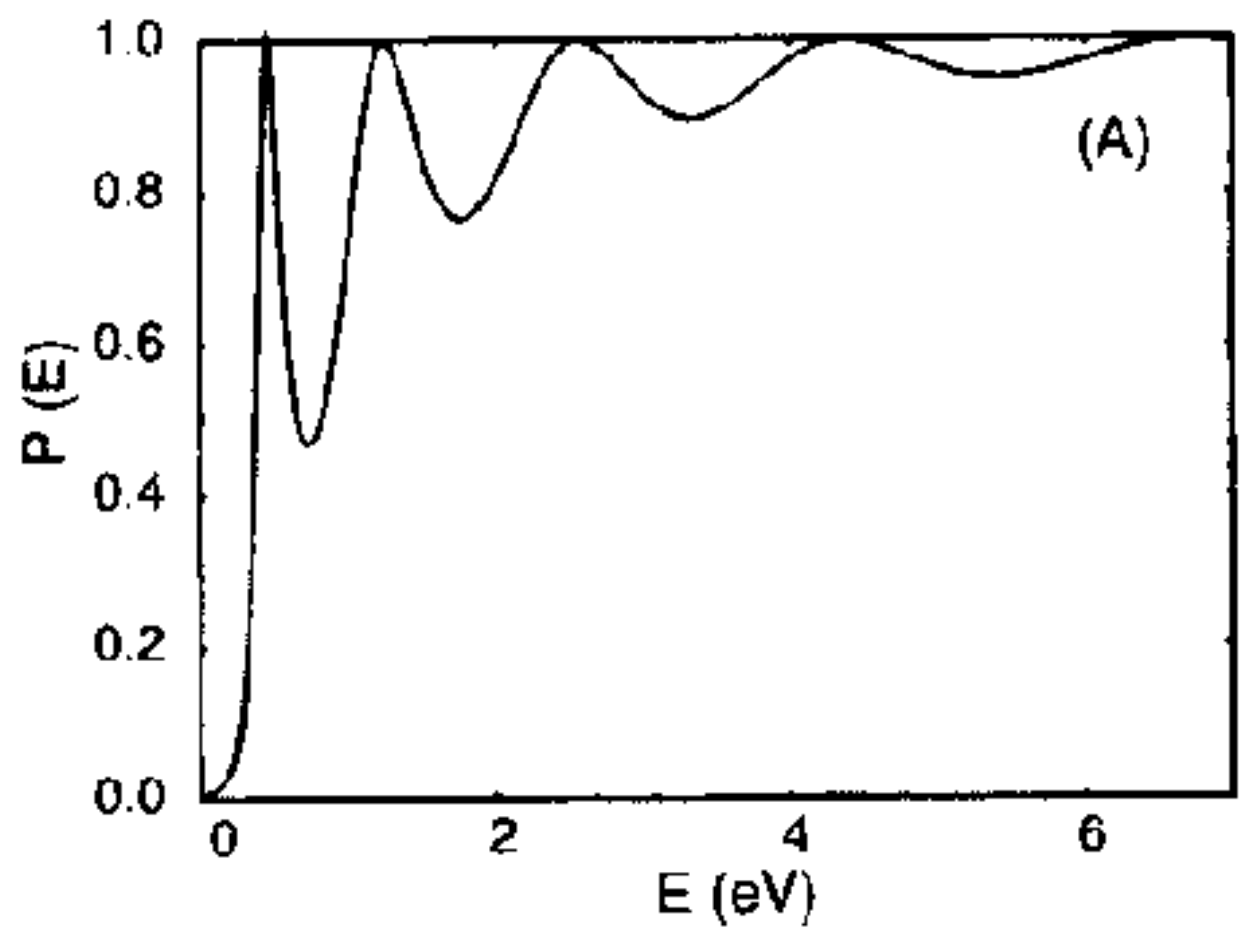
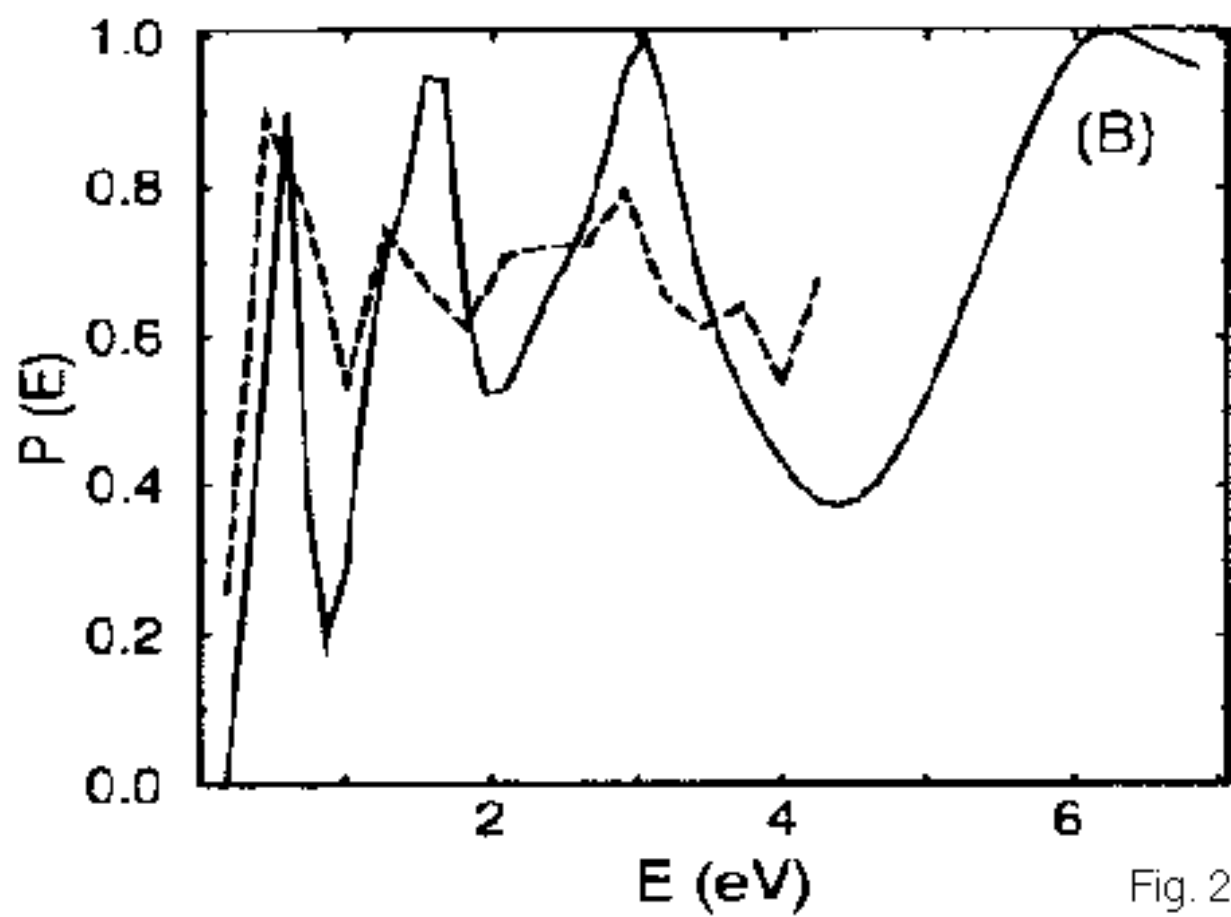

Fig. 20

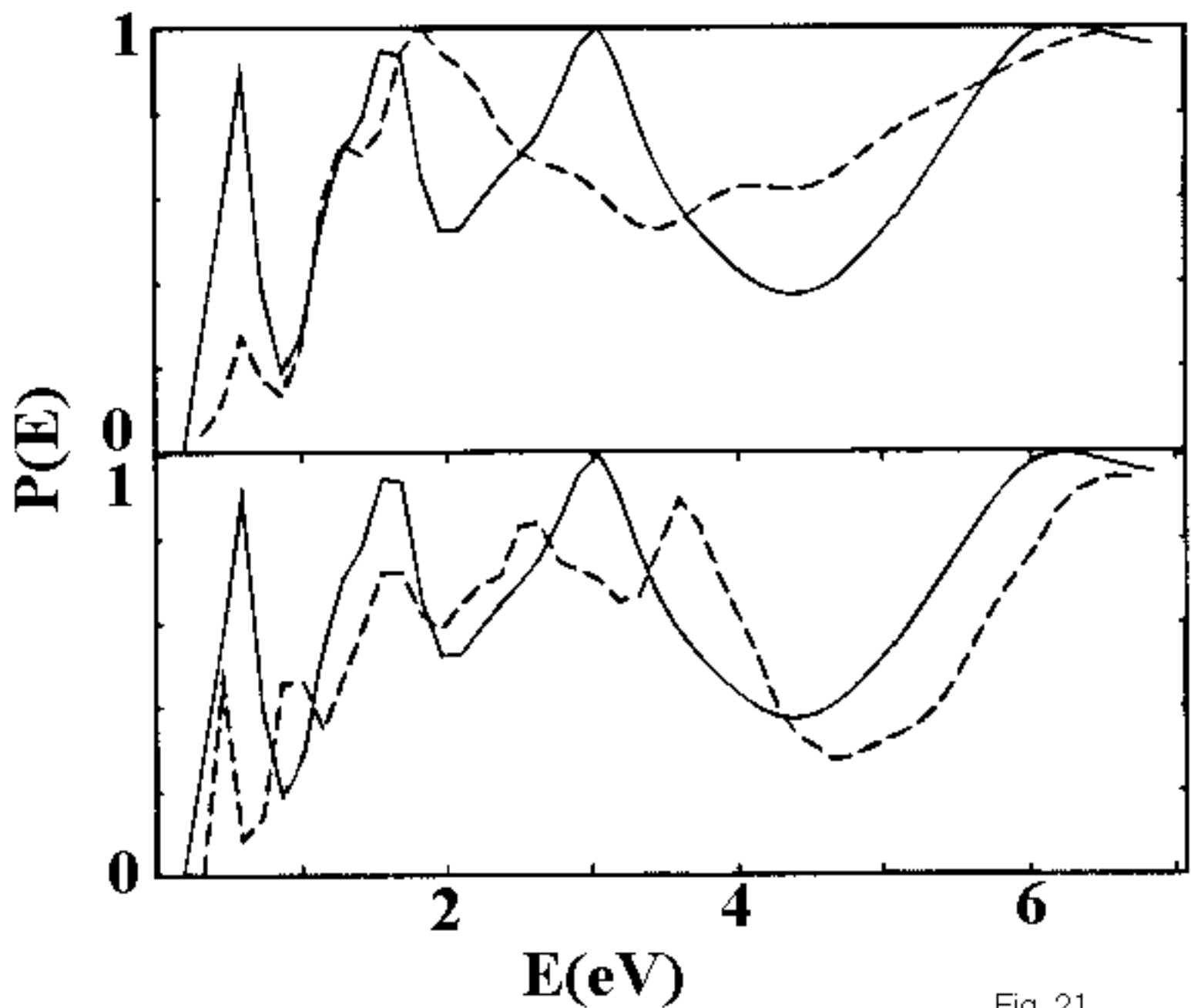

Fig. 21